\documentclass[aps,prc,twocolumn,showpacs,amsmath,amssymb,nofootinbib]{revtex4-1}
\usepackage{graphicx}
\usepackage{color}
\usepackage{times}
\usepackage{amsmath}
\usepackage[normalem]{ulem}
\usepackage{inputenc}
\usepackage{bm}
\usepackage{multirow}
\usepackage{float}
\usepackage{url}
\usepackage{natbib}
\usepackage[colorlinks=true,citecolor=blue,urlcolor=blue,linkcolor=red]{hyperref}
\usepackage{lipsum} 
\usepackage{tikz,xcolor,hyperref}

\newcommand{\apjl}{Astrophys. J. Lett.}
\newcommand{\apjs}{Astrophys. J., Supp.}
\newcommand{\aap}{Astron. Astrophys.}

\newcommand{\mnras}{Mon. Not. R. Astron. Soc.}

\newcommand{\nphysa}{Nuclear Physics A}
\newcommand{\physrep}{Physics Reports}

\newcommand{\Msun}{\,{M}_{\odot}}
\def\fm3{\;\text{fm}^{-3}}

\definecolor{lime}{HTML}{A6CE39}
\DeclareRobustCommand{\orcidicon}{%
	\begin{tikzpicture}
	\draw[lime, fill=lime] (0,0) 
	circle [radius=0.17] 
	node[white] {{\fontfamily{qag}\selectfont \tiny ID}};
	\draw[white, fill=white] (-0.0625,0.095) 
	circle [radius=0.008];
	\end{tikzpicture}
	\hspace{-2mm}
}
\foreach \x in {A, ..., Z}{
	\expandafter\xdef\csname orcid\x\endcsname{\noexpand\href{https://orcid.org/\csname orcidauthor\x\endcsname}{\noexpand\orcidicon}}
}

\begin{document}

\title{Equation of state of nuclear matter and neutron stars: Quark mean-field model versus relativistic mean-field model}

\author{Zhenyu Zhu{\orcidA{}}$^{1}$}
\email{zhenyu.zhu@sjtu.edu.cn}
\author{Ang Li{\orcidB{}}$^{2}$}
\email{liang@xmu.edu.cn}
\author{Jinniu Hu{\orcidC{}}$^{3}$}
\email{hujinniu@nankai.edu.cn}
\author{Hong Shen{\orcidD{}}$^{3}$}
\email{songtc@nankai.edu.cn}
\affiliation{\it $^{1}$ Tsung-Dao Lee Institute, Shanghai Jiao Tong University, Shanghai 201210, China;\\
$^{2}$ Department of Astronomy, Xiamen University, Xiamen 361005, China;\\
$^{3}$ School of Physics, Nankai University, Tianjin 300071, China
}


\begin{abstract}
The equation of state of neutron-rich nuclear matter is of interest to both nuclear physics and astrophysics. We have demonstrated the consistency between laboratory and astrophysical nuclear matter in neutron stars by considering low-density nuclear physics constraints (from $^{208}$Pb neutron-skin thickness) and high-density astrophysical constraints (from neutron star global properties). We have used both quark-level and hadron-level models, taking the quark mean-field (QMF) model and the relativistic mean-field (RMF) model as examples, respectively.
We have constrained the equation of states of neutron stars and some key nuclear matter parameters within the Bayesian statistical approach, using the first multi-messenger event GW170817/AT 2017gfo, as well as the mass-radius simultaneous measurements of PSR J0030+0451 and PSR J0740+6620 from NICER, and the neutron-skin thickness of $^{208}$Pb from both PREX-II measurement and ab initio calculations.
Our results show that, compared with the RMF model, QMF model's direct coupling of quarks with mesons and gluons leads to the evolution of the in-medium nucleon mass with the quark mass correction. This feature enables QMF model a wider range of model applicability, as shown by a slow drop of the nucleon mass with density and a large value at saturation that is jointly constrained by nuclear physics and astronomy.
\end{abstract}

\maketitle

\section{Introduction}

The equation of state (EoS) of cold isospin asymmetric nuclear matter has intrigued considerable attention due to its importance for nuclear physics and astrophysics. Because of the complexity of nonperturbative strong interaction, however, EoS is not able to be derived from its fundamental theory, i.e., quantum chromodynamics (QCD). We have to resort to establishing phenomenological nuclear many-body models by keeping the principles of strong interaction and taking into account experimental or empirical data. Among these models, the relativistic mean-field (RMF) and the quark mean-field (QMF) are two of the successful approaches for describing both nuclei~\cite{1998PhRvC..58.3749T, 2000PhRvC..61d5205S, 2002NuPhA.707..469S, 2014PTEP.2014a3D02H, 2016PhRvC..94d4308X, 2017PhRvC..96e4304H, 2017PhRvC..95e4310X} and neutron stars (NSs)~\cite{2014PhRvC..89b5802H, 2018ApJ...862...98Z, 2018PhRvC..97c5805Z, 2019PhRvC..99b5804Z,2019AIPC.2127b0010L}.

The RMF model was first proposed by Walecka~\cite{1974AnPhy..83..491W} in the scenario of nucleon-meson coupling, where the nucleons and mesons are treated as pointlike particles, and the nucleon-nucleon interaction is mediated by exchanging mesons. Since the intrinsic properties of the nucleon and the phase state of nuclear many-body system are fundamentally non-perturbative QCD problems, the in-medium nuclear interaction would unavoidably depend on the quark substructure and the confinement mechanism.

The first quark-level nuclear model was built by Guichon~\cite{1988PhLB..200..235G} in the name of quark-meson coupling (QMC). In this model, quarks are confined to a spherical bag and directly coupled with the scalar sigma meson, such that the effective quark mass varies, subsequently affects the nucleon mass. The nucleon mass as a function of the scalar field is calculated using the bag model, and the mean-field approximation, which is identical to the RMF approach that is widely applied for the nucleon-nucleon interaction.
While the QMF model inherited most of the QMC approach, it differs in that it uses a potential with a harmonic oscillator form to confine the quarks inside the nucleon, instead of the bag model. Though QMC and QMF models were previously distinguished based on their confining mechanisms~\cite{1996NuPhA.601..349G, 1998PhRvC..58.3749T, 2000PhRvC..61d5205S, 2002PhRvC..65f5206P, 2012PhLB..709..242M, 2012PhRvC..85e5802P}, recent developments~\cite{1986PhRvD..33.1925B, 1989JPhG...15..297F, 2002NuPhA.697..469B, 2013PhRvC..88a5206B, 2014PTEP.2014a3D02H, 2015PhRvC..92d5203M, 2016PhRvC..94c5805M, 2016PhRvC..94d4308X, 2017PhRvC..95e4310X, 2017PhRvC..96e4304H, 2020PhRvC.101b4303W} have evolved both models to the same form. However, in this paper, we use the name QMF and prefer it for future reference.

In both RMF and QMF models, a number of parameters denoting the coupling constants of nucleon-meson (quark-meson in QMF) or meson-meson are introduced and adjusted to fit the experimental or empirical data. The fitted data represents the properties of nuclear matter around saturation density, and thus the models usually give excellent descriptions of the nuclear properties at sub-saturation density and nuclei. However, due to the scarcity of experimental data at high density and high isospin asymmetry, relying solely on nuclear data is currently insufficient for accurate predictions on neutron star matter.
Consequently, enormous attention has been focused on the observations of neutron stars, where the high-density, high-isospin-asymmetry regions of EoS plays an important role.

One of the first breakthroughs was the precise measurement of massive pulsars~\cite{2010Natur.467.1081D,  2013Sci...340..448A, 2016ApJ...832..167F}, whose masses are $\approx 2\Msun$ and require the EoS to be stiff enough to support neutron stars with a mass larger than this value. Later in 2017, the LIGO/Virgo collaborations detected the first gravitational wave (GW) signal from a binary neutron star merger GW170817~\cite{2017PhRvL.119p1101A}. The tidal deformability of the neutron stars, which measures their ability to be deformed by the gravitational field of their companion, was imprinted into the gravitational wave waveform~\cite{2008PhRvD..77b1502F, 2008ApJ...677.1216H, 2010PhRvD..81l3016H, 2011PhRvD..83h4051V, 2018CQGra..35n5010H}. Analysis of the waveforms set an upper limit on tidal deformability and constrained the neutron star radius and EoS~\cite{2018PhRvL.121p1101A, 2018PhRvD..97h3015Z}. The observed electromagnetic counterpart of GW170817, i.e., the kilonova AT2017gfo~\cite{2017PASA...34...69A, 2017Natur.551...64A, 2017Sci...358.1556C, 2017ApJ...848L..17C, 2017ApJ...848L..29D, 2017Sci...358.1570D, 2017Sci...358.1565E, 2017SciBu..62.1433H, 2017Sci...358.1559K, 2017ApJ...850L...1L, 2017Natur.551...67P, 2018ApJ...852L..30P, 2017Sci...358.1574S, 2017Natur.551...75S, 2017ApJ...848L..27T, 2017Natur.551...71T, 2017PASJ...69..101U, 2017ApJ...848L..24V}, is powered by the decay of $r$-process nuclei and its properties are determined or correlated to binary parameters, which rely on the EoS. Several works have discussed the implications of AT2017gfo for the EoS~\cite{2017ApJ...850L..19M, 2018ApJ...852L..29R, 2019MNRAS.489L..91C, 2021MNRAS.505.1661B, 2021ApJ...906...98N, 2022ApJ...926..196H, 2023ApJ...943..163Z}. 
Recently, the NICER collaborations measured the mass-radius of two pulsars, PSR J0030+0451~\cite{2019ApJ...887L..24M, 2019ApJ...887L..21R} and PSR J0740+6620~\cite{2021ApJ...918L..28M, 2021ApJ...918L..27R}. These measurements have set strong constraints on the EoS, as it is encoded into the neutron star mass-radius relation.
Furthermore, the radii and tidal deformabilities of neutron stars are highly sensitive to the symmetry energy slope ($L_0$) at the nuclear saturation density, which can be extracted from neutron-skin thickness~\cite{2021PhRvL.126q2502A,2022NatPh..18.1196H,2022PhRvL.129d2501A}. Several studies have discussed the implications and correlations between neutron star observations and $L_0$~\cite{2005PhRvL..95l2501T, 2009PhRvL.102l2502C, 2010PhRvC..82b4321C, 2011PhRvL.106y2501R,2018PhRvL.120q2702F, 2021PhRvL.126q2503R, 2021PhRvL.127s2701E, 2022PhLB..83037098L, 2023PhRvC.107a5801M, 2023PhRvC.107d5802S, 2023PhRvC.107c5805C, 2023arXiv230314763M}.

In this paper, we undertake a systematic investigation and comparison of the performance of the nucleon-meson (RMF) and quark-meson (QMF) models. To achieve this, we adopt the same functional form of the many-body Lagrangian and maintain the same saturation properties for both models. Specifically, the coupling constants of quark-meson or nucleon-meson interactions are obtained by fitting against empirical data at the saturation density. Consequently, any differences between the RMF and QMF models arise from their respective treatments of confined quarks and nucleons.
We undertake extensive computations and comparisons of various properties of nuclear matter, including the considered experimental data on pressure and symmetry energy for both RMF and QMF models. Additionally, we perform Bayesian inference for both models and compare their respective results. To this end, we take into account several astrophysical and experimental observations as likelihood functions, including the multi-messenger event GW170817/AT2017gfo, NICER mass-radius measurements of pulsars, PREX-II experiment for the neutron-skin of $^{208}$Pb, and their predictions from ab initio calculations.
Using the Bayesian framework, we report and discuss our findings on the most preferred maximum masses, radii, and tidal deformabilities of $1.4\Msun$ stars for both RMF and QMF models.
It should be noted that our analysis does not take into account the possible non-nucleon degrees of freedom that may exist in massive neutron stars, for example, deconfined quarks, 
hyperons, and Delta isobars~\cite{2020ApJ...904..103M, 2021ApJ...913...27L, 2021MNRAS.506.5916L, 2022MNRAS.515.5071M, 2022ApJ...936...69M,  2023ApJ...942...55S, 2022ApJ...935...88H, 2023arXiv230503323Z, 2023arXiv230508401M}. The appearance of these non-nucleon degrees of freedom can considerably impact the EoS, softening it.
However, the interactions between quarks, as well as between hyperons and nucleons are subject to large uncertainties. Including these additional degrees of freedom in our analyses would complicate them considerably. Therefore, we focus on nucleons in our analyses and concentrate on the results pertaining to neutron stars.

\section{Theoretical framework}\label{sec:model}

In this section, we will introduce the widely used RMF approach~\citep{2011PrPNP..66..519N}, which is based on an effective Lagrangian incorporating meson fields to mediate strong interactions between hadrons or quarks. The latter approach is known as the QMF model~\citep{2000PhRvC..61d5205S,1998PhRvC..58.3749T}. The QMF model self-consistently relates the internal quark structure of a nucleon and hyperon to the RMFs that arise in nuclear and hyperonic matter, respectively. It has been employed extensively in calculations of both finite (hyperon-)nuclei and infinite dense matter~\citep{2000PhRvC..61d5205S, 2002NuPhA.707..469S, 2019PhRvC..99b5804Z,
2014PTEP.2014a3D02H, 2014PhRvC..89b5802H,
 2017PhRvC..96e4304H, 2016PhRvC..94d4308X, 2017PhRvC..95e4310X,
 2018PhRvC..97c5805Z, 2018ApJ...862...98Z, 2019PhRvC..99b5804Z,
 2019AIPC.2127b0010L, 2020ApJ...904..103M}.

\subsection{The RMF model}\label{sec:RMF}

For describing nuclear matter, we consider the $\sigma,~\omega$ and
$\rho$ mesons exchanging in the RMF Lagrangian:
\begin{eqnarray}
  \label{eq:lagrangian}
\mathcal{L}& = & \overline{\psi}\left(i\gamma_\mu \partial^\mu - M_N^\ast - g_{\omega_N}\omega\gamma^0 - g_{\rho_N}\rho\tau_{3}\gamma^0\right)\psi  \nonumber \\
           && -\frac{1}{2}(\nabla\sigma)^2 - \frac{1}{2}m_\sigma^2 \sigma^2 - \frac{1}{3} g_2\sigma^3 - \frac{1}{4}g_3\sigma^4 \nonumber \\
           && + \frac{1}{2}(\nabla\omega)^2 + \frac{1}{2}m_\omega^2\omega^2 + \frac{1}{2}g_{\omega N}^2\omega^2 \Lambda_v g_{\rho N}^2\rho^2 \nonumber \\
           & & + \frac{1}{2}(\nabla\rho)^2 + \frac{1}{2}m_\rho^2\rho^2\ ,
\label{eq:L}
\end{eqnarray}
where $M_N^\ast=M_N-g_{\sigma_N} \sigma$ is the effective nucleon
mass, and the nucleon mass in free space $M_N=939$ MeV is adopted. 
In the nuclear medium, the nucleon mass drops from its coupling
to the in-medium-modified chiral condensate.
The $\sigma$, $\omega$, $\rho$, and $\psi$ denote the $\sigma$, $\omega$, $\rho$ meson, and nucelon field operators, respectively. 
$m_{\sigma} = 510~\rm{MeV}$,~$m_{\omega}=783~\rm{MeV}$, and $m_{\rho}=770~\rm{MeV}$ are the meson masses.
The $g_{\sigma_N}$, $g_{\omega_N}$ and $g_{\rho_N}$ are the nucleon-meson coupling constants for $\omega$ and $\rho$ mesons. 
There are six parameters
($g_{\sigma_N}, g_{\omega_N}, g_{\rho_N}, g_2, g_3, \Lambda_v$) in the
Lagrangian. 
The last parameter $\Lambda_v$, which represents the
coupling constant of $\omega$-$\rho$ coupling, is introduced to reduce
the symmetry energy slope (see e.g., Refs.~\cite{2008PhR...464..113L,
  2018ApJ...862...98Z, 2019PhRvC..99b5804Z}).
The EoS for a neutron star is determined by these six parameters, which may be obtained through the process of fitting empirical data at saturation density.

The equation of motion for each meson could be obtained after the
variation of the Lagrangian and applying the mean-field approximation:
\begin{eqnarray}
  \label{eq:eqs_mot1}
  m_\sigma^2 \sigma + g_2 \sigma^2 + g_3 \sigma^3 & = & g_{\sigma_N} n_{_{\rm S}}\ , \\
  \label{eq:eqs_mot2}
  (m_\omega^2 + \Lambda_v g_\omega^2 g_\rho^2 \rho^2) \omega & = & g_{\omega_N} n_{_{\rm B}}\ , \\
  \label{eq:eqs_mot3}
  (m_\rho^2 + \Lambda_v g_\omega^2 g_\rho^2 \omega^2) \rho & = & g_{\rho_N} n_{3}\ ,
\end{eqnarray}
where
\begin{eqnarray}
  n_{_{\rm S}} & = & \sum_{i=n,p}\frac{1}{\pi^2} \int_0^{p_{\rm F_{i}}} \frac{M_{\rm N}^\ast}{E_{\rm F_{i}}} p_{\rm F_{i}}^2 dp_{\rm F} \\
  n_{_{\rm B}} & = & n_{\rm p} + n_{\rm n} = \sum_{i=n,p} \frac{p_{\rm F_{i}}^3}{3\pi^2} \\
  n_{3} & = & n_{\rm p} - n_{\rm n} = \frac{p_{\rm F_{p}}^3}{3\pi^2} - \frac{p_{\rm F_{n}}^3}{3\pi^2} \\
\end{eqnarray}
is the scalar density, vector (baryonic) density and isovector density, respectively. $p_{\rm F_{i}}$ and
$E_{\rm F_{i}}=\sqrt{M_N^{\ast 2}+p_{\rm F_{i}}^2}$ denote the Fermi momentum
and Fermi energy, respectively. The number density of proton and
neutron are represented by $n_{\rm p}$ and $n_{\rm n}$,
respectively. Similarly, the equation of motion for the single nucleon
is yielded by varying $\overline{\psi}$:
\begin{eqnarray}
  \label{eq:eqs_mot_n}
  [\vec{\gamma}\cdot\vec{p} + (M_N + U_{\rm S}) - \gamma^0(\epsilon - U_{\rm V})]\psi=0,
\end{eqnarray}
where $U_{\rm S}=M_N^\ast - M_N$ and
$U_{\rm V}=g_{\omega_N}\omega + g_{\rho_N}\rho \tau_3$ denote the
scalar and vector potential of an in-medium nucleon, respectively, and $\epsilon$ is the single
nucleon energy.

These equations of motion will be solved with the
$\beta$-equilibrium and charge-neutrality conditions simultaneously for the study of neutron stars. Once the meson fields are known after solving Eqs.~(\ref{eq:eqs_mot1})--(\ref{eq:eqs_mot3}), the energy density and pressure contributed from nucleons can be computed by:
\begin{eqnarray}
  \label{eq:ener}
  e_{_{\rm N}} & = & \sum_{i=n,p} e_{\rm kin}^i + \frac{1}{2} m_{\sigma}^2 \sigma^2 + \frac{1}{3} g_2 \sigma^3 + \frac{1}{4}g_3 \sigma^4 \nonumber \\
    & & - \frac{1}{2} m_\omega^2 \omega^2 - \frac{1}{2} m_\rho^2 \rho^2 - \frac{1}{2} \Lambda_v (g_{\omega_N} g_{\rho_N} \omega \rho)^2 \nonumber \\
    & & + g_{\omega_N} \omega (n_{\rm n} + n_{\rm p}) + g_{\rho_N} \rho (n_{\rm p} - n_{\rm n})\ , \\
  \label{eq:press}
  p_{_{\rm N}} & = & \sum_{i=n,p} p_{\rm kin}^i - \frac{1}{2} m_{\sigma}^2 \sigma^2 - \frac{1}{3} g_2 \sigma^3 - \frac{1}{4}g_3 \sigma^4 \nonumber \\
  & & + \frac{1}{2} m_\omega^2 \omega^2 + \frac{1}{2} m_\rho^2 \rho^2 + \frac{1}{2} \Lambda_v (g_{\omega_N} g_{\rho_N} \omega \rho)^2\ ,
\end{eqnarray}
where
\begin{eqnarray}
  \label{eq:ekin}
  e_{\rm kin} & = & \frac{1}{8\pi^2} \biggl[2p_{\rm F}^3E_{\rm F} + M_{\rm N}^{\ast 2} p_{\rm F} E_{\rm F} \nonumber \\
              & & \quad\quad\ - M_{\rm N}^{\ast 4} \ln\left(\frac{p_{\rm F}+E_{\rm F}}{M_{\rm N}^\ast}\right)\biggr]\ , \\
  p_{\rm kin} & = & \frac{1}{24\pi^2} \biggl[(2p_{\rm F}^3 -3 M_{\rm N}^{\ast 2} p_{\rm F}) E_{\rm F} \nonumber \\
              & & \quad\quad\ \ \ + 3M_{\rm N}^{\ast 4} \ln\left(\frac{p_{\rm F}+E_{\rm F}}{M_{\rm N}^\ast}\right)\biggr]\ ,
\end{eqnarray}
are the kinetic contributions to the energy and pressure, respectively.

\subsection{The QMF model}\label{sec:QMF}

In the 1988 article, \citet{1988PhLB..200..235G} developed a novel model for nuclear matter to account for the changes of nucleon properties in nuclear matter, specifically those related to the European Muon Collaboration (EMC) effects. The model is similar to the RMF model, but instead of coupling the scalar and vector meson fields with the nucleons, they couple directly with the quarks. The model uses a potential model~\citep{1978PhRvD..18.4187I} for the nucleon, and quarks are confined by a phenomenological confinement potential, typically in the form of a polynomial. A harmonic oscillator potential is generally adopted, facilitating analytic solutions to the Dirac equation.
The Dirac equation of the confined quarks is written as:
\begin{eqnarray}
&& [\gamma^{0}(\epsilon_{q}-g_{\omega}^q\omega-\tau_{3}^qg_{\rho}^q\rho) \nonumber \\
&&
-\vec{\gamma}\cdot\vec{p} -(m_{q}-g_{\sigma}^q\sigma)-U(r)]\psi_{q}(\vec{r})=0\ ,
\end{eqnarray}
where the scalar-vector form of the Dirac structure is chosen for the quark confinement
potential.
We adopt the constitute quark mass as the value of $m_q$, which is
$m_q=300$ MeV.
$\psi_{q}(\vec{r})$ is the quark field, $\sigma$, $\omega$, and
$\rho$ are again the classical meson fields. $g_{\sigma}^q$, $g_{\omega}^q$,
and $g_{\rho}^q$ are the coupling constants of $\sigma, ~\omega$, and
$\rho$ mesons with quarks, respectively. $\tau_{3q}$ is the third
component of the isospin matrix. The $U(r)$ represents the harmonic oscillator potential, which has a form of:
\begin{eqnarray}
U(r)=\frac{1}{2}(1+\gamma^0)(ar^2+V_0)\ .
\end{eqnarray}
The potential constants $a$ and $V_0$ will be determined in the
following steps by reproducing the nucleon in-vacuum properties (mass and
radius).
Here, we have made the assumption that constituent quarks move within a nonperturbative vacuum. However, it is possible that the confinement mechanism generates a small region where quarks are most likely to be present, forming a ``bag" in which chiral symmetry is restored. For a detailed discussion on how both the potential and bag models can be used to describe the interaction of constituent quarks in nucleons, please refer to Ref.~\cite{2019PhRvC..99b5804Z}.

This Dirac equation can be solved exactly and its ground state
solution for energy is:
\begin{eqnarray}
  \label{eq:qmf_e}
(\mathop{\epsilon'_q-m'_q})\sqrt{\frac{\lambda_q}{a}}=3\ ,
\end{eqnarray}
where
$\lambda_q=\epsilon_q^\ast+m_q^\ast,\
\mathop{\epsilon'_q}=\epsilon_q^\ast-V_0/2,\
\mathop{m'_q}=m_q^\ast+V_0/2$. The effective single quark energy is
given by
$\epsilon_q^*=\epsilon_{q}-g_{\omega}^q\omega-\tau_{3q}g_{\rho}^q\rho$
and the effective quark mass by $m_q^\ast  = m_q-\delta m_q$, with the quark mass reduction defined as $\delta m_q = g_{\sigma}^q\sigma$.
It is worth mentioning that, within QMF model, the mass of the $\sigma$ meson ($m_{\sigma} = 510~\rm{MeV}$) is chosen~\cite{2000PhRvC..61d5205S} to reproduce the charge radius of $^{40}\text{Ca}$ to be around $3.45\;\text{fm}$. For both RMF and QMF models, it determines the range of the attractive interaction, such that decreasing $m_{\sigma}$ results in a reduction of $g_{\sigma_N}$.

The solution for the wave function is:
\begin{eqnarray}
\Psi(r,\theta,\phi)=\frac{1}{r} \left(
\begin{array}{c}
F(r)Y_{1/2 m}^0(\theta,\phi) \\
iG(r)Y_{1/2 m}^1(\theta,\phi)
\end{array}
\right)\ ,
\end{eqnarray}
where
\begin{eqnarray}
F(r)=\mathcal{N}\left(\frac{r}{r_{0}}\right)\exp(-r^2/2r_0^2)\ ,  \\
G(r)=-\frac{\mathcal{N}}{\lambda_qr_0}\left(\frac{r}{r_{0}}\right)^2\exp(-r^2/2r_0^2)\ ,
\end{eqnarray}
\begin{eqnarray}
r_0=(a\lambda_q)^{-1/4},\ \ \ \mathcal{N}^2=\frac{8\lambda_q}{\sqrt{\pi}r_0}\frac{1}{3\mathop{\epsilon'_q}+\mathop{m'_q}}\ .
\end{eqnarray}
The radius of the nucleon in the ground state could be calculated with the
wave function by:
\begin{eqnarray}
\langle r_N^2\rangle = \frac{\mathop{11\epsilon'_q + m'_q}}{\mathop{(3\epsilon'_q + m'_q)(\epsilon'^2_q-m'^2_q)}}\ .
\end{eqnarray}

The energy obtained by solving Eq.~(\ref{eq:qmf_e}), e.g., $E_N^0=\sum_q\epsilon_q^\ast$, represents only the zeroth order and further corrections are required to accurately represent the nucleon properties. In the QMF model, corrections such as the center-of-mass correction, pionic correction, and gluonic correction are typically taken into account.

The center-of-mass correction is expressed as:
\begin{eqnarray}
\epsilon_{\rm c.m.}=\frac{\mathop{77\epsilon'_q + 31m'_q}}{3(\mathop{3\epsilon'_q + m'_q})^2r_0^2}\ .
\end{eqnarray}
This correction ensures that the three constituent quarks move independently in the confining potential. 
The pion correction arises due to the chiral symmetry of QCD theory, whereas the gluon correction is a result of the short-range exchange interaction between quarks.
The pionic correction has the form of
\begin{eqnarray}
\delta M_N^\pi=-\frac{171}{25}I_\pi f^2_{NN\pi}\ ,
\end{eqnarray}
where
\begin{eqnarray}
I_\pi=\frac{1}{\pi m_\pi^2}\int_0^{\infty}dk\frac{k^4u^2(k)}{k^2+m_\pi^2}\ , \nonumber
\end{eqnarray}
\begin{eqnarray}
u(k)=\left[1-\frac{3}{2}\frac{k^2}{\lambda_q(\mathop{5\epsilon'_q + 7m'_q})}\right]\exp\left(-\frac{1}{4}r_0^2k^2\right)\ .\nonumber
\end{eqnarray}
The constants $m_\pi=140$ MeV and $f_\pi=93$ MeV are the mass of $\pi$
meson and the phenomenological pion decay constant, respectively. For
gluonic correction, it reads as:
\begin{eqnarray}
(\Delta E_N)_g=-\alpha_c\left(\frac{256}{3\sqrt{\pi}}\frac{1}{R_{uu}^3}\frac{1}{(\mathop{3\epsilon'_q + m'_q})^2}\right),
\end{eqnarray}
where
\begin{eqnarray}
R_{uu}^2=\frac{6}{\mathop{\epsilon'^2_q - m'^2_q}} \ ,\nonumber
\end{eqnarray}
and $\alpha_c=0.58$ is a constant.

Finally, the effective nucleon mass is yielded by combining all the
contributions of energy, and it is written as:
\begin{eqnarray}
  \label{eq:qmf_meff}
M^\ast_N=E^{0}_N-\epsilon_{\rm c.m.}+\delta M_N^\pi+(\Delta E_N)_g\ .
\end{eqnarray}
Note that in the quark-level model of QMF, unlike the hadron-level model of RMF, the nucleon properties vary based on the strength of the mean fields, and this variation is exclusively expressed through $M^\ast_N$, which is associated with the $\sigma$ mean field. Meanwhile, the $\omega$ and $\rho$ mean fields do not induce any change in the nucleon properties.

Once the expressions of the nucleon mass and radius are known, the potential parameters ($a$ and $V_0$) can be determined by reproducing the nucleon mass and radius in free space, i.e., $M_N = 939$ MeV and $r_N = 0.87$ fm. Once these parameters are obtained, the relation between the effective nucleon mass $M^\ast_N$ and the strength of the $\sigma$ field can be determined.

The QMF Lagrangian and the equations of motion of the many-body system have the same form as Eqs.~(\ref{eq:lagrangian})--(\ref{eq:eqs_mot3}) if we convert $g_{\omega}^q$ and $g_{\rho}^q$ to $g_{\omega_N}$ and $g_{\rho_N}$ using the quark counting rule, i.e., $g_{\omega_N}=3g_{\omega}^q$ and $g_{\rho_N}=g_{\rho}^q$, and define $g_{\sigma_N}$ in QMF model as $g_{\sigma_N}=-\partial M_{N}^\ast / \partial \sigma$. Using this Lagrangian, the energy density and pressure can be obtained, and their expressions are identical to Eqs.~(\ref{eq:ener})--(\ref{eq:press}).

\subsection{The coupling constants}
\begin{table}   \vspace{-0.2cm}
\begin{center}
  \caption{Saturation properties uniformly chosen in the empirical ranges and used in this study for the fitting of RMF/QMF meson coupling parameters and the Bayesian inference of the (binary) neutron star observations: The saturation density $n_0$ (in fm$^{-3}$) and the corresponding values at saturation point for the binding energy $E/A$ (in MeV), the incompressibility
    $K_0$ (in MeV), the symmetry energy $J_0$ (in MeV), the symmetry energy slope $L_0$ (in MeV) and the ratio between the effective mass and free nucleon mass $M_N^\ast/M_N$.
    }\label{tab:sat}
\setlength{\tabcolsep}{3pt}
\renewcommand{\arraystretch}{1.2}
\begin{tabular}{cccccc}\hline
$n_0$ & $E/A$  & $K_0$ & $J_0$& $L_0$ & $M_N^\ast/M_N$ \\ 
$[{\rm fm}^{-3}]$ & [MeV] & [MeV] & [MeV] & [MeV] & / \\ \hline
$0.16$ & $-16$ & U(220,340) & U(28,45) & U(20,150) & U(0.55,0.80) \\ \hline
\end{tabular}
\end{center}
  \vspace{-0.5cm}
\end{table}

It should be noted that the values of the coupling constants in the Lagrangian are not predetermined and must be obtained by fitting experimental data of symmetric nuclear matter at saturation density. Table \ref{tab:sat} contains six quantities and their prior distributions that will be utilized in the Bayesian inference process. The saturation density $n_0$ and energy per nucleon $E/A$ are constants, whereas the incompressibility $K_0$, symmetry energy $J_0$, symmetry energy slope $L_0$, and the ratio between the effective mass and free nucleon mass $M_N^\ast/M_N$ have uniform prior distributions with reasonably wide ranges that are presented in the table.

The energy density and pressure at saturation density are given by
Eqs.~(\ref{eq:ener})--(\ref{eq:press}), and are determined as
$147.68\ {\rm MeV/fm^3}$ and $0$ from the Table~\ref{tab:sat}. Furthermore, we can derive the expressions for the symmetry energy $J_0$, compressibility $K_0$, and symmetry energy slope $L_0$ of symmetric nuclear matter at saturation density, which are presented below:
\begin{eqnarray}
  \label{eq:esym}
  J_0 & = & \frac{p_{\rm F}^2}{6E_{\rm F}} + \frac{g_{\rho_N}^2}{2[m_\rho^2 + \Lambda_v (g_{\omega_N} g_{\rho_N} \omega)^2]} (n_{\rm p} + n_{\rm n})\ , \\
  \label{eq:k0}
  K_0 & = & \frac{3p_{\rm F}^2}{E_{\rm F}} + \frac{3M_{\rm N}^\ast p_{\rm F}}{E_{\rm F}} \frac{dM_{\rm N}^\ast}{dp_{\rm F}} + \frac{9g_{\omega_N}^2}{m_\omega^2 }n_0\ , \\
  \label{eq:l0}
  L_0 & = & 3J_0 + \frac{1}{2}\left(\frac{3\pi^2}{2}n_0 \right)^{2/3} \frac{1}{E_{\rm F}} \times \nonumber \\
  & & \left(\frac{g_{\omega_N}^2}{m_\omega^2}\frac{n_0}{E_{\rm F}}
            - \frac{K_0}{9E_{\rm F}} - \frac{1}{3} \right) \nonumber \\
        & & - \left(\frac{3g_\rho^2}{m_\rho^2 + \Lambda_v (g_{\omega_N} g_{\rho_N} \omega)^2}\right)^2 \frac{g_{\omega_N}^3 \Lambda_v \omega n_0^2}{m_\omega^2}\ .
\end{eqnarray}
Note that the expressions are simplified for symmetric nuclear matter by vanishing isovector meson field $\rho$.

Once the values of the six quantities presented in Table~\ref{tab:sat} are fixed, the six parameters, namely $g_{\sigma_N}$, $g_{\omega_N}$, $g_{\rho_N}$ (or $g_\sigma^q$, $g_\omega^q$, $g_\rho^q$ for QMF), $g_2$, $g_3$, and $\Lambda_v$, will be uniquely determined. In the following paragraphs, we outline our strategy for determining these parameters when the saturation properties are known.

We sum up the energy density and pressure expressions~(\ref{eq:ener})
-- (\ref{eq:press}) to yield a simplified expression:
\begin{eqnarray}
  \label{eq:eplusp}
  e_{_{\rm N}} + p_{_{\rm N}} = e_{_{\rm N}} = \sum_{i=n,p} e_{\rm kin}^i + \sum_{i=n,p} p_{\rm kin}^i + g_\omega \omega n_0\ .
\end{eqnarray}
The left-hand-side is known from $E/A$ and $n_0$, and the kinetic
terms only depends on Fermi momentum $p_{\rm F}$ and effective
mass. Combine this equation with the $\omega$ equation of motion, we
can express $\omega$ and $g_\omega$ in terms of the known
quantities as follows:
\begin{eqnarray}
  \label{eq:omega}
  \omega  & = & \sqrt{\frac{(E/A + M_{\rm N}) n_0 - \sum_i (e_{\rm kin} + p_{\rm kin})}{m_\omega^2}}\ , \\
  \label{eq:gomega}
  g_{\omega_N} & = & \frac{m_\omega^2 \omega}{n_0}\ .
\end{eqnarray}
The expressions for $\Lambda_v$ and $g_\rho$ (where $\rho$ vanishes for symmetric nuclear matter) can also be obtained in the same manner by combining Eqs.~(\ref{eq:esym}) and (\ref{eq:l0}):
\begin{eqnarray}
  \label{eq:lambda_c}
  \Lambda_v & = & - \frac{m_\omega^2 \alpha}{3 \beta^2 g_{\omega_N}^3 \omega n_0^2}\ , \\
  \label{eq:grho}
  g_{\rho_N} & = & \sqrt{\frac{m_\rho^2}{\beta^{-1} - \Lambda_v (g_{\omega_N} \omega)^2}}\ ,
\end{eqnarray}
where $\alpha$ and $\beta$ are written as:
\begin{eqnarray}
  \label{eq:beta}
  \alpha & = & L_0 - 3J_0 -  \frac{1}{2}\left(\frac{3\pi^2}{2}n_0 \right)^{2/3} \frac{1}{E_{\rm F}} \times \nonumber \\
         & & \left(\frac{g_{\omega_N}^2}{m_\omega^2}\frac{n_0}{E_{\rm F}}
             - \frac{K_0}{9E_{\rm F}} - \frac{1}{3} \right)\ , \\
  \beta & = & \frac{2J_0}{n_0} - \frac{p_{\rm F}^2}{3E_{\rm F} n_0}\ .
\end{eqnarray}

The determination of the last three parameters relies on Eqs.~(\ref{eq:press}), (\ref{eq:eqs_mot1}), and the derivative of Eq.~(\ref{eq:eqs_mot1}) with respect to $\sigma$. In the QMF model, the effective nucleon mass $M_{\rm N}^\ast$ is a function of the effective quark mass $m_q^\ast$, which solely depends on $g_\sigma^q \sigma$. Thus, $M_{\rm N}^\ast$ can be treated as a function of $\delta m^q = g_\sigma^q \sigma$. On the other hand, in the RMF model, $M_{\rm N}^\ast$ is a linear function of $\delta m = -g_{\sigma_N} \sigma$. The value of $\delta m^q$ or $\delta m$ is known once the effective nucleon mass is determined in both cases.

Equations.~(\ref{eq:press}), (\ref{eq:eqs_mot1}), and the derivative of Eq.~(\ref{eq:eqs_mot1}) are linear with respect to the variables $m_\sigma^2 \sigma^2$, $g_2 \sigma^3$, and $m_3 \sigma^4$. Solving them is a straightforward process, and the solution is written as:
\begin{eqnarray}
  \label{eq:sigma}
  \sigma & = & \sqrt{\frac{C - 6B + 12A}{m_\sigma^2}}\ , \\
  g_2 & = & \frac{-3C + 15B - 24A}{\sigma^3}\ , \\
  g_3 & = & \frac{2C - 8B + 12A}{\sigma^4}\ ,
\end{eqnarray}
where we have introduced three parameters $A$, $B$, and $C$ to simplify the expressions, and they are written as:
\begin{eqnarray}
  \label{eq:a}
  A & = & \sum_{i=n,p} p_{\rm kin}^i + \frac{1}{2} m_\omega^2 \omega^2, \\
  \label{eq:b}
  B & = & -\delta m_q  \frac{d M_{\rm N}^\ast}{d(\delta m_q)} n_s, \\
  \label{eq:c}
  C & = & -(\delta m_q)^2\left[\frac{d^2M_{\rm N}^\ast}{d(\delta m_q)^2}n_s + \left(\frac{dM_{\rm N}^\ast}{d(\delta m_q)}\right)^2 n_s^\prime \right. \nonumber \\
  & & \quad\quad\quad\quad\ \left. + \frac{dM_{\rm N}^\ast}{d(\delta m_q)} \frac{dp_{\rm F}}{d(\delta m_q)} \tilde{n}_s^\prime \right].
\end{eqnarray}
The $n_s^\prime$ and $\tilde{n}_s^\prime$ denote the derivatives of
$n_s$ respect to $M_{\rm N}^\ast$ and $p_{\rm F}$, respectively,
\begin{eqnarray}
  \label{eq:nsp}
  n_s^\prime & = & \frac{1}{\pi^2} \left(p_{\rm F}E_{\rm F} + 2\frac{p_{\rm F}}{E_{\rm F}} M_{\rm N}^{\ast 2}
                   - 3 M_{\rm N}^{\ast 2} \ln\left|\frac{p_{\rm F} + E_{\rm F}}{M_{\rm N}^\ast} \right|  \right),
                   \nonumber \\ \\
  \tilde{n}_s^\prime & = & \frac{2}{\pi^2} \frac{M_{\rm N}^\ast p_{\rm F}^2}{E_{\rm F}}.
\end{eqnarray}
In the case of the RMF model, $\delta m^q$ in Eqs.~(\ref{eq:b})--(\ref{eq:c}) is substituted by $\delta m$, and the expressions of $A$, $B$, and $C$ can be further simplified by taking into account that $d M_{\rm N}^\ast/d(\delta m) = 1$ and $d^2 M_{\rm N}^\ast/d(\delta m)^2 = 0$. The last parameters, $g_{\sigma_N}$ in RMF or $g_\sigma^q$ in QMF, can easily be obtained as $-\delta m^q / \sigma$ or $-\delta m / \sigma$, respectively.

\subsection{Neutron stars}\label{sec:ns} 
Once the coupling constants are known from the saturation properties of symmetric nuclear matter, one can calculate the energy and pressure using Eqs.~(\ref{eq:ener})--(\ref{eq:press}) for any baryonic number density $n$ and neutron fraction. However, in a neutron star, the contributions from leptons cannot be neglected. Therefore, the fractions of leptons must also be considered. These fractions are determined by the $\beta$-equilibrium condition:
\begin{eqnarray}
  \label{eq:beta}
  \mu_{\rm n} & = & \mu_{\rm p} + \mu_{\rm e}\ ,
\end{eqnarray}
and the charge-neutrality condition:
\begin{eqnarray}
  \label{eq:beta}
  n_{\rm p} & = & n_{\rm e} + n_{\rm \mu}\ ,
\end{eqnarray}
where $\mu_{\rm n, p, e}$ denote the chemical potentials of the corresponding components and ${\rm e}$ and $\mu$ denote electrons and muons, respectively.
The electromagnetic interaction in neutron stars can be neglected because of the local charge neutrality. Therefore the leptons can be treated as the ideal gas and their contributions to the energy and pressure read as:
\begin{eqnarray}
  \label{eq:lep_e}
  e_{i} & = & \frac{1}{8\pi^2} \biggl[2p_{{\rm F}_i}^{3} e_{{\rm F}_i} + m_{i}^{2} p_{{\rm F}_i} e_{{\rm F}_i}  - m_{i}^{4} \ln\left(\frac{p_{{\rm F}_i}+e_{{\rm F}_i}}{m_{i}}\right)\biggr]\ , \nonumber \\  \\
  \label{eq:lep_p}
  p_{i} & = & \frac{1}{24\pi^2} \biggl[(2p_{{\rm F}_i}^3 -3 m_{i}^{2} p_{{\rm F}_i}) e_{{\rm F}_i} + 3m_{i}^{4} \ln\left(\frac{p_{{\rm F}_i}+e_{{\rm F}_i}}{m_{i}}\right)\biggr]\ . \nonumber \\
\end{eqnarray}
In conclusion, the total energy density and pressure of the neutron star can be obtained through the addition of the energy and pressure contributions of both nucleons and leptons:
\begin{eqnarray}
  \label{eq:tot_e}
  e & = & e_{_{\rm N}} + e_{\rm e} + e_\mu\ , \\
  \label{eq:tot_p}
  p & = & p_{_{\rm N}} + p_{\rm e} + p_\mu\ .
\end{eqnarray}

The previously mentioned calculations using RMF or QMF provide the EoS specific to the neutron star core. However, to fully describe the properties of neutron stars, the EoS of the crust must also be taken into account. Unfortunately, the crust EoS cannot be obtained through aforementioned methods. Therefore, we utilize the BPS + NV EoS~\cite{1971ApJ...170..299B, 1973NuPhA.207..298N} for the crust and combine it with our previously obtained core EoSs.

After obtaining the EoSs, the next step is to solve the TOV equation, given by:
\begin{eqnarray}
\frac{dm}{dr} & = &  4\pi r^2e\,, \label{eq:tov1} \\
\frac{dp}{dr} & = &  -(e+p)\frac{m+4\pi r^3 p}{r(r - 2m)}\,, \label{eq:tov2} \\
\frac{d \nu}{dr} & = & -\frac{2}{e+p} \frac{dp}{dr},
\end{eqnarray}
Note that the equations are expressed in natural units with $C=G=1$. These units are also applied in the subsequent equations of tidal deformability. This allows for the computation of the mass-radius relationship for neutron stars based on the corresponding EoS. To solve the TOV equation, boundary conditions are specified at the surface of the star as $m(R)=M$, $p(R)=0, \nu(R)=\ln(1-2M/R)$, where $M$ and $R$ represent the mass and radius of the star, respectively.

We will conclude the section by describing the tidal deformability of neutron stars, which is a crucial quantity for probing the EoS through the emission of gravitational waves.

The concept of tidal deformability pertains to a star's susceptibility to deformation owing to its companion's tidal force. It is quantified as the ratio of the induced multipole moment of the star to the inducing tidal field from its companion. The tidal field, being significantly smaller than the gravitational field of the star itself, can be treated as a perturbation.
The main equation of the perturbed equation is written as~\cite{2008ApJ...677.1216H, 2020PhRvD.102h4058Z}:
\begin{eqnarray}
& & \frac{d^2 H_0}{dr^2} + \left[\frac{2}{r} + \frac{2m}{r^2}e^\lambda + 4\pi r(p-e)e^\lambda \right] \frac{d H_0}{dr} \nonumber \\
& & + \biggl[4\pi e^\lambda\left(4e+8p+(p+e)\left(1+\frac{1}{c_s^2}\right)\right) \nonumber \\
& & \quad\ \ -\frac{6e^\lambda}{r^2} - \left(\frac{d\nu}{dr}\right)^2 \biggr] H_0  = 0\,,  \label{eq:tidal}
\end{eqnarray}
where $H_0$, $\nu$ and $\lambda$ are metric functions and $\lambda=-\ln(1-2m/r)$. This equation is solved with the TOV equation, from the center of the star to its surface, while imposing the interior boundary condition $H_0(r) \rightarrow \alpha_t r^2$ as $r$ approaches 0, where $\alpha_t$ is a constant. 
On the surface of the star, the exterior boundary is:
\begin{eqnarray}
H_0^{\rm int}(R) & = & H_0^{\rm ext}(R)\ , \label{eq:matching1n1}\\
(H_0^{\rm int})^\prime(R) & = & (H_0^{\rm ext})^\prime(R)\ , \label{eq:matching1n2}
\end{eqnarray}
where the exterior solution of $H_0$ reads as:
\begin{eqnarray}
  & & H_0^{\rm ext} = c_1 Q_2^2(z) + c_2 P_2^2(z)\ .
  \label{eq:solt}
\end{eqnarray}
The $Q_2^2$ and $P_2^2$ are the associated Legendre functions of the first and second kind, and $z=r/M - 1$. 
In practice, the solution process for equations (\ref{eq:tov1})--(\ref{eq:tidal}) involves initially setting an arbitrary value for $\alpha_t$ as the interior boundary. Following this step, we match the exterior solution $H_0^{\rm ext}$ at the surface of the star to obtain $c_1$ and $c_2$. Finally, the love number and tidal deformability are calculated as follows:
\begin{eqnarray}
  k_2 & = & \frac{4}{15}  \frac{c_1}{c_2} \left(\frac{M}{R}\right)^5
  \,,  \label{eq:lovenum1} \\
  \Lambda^{\rm T} & := & \frac{2}{3}k_2\left(\frac{M}{R}\right)^{-5}\,. \label{eq:lovenum2}
\end{eqnarray}
It is worth noting that the love number $k_2$ is solely dependent on the ratio $c_1/c_2$, and that different choices of $\alpha_t$ will yield the same value for this ratio, thus resulting in the correct tidal deformability value.

\begin{figure*}
\vspace{-0.3cm}
{\centering
  \includegraphics[width=0.33\textwidth]{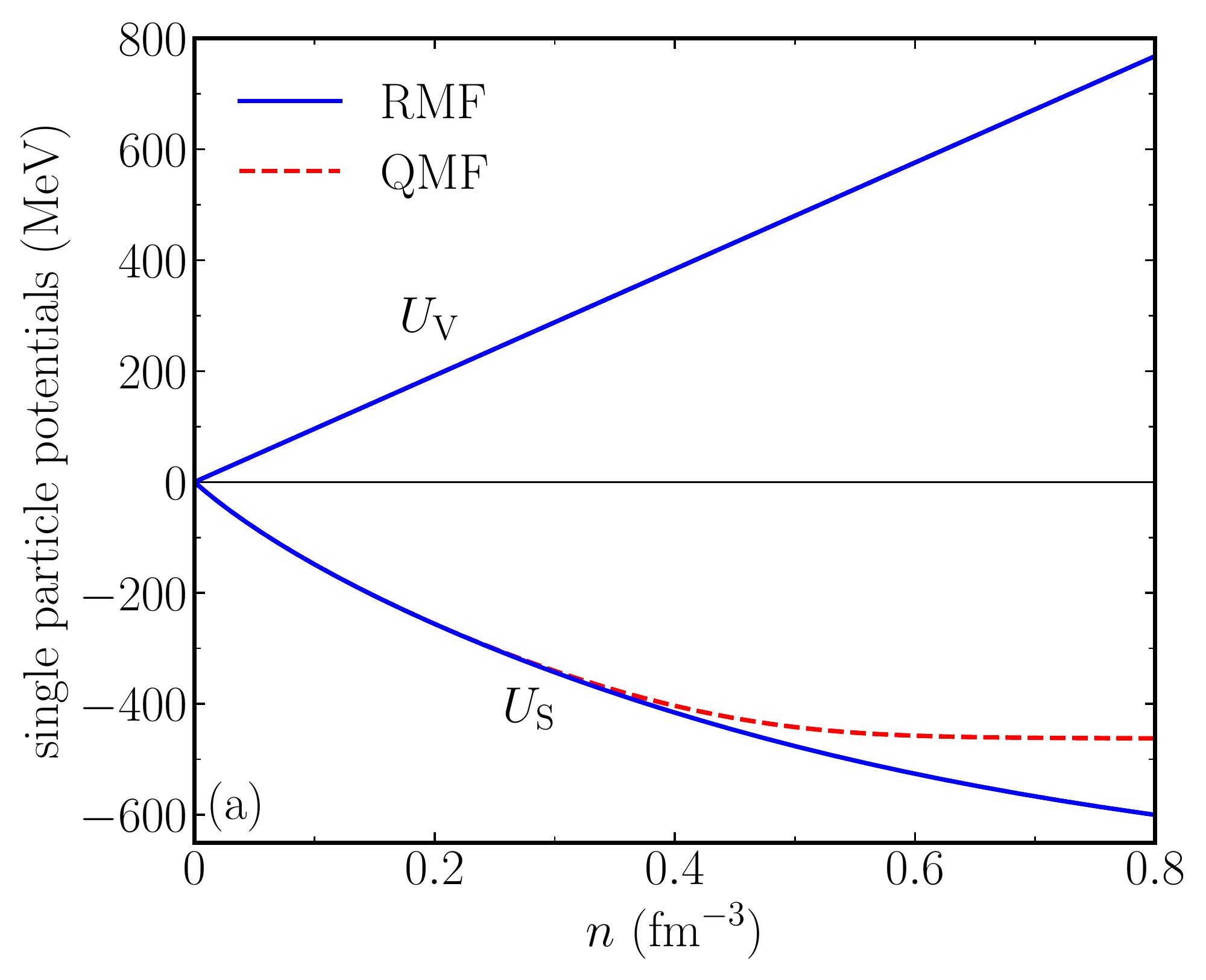}
  \includegraphics[width=0.33\textwidth]{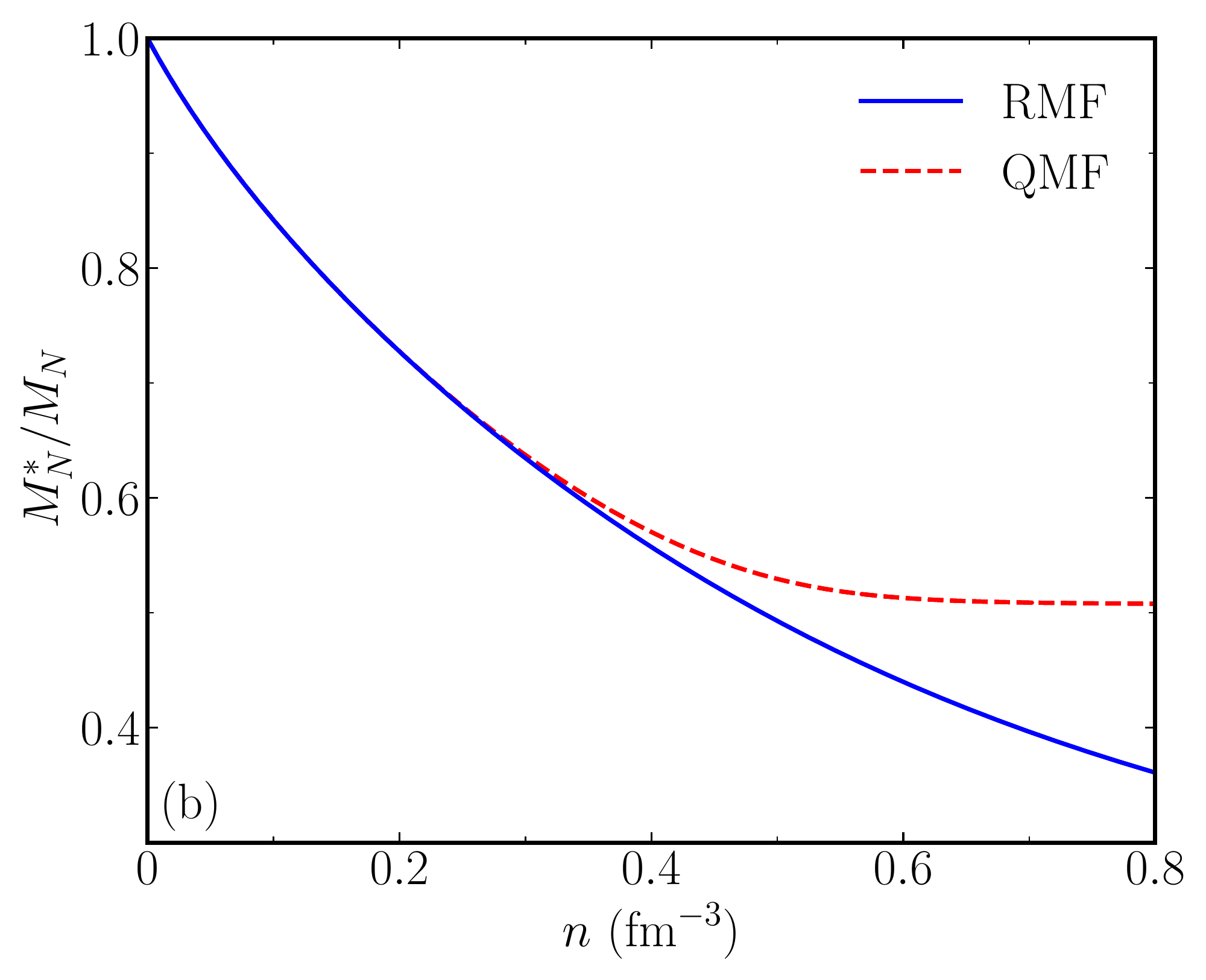}
    \includegraphics[width=0.33\textwidth]{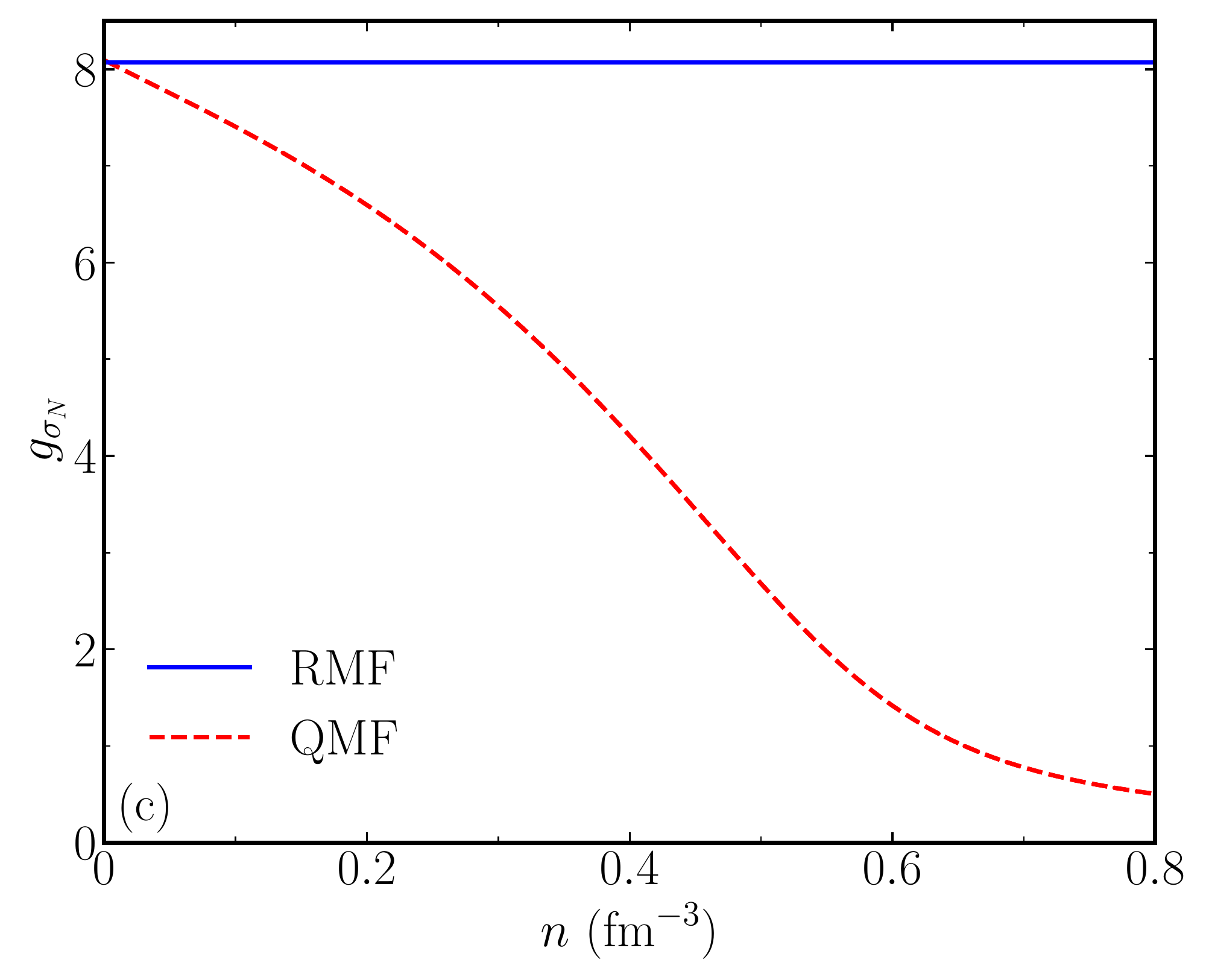}}   \vspace{-0.3cm}
  \caption{Scalar (vector) potential $U_{\rm S}$ ($U_{\rm V}$) (left panel), effective mass $M_N^\ast$ (middle panel) and nucleon-$\sigma$ coupling constants (right panel) as functions of nucleon number density, for the representative case of $K_0$ = 240 MeV,  $J_0$ = 31 MeV, $M_N^*/M_N$ = 0.77 in symmetric nuclear matter.
  The calculations are done for both RMF (solid lines) and QMF (dash lines) models.
   The left panel shows only one line for $U_{\rm V}$ as its results from RMF and QMF models are identical.
  }\label{fig:snm}
  \vspace{-0.5cm}
\end{figure*}

\subsection{The Bayesian analysis}
\label{sec:data} 
When considering a model hypothesis with a set of parameters denoted by ${\bm \theta}$ and some associated data $d$, the posterior probability can be obtained using Bayes' theorem, given by:
\begin{eqnarray}
  \label{eq:bayes}
  p({\bm \theta}|d) = \frac{\mathcal{L}(d|{\bm \theta}) p({\bm \theta})}{\int
  \mathcal{L}(d|{\bm \theta}) p({\bm \theta}) d{\bm \theta}}\ ,
\end{eqnarray}
where $\mathcal{L}(d|{\bm \theta})$ represents the likelihood of observing the data $d$ with a given set of parameters ${\bm \theta}$, and $p({\bm \theta})$ represents the prior distribution of these parameters. The denominator of Eq. (\ref{eq:bayes}) is the probability of the data given all possible parameter values and acts as a normalization factor.

In this paper, we employed the Python package \texttt{BILBY}~\citep{2019ApJS..241...27A, 2020MNRAS.499.3295R} and the nested sampler \texttt{pymultinest}~\citep{2014A&A...564A.125B} to conduct the Bayesian inference and generate posterior samples.
The present analysis considers a total likelihood composed of four parts, given by:
\begin{eqnarray}
  \label{eq:bayes1}
 \mathcal{L}(d|{\bm \theta}) & = & \mathcal{L}_{\rm AT2017gfo} \times \mathcal{L}_{\rm GW170817} \times \mathcal{L}_{\rm NICER} \nonumber \\
 & & \times \mathcal{L}_{{\rm skin}}\ ,
\end{eqnarray}
where $\mathcal{L}_{\rm AT2017gfo}$, $\mathcal{L}_{\rm GW170817}$, $\mathcal{L}_{\rm NICER}$, and $\mathcal{L}_{\rm skin}$ represent the likelihood of observing the kilonova light curves of AT2017gfo, the gravitational wave event GW170817, the NICER's measurement of mass and radius of two pulsars, and the neutron-skin data, respectively. The neutron-skin data includes both the PREX-II experiments~\cite{2021PhRvL.126q2502A} and the {\it ab initio} predictions~\cite{2022NatPh..18.1196H}.

\begin{itemize}

\item The light curve of AT2017gfo was obtained using a radiation transfer model that incorporates various input parameters of the binary neutron star merger ejecta. For more details on this model and the corresponding likelihood for AT2017gfo, please refer to our previous work~\cite{2023ApJ...943..163Z}.

\item  The GW170817 likelihood was computed by implementing a
high-precision interpolation that developed in Ref.~\cite{2020MNRAS.499.5972H}, which is encapsulated in the python package \textsf{toast}~\footnote{\url{https://git.ligo.org/francisco.hernandez/toast}},
and the likelihood function is given by:
\begin{eqnarray}
  \label{eq:llh_gw}
  \mathcal{L}_{\rm GW170817} = F(\Lambda_1, \Lambda_2, \mathcal{M}, q)\ ,
\end{eqnarray}
where $\mathcal{M}$ and $q$ denotes the chirp mass and mass ratio, and
$\Lambda_1$ and $\Lambda_2$ denote the tidal deformabilities of the individual
stars. $F(\cdot)$ is the interpolation function. 
This interpolation table is obtained by fitting the strain data using the gravitational wave waveform from the component masses and their corresponding tidal deformabilities. 

\item The NICER Collaboration has measured the mass and radius of two pulsars, namely PSR J0030+0451~\cite{2019ApJ...887L..24M, 2019ApJ...887L..21R} and PSR J0740+6620~\cite{2021ApJ...918L..28M, 2021ApJ...918L..27R}. To generate the likelihoods for these pulsars in our analysis, we use the ST+PST model samples for PSR J0030+0451~\citep{Riley2019b} and the NICER x XMM samples for PSR J0740+6620~\citep{Riley2021b}. We employ the KDE method to generate the posterior distributions, which are treated as the likelihoods in our analysis.

\item The PREX-II experimental data has provided the value of neutron-skin $\Delta r_{\rm np}$ of $^{208}$Pb as $0.283\pm 0.071$ fm~\cite{2021PhRvL.126q2502A}, which can be used to obtain the value of $L=106\pm 37$ MeV through the known correlation between $\Delta r_{\rm np}$ $^{208}$Pb and $L$. We incorporate this constraint on $L$ in our analysis by treating the likelihood of PREX-II as:
\begin{eqnarray}
  \label{eq:llh_prex}
  \mathcal{L}_{\rm PREX-II} = \frac{1}{\sqrt{2\pi \sigma_L}} \exp\left[-\frac{(L-L_0)^2}{2\sigma_L^2} \right]\ ,
\end{eqnarray}
where $L_0=106$ MeV and $\sigma_L=37$ MeV denote the mean and deviation, respectively. Furthermore, {\it ab initio} calculations~\cite{2022NatPh..18.1196H} have predicted $\Delta r_{\rm np}$ for $^{208}$Pb and yielded the posterior distributions of both neutron-skin and saturation properties based on $J_0$, $K_0$ and $L_0$. We incorporate these distributions in our analysis by treating the likelihood as a multivariate Gaussian function:
\begin{eqnarray}
\label{eq:llh_eft}
\mathcal{L}_{\rm EFT} = \frac{1}{\sqrt{(2\pi)^k |\bm{\Sigma}|}} \exp\left[-\frac{ (\bm{x}-\bm{\mu})^{\rm T} \bm{\Sigma}^{-1} (\bm{x}-\bm{\mu})}{2} \right] \ , \nonumber \\
\end{eqnarray}
where $\bm{x}$ represents the vector of $[J_0\ L_0\ K_0]$, and the means and covariance matrix are given in the supplemental material of Ref.~\cite{2022NatPh..18.1196H}.
We combine these results for neutron-skin and obtain the likelihood of skin data as:
\begin{eqnarray}
  \label{eq:llh_skin}
  \mathcal{L}_{\rm skin} = \mathcal{L}_{\rm PREX-II} \times \mathcal{L}_{\rm EFT}\ .
\end{eqnarray}

It is noteworthy that we merely included the data of PREX-II experiment and {\it ab initio} calculations in our analyses, despite the availability of other experimental data~\cite{2022PhRvL.129d2501A} and analyses~\cite{2022arXiv220703328Z, 2021PhRvL.127s2701E}. 
The PREX-II experiment inferred a large value for the symmetry energy slope, representing an extreme case. In contrast, the {\it ab initio} calculations are consistent with the results obtained from other analyses. We believe that incorporating neutron skin data from PREX-II and {\it ab initio} calculations adequately encompasses the diverse range of results obtained from neutron skin analyses.

\end{itemize}

\section{Results and discussion}\label{sec:res}

\subsection{Nuclear matter EoS: RMF vs. QMF}\label{sec:comp_theory}

In this section, we analyze and compare the performances of the RMF and QMF models in describing nuclear matter. 
It should be noted that the primary difference between these two models is the effective nucleon mass $M_N^\ast$. We will review their respective formulations and provide a contrast between them in the following paragraphs:
\begin{eqnarray}
  \label{eq:rmf_meff}
  M^\ast_N & = & M_N - g_{\sigma_N} \sigma \quad\quad\quad\quad\quad\quad\quad\ \ \ \,({\rm RMF})\ , \\
  \label{eq:qmf_meff2}
  M^\ast_N & = & E^{0}_N-\epsilon_{\rm c.m.}+\delta M_N^\pi+(\Delta E_N)_g\ ({\rm QMF})\ .
\end{eqnarray}

In Fig.~\ref{fig:snm}, we present the scalar (vector) potential of the single nucleon, the effective mass ratio and the nucleon-$\sigma$ coupling constants $g_{\sigma_N}$ of symmetric nuclear matter, for both the RMF and QMF models, shown as functions of nucleon number density. 
Note that the formulations for calculating $g_{\omega_N}$, $\omega$, $g_{\rho_N}$, and $\rho$ in the parameter fitting procedure, as outlined in equations (\ref{eq:omega}) through (\ref{eq:grho}), are identical for both models. 
 As a result, the coupling constants $g_{\omega_N}$ and $g_{\rho_N}$, along with the equations of motion for the corresponding mesons, given by equations (\ref{eq:eqs_mot2}) through (\ref{eq:eqs_mot3}), remain the same for any specific density, provided the saturation properties are fixed.
Meanwhile, we recall the expressions for the scalar and vector potentials: 
\begin{eqnarray}
  \label{eq:potential}
  U_{\rm S} & = & M_{\rm N}^\ast - M_{\rm N}\ , \\
  U_{\rm V} & = & g_{\omega_N} \omega + g_{\rho_N} \rho \tau_3\ .
\end{eqnarray}
It is worth noting that $U_{\mathrm{V}}$ depends solely on the $\omega$ and $\rho$ couplings, while $U_{\mathrm{S}}$ only depends on the nucleon effective mass. As a result, the vector potentials are identical for both models in the left panel.
Differences between the two models in the scalar potential $U_{\mathrm{S}}$ and the effective mass $M_{\mathrm{N}}$ of the single nucleon (in the middle panel) are negligible for densities below $0.25\ \mathrm{fm}^{-3}$ (or $\approx 1.5 n_0$), for that the same $M_N$ value at saturation density are fitted. 
However, as the density increases, the RMF and QMF results begin to diverge, with the QMF predictions exhibiting larger values. Notably, the QMF curves for the scalar potential and effective mass become increasingly flat when the density exceeds $0.6\ \mathrm{fm}^{-3}$ (or $\approx 3.7 n_0$),.

\begin{figure}
\vspace{-0.2cm}
{\centering
\resizebox*{0.48\textwidth}{0.3\textheight}
{\includegraphics{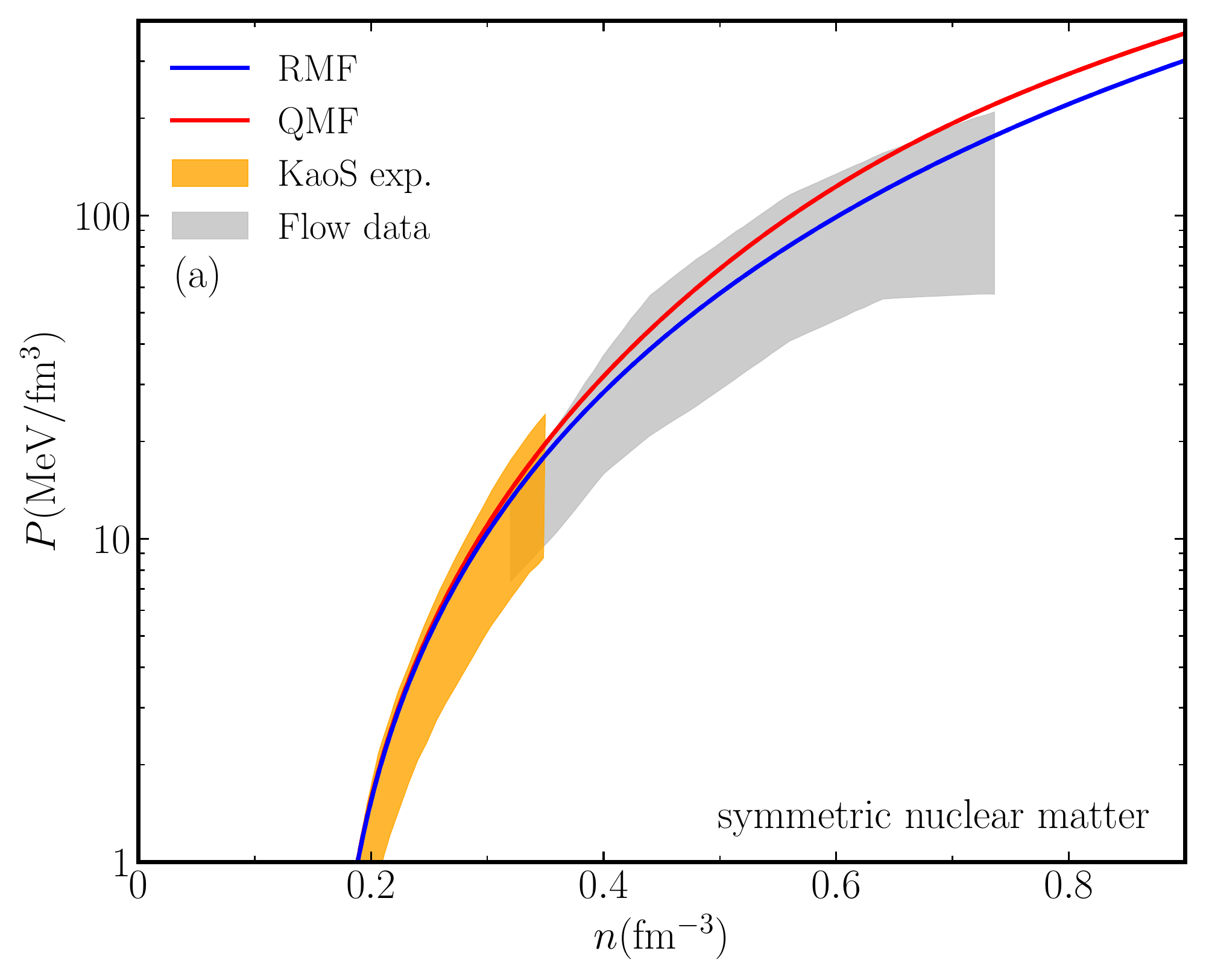}}}
{\centering
\resizebox*{0.48\textwidth}{0.3\textheight}
{\includegraphics{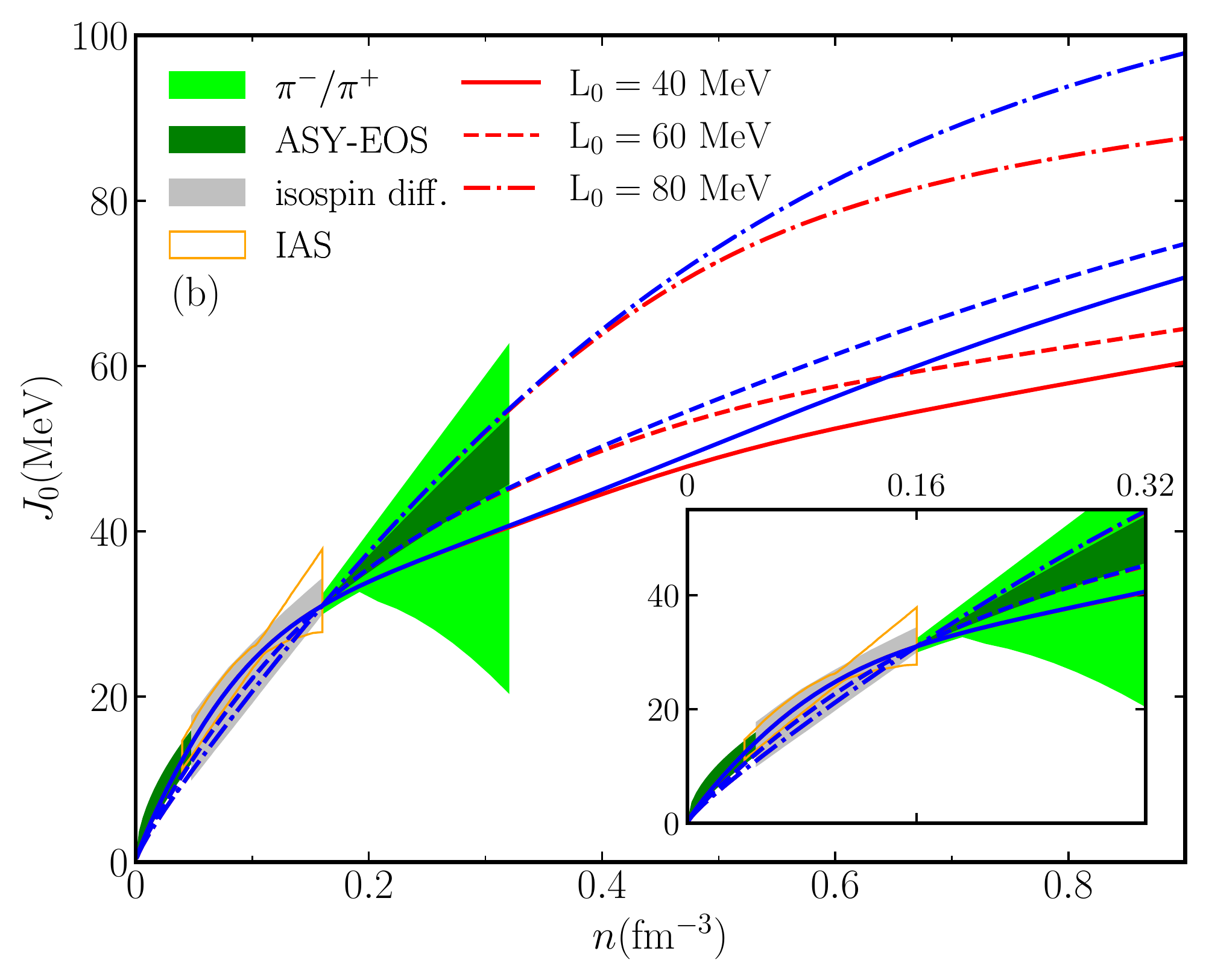}}}   \vspace{-0.3cm}
\caption{Pressure (upper panel) and symmetry energy (lower panel) as functions of density for both the RMF (blue) and QMF (red) models for fixed $K_0$ (= 240 MeV), $J_0$ (= 31 MeV), $M_N^*/M_N$ (= 0.77) and three different values of $L_0$ (= 40, 60, 80 MeV), along with experimental constraints.
  The orange and grey shaded areas in the upper panel represent the flow~\citep{2002Sci...298.1592D} and kaon production~\citep{2006PrPNP..56....1F} analyses, respectively, of the heavy-ion experimental data.
  Meanwhile, in the lower panel, various experimental regions from heavy-ion experiments are included, including the $\pi^-/\pi^+$ ratio~\citep{2021PhRvC.103a4616L} (lime), the ASY-EOS data with a consistent value of $J_0~$\citep{2016PhRvC..94c4608R_etal} (green), and the isospin diffusion data~\citep{2009PhRvL.102l2701T} (grey). The orange contour indicates the results of the isobaric analog states (IAS)~\citep{2014NuPhA.922....1D}.
}\label{fig:snm_pesym}
\vspace{-0.5cm}
\end{figure}

As seen in Eq.~(\ref{eq:rmf_meff}), the effective mass in the RMF model is a linear function of the scalar meson field $\sigma$, while in the QMF model, its dependency is more complicated. Due to this nonlinear behavior, the nucleon-$\sigma$ coupling constant $g_{\sigma_N}$ varies with the $\sigma$ field in the QMF model, in contrast to being a constant in the RMF model.
Within the QMF model, the parameter $g_{\sigma_N}$ decreases as the density increases, eventually approaching zero in the high-density region. This trend is in agreement with the observed flatness of the effective mass curve after $0.6\ {\rm fm}^{-3}$ in the middle panel.
The density-dependence of $g_{\sigma_N}$ arises from the fact that the effective nucleon mass $M^\ast_N$ is not solely determined by the background scalar field, or more specifically, the chiral condensate, but also takes into account other contributions such as the pion cloud and quark mass [refer to Eq.~(\ref{eq:qmf_meff})]. As we will show below, the results obtained for nuclear matter are highly dependent on the behavior of $M^\ast_N (n)$ (or equivalently $g_{\sigma_N}(n)$). Therefore, the effects of nucleon substructure, which are highly dependent on the confinement mechanism, play a crucial role. Interested readers may consult Ref.~\cite{2018PhRvC..97c5805Z} for further comparisons of the resulting nucleon and nuclear matter properties obtained from different realizations of confinement.
\begin{table}   \vspace{-0.4cm}
\begin{center}
  \caption{Coupling constants in the Lagrangian [Eq.~(\ref{eq:L})] for the RMF and QMF results presented in Fig.~\ref{fig:snm} and Fig.~\ref{fig:snm_pesym}, namely from reproducing the saturation properties of $K_0$ = 240 MeV,  $J_0$ = 31 MeV, $M_N^*/M_N$ = 0.77 and three values of $L_0$ = 40, 60, 80 MeV. }\label{tab:coup}
\setlength{\tabcolsep}{3pt}
\renewcommand{\arraystretch}{1.2}
\begin{ruledtabular}
\begin{tabular*}
{\hsize}{@{}@{\extracolsep{\fill}}cccc@{}}
\multicolumn{4}{c}{RMF} \\ 
$L_0\ [{\rm MeV}]$ & $40$ & $60$ & $80$ \\ 
$g_{\sigma_N}$ & \multicolumn{3}{c}{$8.07014$} \\ 
$g_{\omega_N}$  & \multicolumn{3}{c}{$8.75245$} \\ 
$g_{\rho_N}$ & $5.41294$ & $4.58306$ & $4.04596$ \\ 
$g_2\ [{\rm fm}^{-1}]$ & \multicolumn{3}{c}{$20.41105$} \\ 
$g_3$ & \multicolumn{3}{c}{$-15.25089$} \\ 
$\Lambda_v$ & $0.76937$ & $0.43067$ & $0.09197$ \\ 
\end{tabular*}
\end{ruledtabular}
\vspace{12pt}
\begin{ruledtabular}
\begin{tabular*}
{\hsize}{@{}@{\extracolsep{\fill}}cccc@{}}
\multicolumn{4}{c}{QMF} \\ 
$L_0\ [{\rm MeV}]$ & $40$ & $60$ & $80$ \\ 
$g^q_\sigma$ & \multicolumn{3}{c}{$3.86208$} \\ 
$g^q_\omega$  & \multicolumn{3}{c}{$8.75245$} \\ 
$g^q_\rho$ & $5.41294$ & $4.58306$ & $4.04596$ \\ 
$g_2\ [{\rm fm}^{-1}]$ & \multicolumn{3}{c}{$-14.62174$} \\ 
$g_3$ & \multicolumn{3}{c}{$-66.36244$} \\ 
$\Lambda_v$ & $0.76937$ & $0.43067$ & $0.09197$ \\ 
\end{tabular*}
\end{ruledtabular}
\end{center}   \vspace{-0.5cm}
\end{table}

We display the pressure and symmetry energy in Fig.~\ref{fig:snm_pesym} in the upper and lower panel, respectively. 
Several experimental constraints from the heavy-ion collision and nuclear structure studies are included for comparison.  
Both models produce reasonable pressure-density relations for symmetric nuclear matter, with their lines in the upper panel passing through the shaded region and being consistent with experimental constraints. However, in the high-density region, QMF predicts larger pressures than the RMF model due to its larger effective mass and kinetic term for pressure. In the lower panel, the ASY-EOS constraint does not overlap with the lines representing our model predictions in the supra-saturation density.
In contrast to the pressure, the RMF model predicts larger symmetry energy than QMF in the high-density region, as evident from the formulation of symmetry energy [Eq.~(\ref{eq:esym})], where the difference between the two models is the value of the Fermi energy $E_{\rm F}$ in the first term. The RMF model yields a smaller effective mass and thus a larger first term than the QMF model.
We collect in Table~\ref{tab:coup} the coupling constants corresponding to the representative EoS cases presented in both Fig.~\ref{fig:snm} and Fig.~\ref{fig:snm_pesym}.
\begin{figure}
{\centering
\resizebox*{0.48\textwidth}{0.3\textheight}
{  \includegraphics{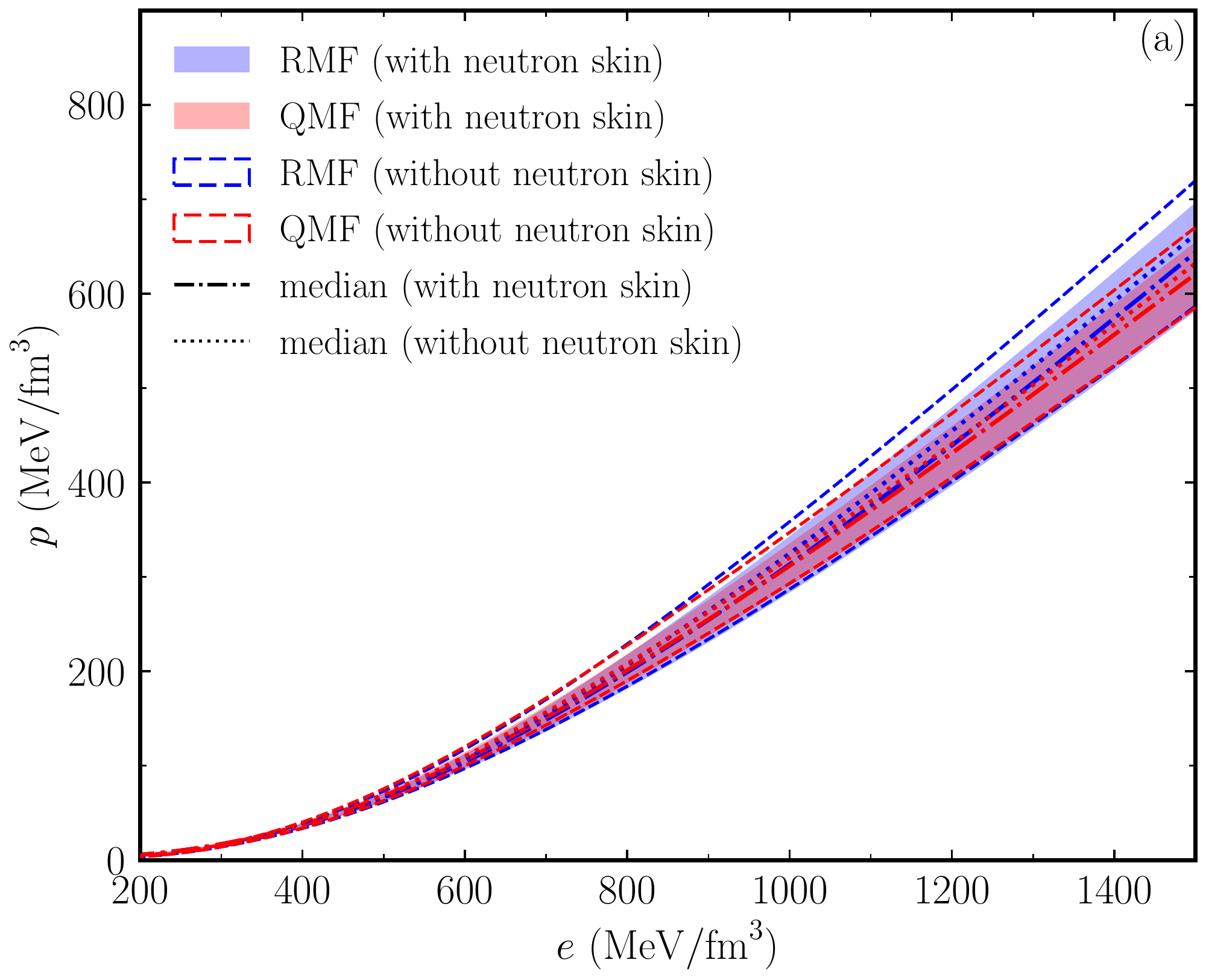}}}
{\centering
\resizebox*{0.48\textwidth}{0.3\textheight}
{\includegraphics{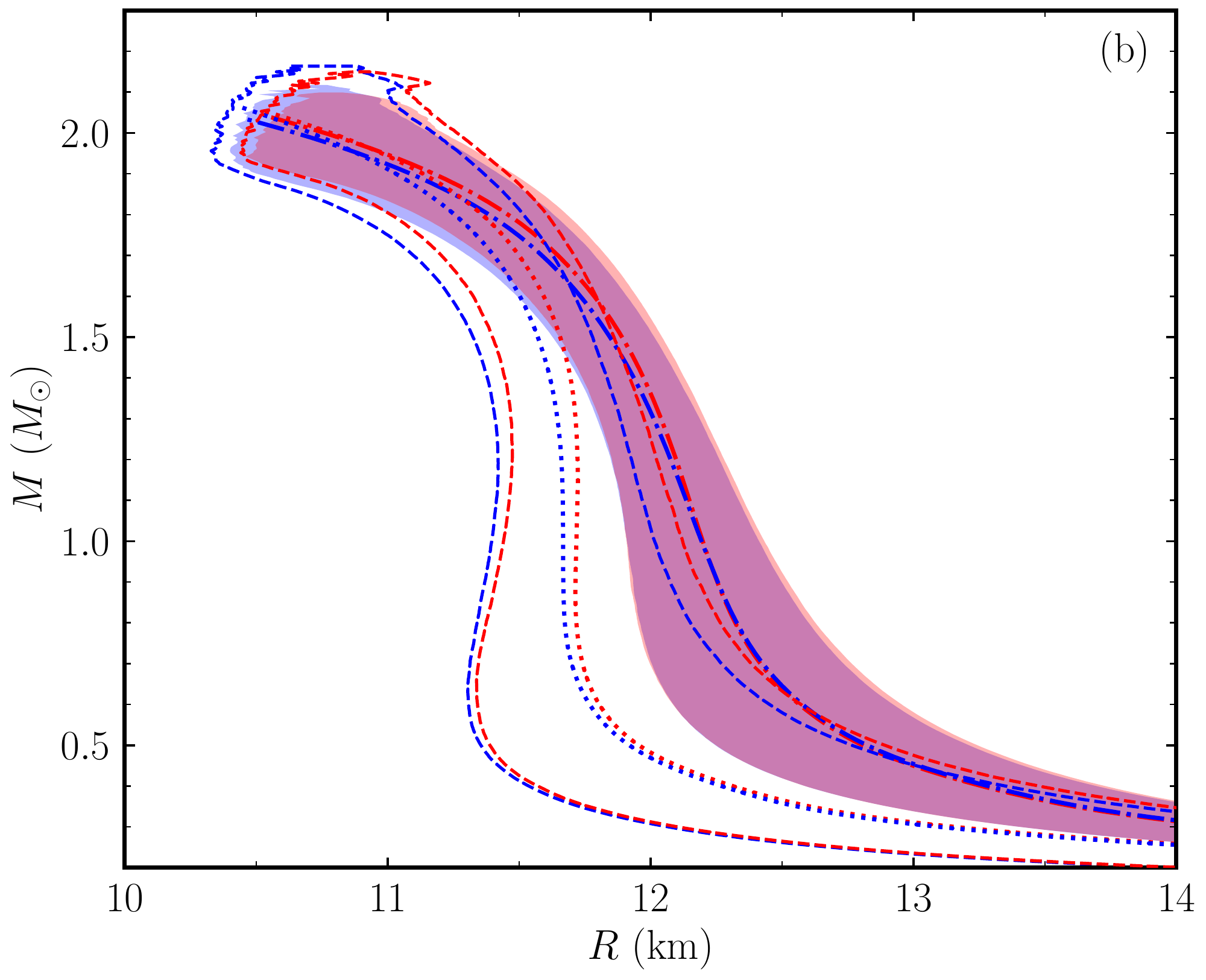}}}   \vspace{-0.3cm}
  \caption{Posteriors of the neutron star EoS and mass-radius relation for both the RMF (blue) and QMF (red) models. In all cases, the likelihood is based on observational data (GW170817/AT2017gfo + NICER). The shaded regions and dashed lines represent the 90\% confidence interval for the analyses with and without neutron-skin data (PREX-II + EFT), respectively. The dash-dot and dot lines indicate the median results for each analysis.
  }\label{fig:eos_mr}
  \vspace{-0.5cm}
\end{figure}

\begin{figure*}
{\centering
  \includegraphics[width=0.245\textwidth]{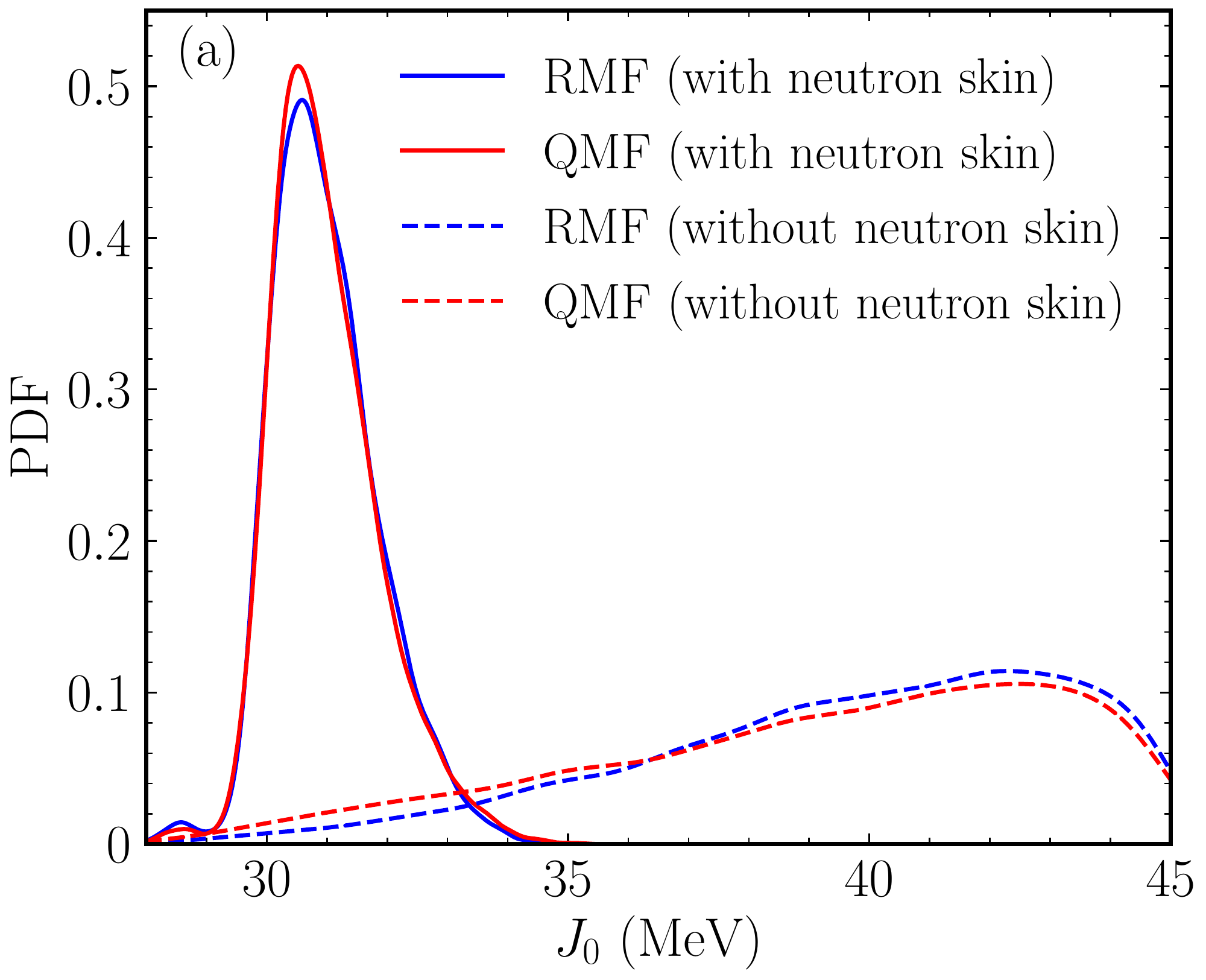}
  \includegraphics[width=0.245\textwidth]{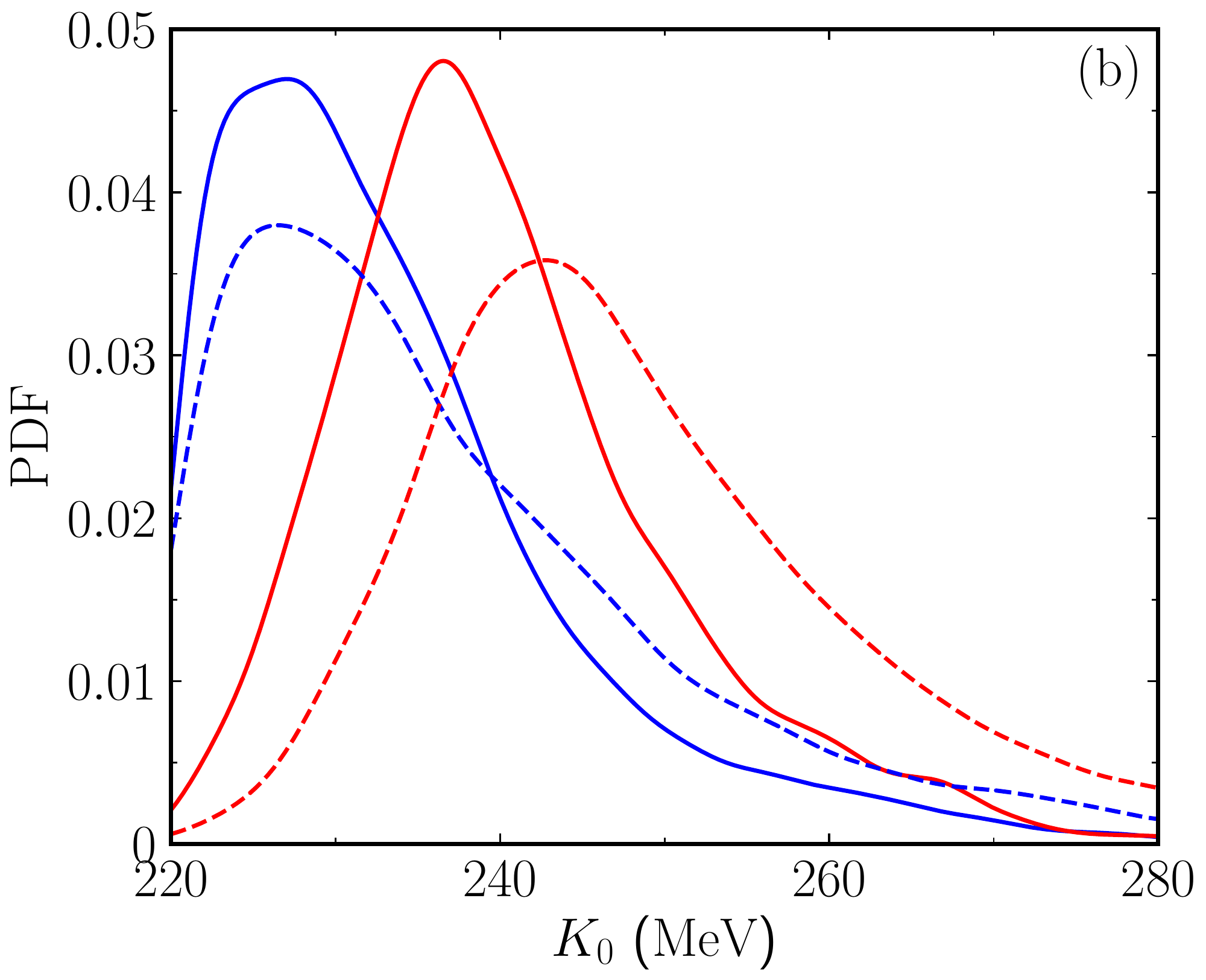}}
  \includegraphics[width=0.245\textwidth]{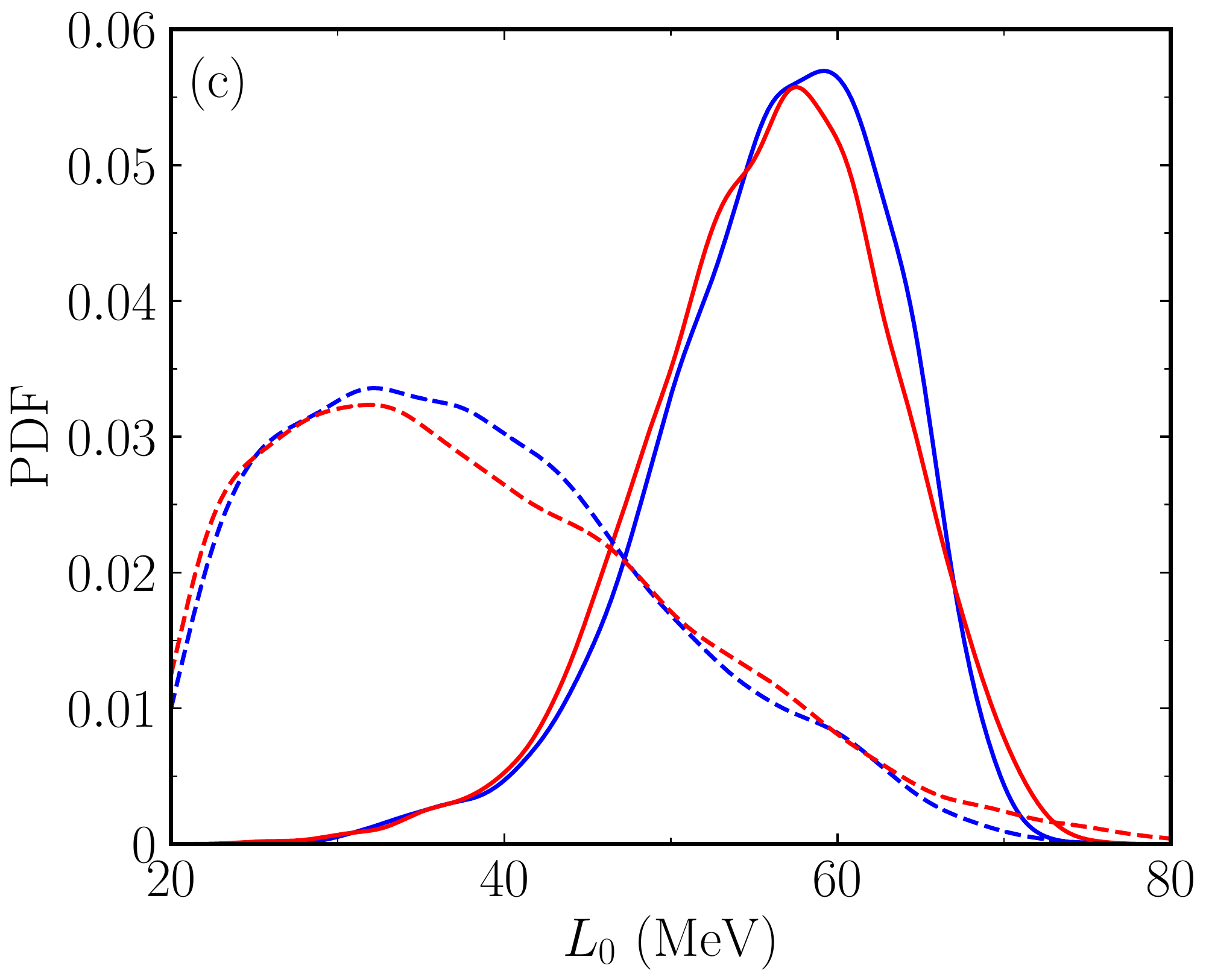}
  \includegraphics[width=0.245\textwidth]{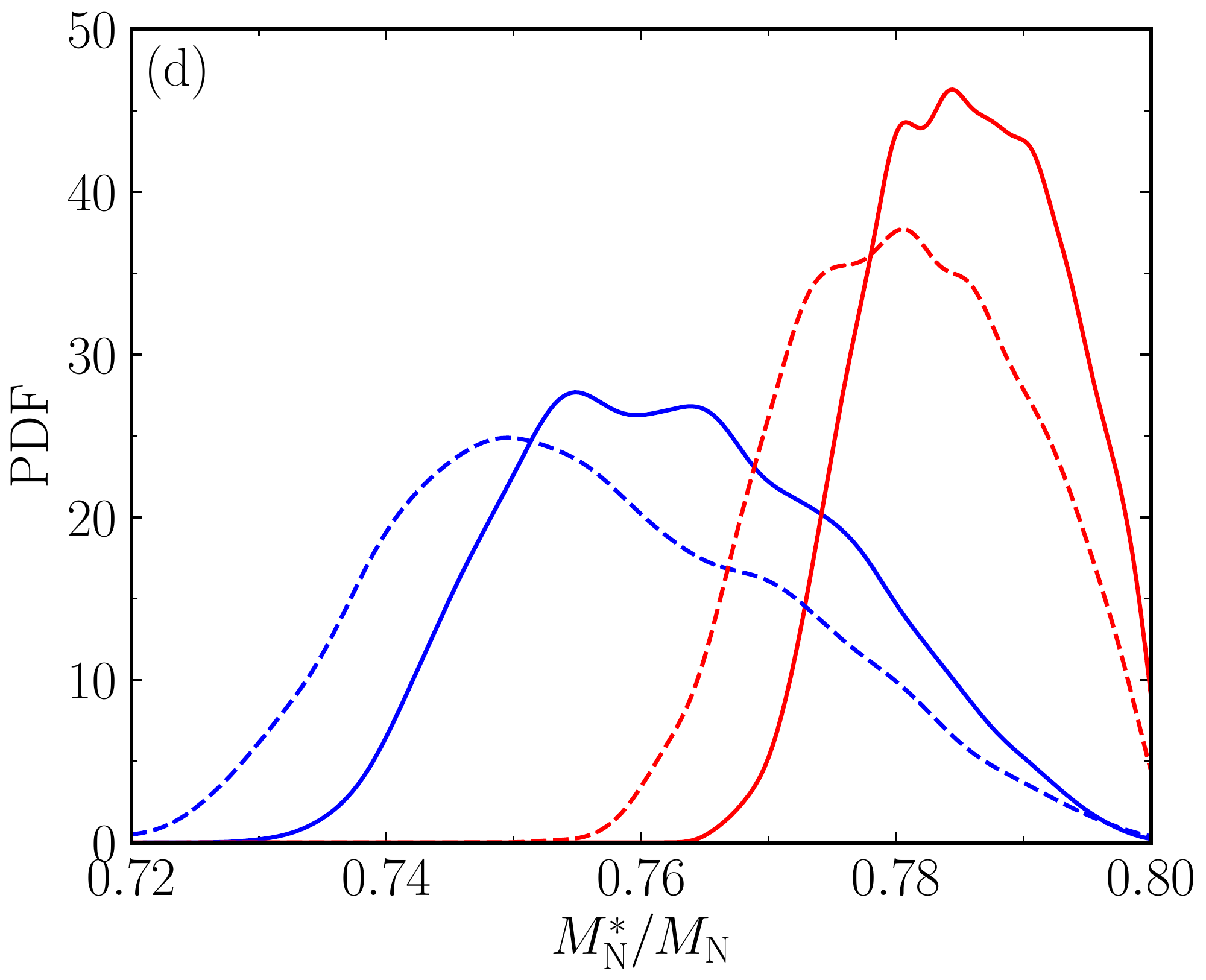}
  \caption{Posterior probability distribution functions of the quantities at saturation density for each analysis (see details in Sec.~\ref{sec:data}). The RMF and QMF models are denoted by the blue and red lines, respectively, while solid and dashed lines denote the analyses conducted with and without neutron-skin data.
  }\label{fig:sat_dis}
  \vspace{-0.1cm}
\end{figure*}
\begin{table*}
\centering
   \vskip-3mm
  \caption{Most probable intervals of various properties for nuclear matter ($90\%$ confidence interval) within the RMF and QMF framework, constrained by two different analyses with or without the skin data (from both the PREX-II experiments and the ab initio predictions).
  }
  \setlength{\tabcolsep}{0.8pt}
\renewcommand\arraystretch{1.7}
  \begin{ruledtabular}
  \begin{tabular}{lccc}      
   Parameters & 
   &  GW170817/AT2017gfo + NICER & GW170817/AT2017gfo + NICER~+ neutron-skin\\
    \hline \hline 
    \multirow{2}{*}
    {$J_0/\rm MeV$} 
    & RMF &  $40.315_{-7.253}^{+4.163} $  &  $30.867_{-1.059}^{+1.758}$
    \\
    & QMF &  $39.919_{-8.132}^{+4.454} $    &  $30.842_{-1.025}^{+1.851}$
    \\
    \hline
     \multirow{2}{*}
     {$K_0/\rm MeV$}
     & RMF  &  $233.501_{-12.040}^{+33.648} $  &  $230.860_{-9.601}^{+25.270}$
     \\
     & QMF  &  $246.219_{-15.390}^{+29.475} $  &  $238.606_{-12.161}^{+22.002}$
     \\
    \hline    
    \multirow{2}{*}
    {$L_0/\rm MeV$} 
    & RMF &  $36.901_{-14.368}^{+22.028}$  & $57.015_{-13.372}^{+9.184}$
    \\
    & QMF &  $36.723_{-14.595}^{+24.303}$  & $56.453_{-13.225}^{+10.634}$ 
    \\
    \hline    
    \multirow{2}{*}
    {$M_N^*/M_N$} 
     & RMF &  $0.755_{-0.022}^{+0.028}$ &  $0.763_{-0.019}^{+0.023}$
     \\
     & QMF &  $0.781_{-0.015}^{+0.015}$ &  $0.786_{-0.012}^{+0.012}$
     \\
    \hline    
    \multirow{2}{*}
    {  $M_{\rm TOV}\ (\Msun)$} 
     & RMF & $2.014_{-0.099}^{+0.084}$ & $1.990_{-0.082}^{+0.074}$
     \\
     & QMF & $2.011_{-0.076}^{+0.074}$  & $1.989_{-0.061}^{+0.064}$
     \\
    \hline   
    \multirow{2}{*}    
    { $R_{1.4}\ ({\rm km})$ } 
    & RMF & $11.618_{-0.239}^{+0.217}$ & $11.936_{-0.209}^{+0.167}$
    \\  
    & QMF & $11.695_{-0.250}^{+0.225}$ & $11.974_{-0.233}^{+0.155}$
    \\    
    \hline    
    \multirow{2}{*}    
    {$\Lambda_{1.4} $ } 
    & RMF & $292.187_{-37.133}^{+29.647}$ & $323.711_{-35.490}^{+26.277}$
    \\  
    & QMF  & $306.259_{-41.607}^{+30.381}$ & $332.145_{-43.056}^{+26.443}$
   \\         
  \end{tabular}
  \end{ruledtabular}
  \label{tb:2}   \vspace{-0.5cm}
\end{table*}

\subsection{Bayesian inference of the EoSs from nuclear measurements and astrophysical observations: RMF vs. QMF}

In this section, we compare the statistical performance of the RMF and QMF models using available data from gravitational wave and kilonova observation GW170817/AT2017gfo, NICER mass-radius measurements of two pulsars PSR J0030+0451 and PSR J0740+6620, PREX-II measurements of $^{208}$Pb neutron-skin, and ab initio calculations for several saturation quantities. We employ Bayesian inference to evaluate the performance of both models, using the aforementioned data to calculate the likelihood. The crucial input parameters are the nuclear properties at saturation density, namely, the symmetry energy $J_0$, incompressibility $K_0$, symmetry energy slope $L_0$, and nucleon effective mass $M_{\rm N}^\ast$.

In Fig.~\ref{fig:eos_mr}, we present the posteriors of our analyses, with the shaded regions and dashed lines representing the 90\% confidence interval of the posterior distributions, and the dash-dot and dot lines representing the median values. The RMF and QMF models are represented by red and blue regions or lines, respectively. The posteriors denoted by the shaded regions have taken into account both the observational data (GW170817/AT2017gfo + NICER) and neutron-skin data (PREX-II + EFT), while those denoted by dashed lines merely considered the observational data in their likelihoods. We note that the RMF and QMF models show consistent results in mass-radius relations for both cases, with and without neutron-skin data. Such consistency is also displayed in Table~\ref{tb:2}, where we list the median value and 90\% confidence interval for the maximum mass, radius, and tidal deformability of $1.4\Msun$ neutron star.
We also note that the analyses with neutron-skin data favor larger radii and tidal deformability of neutron stars. The results including neutron-skin data give a radius of $\approx 11.9$ km and a tidal deformability of $\approx 330$ for a $1.4\Msun$ neutron star, while the value from analyses without neutron-skin data is $\approx 11.6$ km and $\approx 300$. This difference is mainly contributed by the large value of the symmetry energy slope $L_0$ (as seen in Fig.~\ref{fig:sat_dis}), which is positively correlated with the radii and tidal deformabilities of neutron stars \cite{2018PhRvC..97c5805Z}.

We presented the posterior distributions of these saturation properties in Fig.~\ref{fig:sat_dis} and their most probable intervals in Table~\ref{tb:2}, following the same conventions as Fig.~\ref{fig:eos_mr}. 
As noted in Fig.~\ref{fig:sat_dis}, neutron-skin data favor a larger value of $L_0$. Conversely, the observational data impose a weak constraint on the symmetry energy $J_0$, which exhibits a large uncertainty for both models that do not consider neutron-skin data. However, the analyses incorporating neutron-skin data yield a smaller uncertainty and a value of $J_0$ ($\approx 31$ MeV) that agrees with other experimental measurements, see e.g., Refs.~\cite{2009PhRvL.102l2701T, 2013PhLB..727..276L, 2013ApJ...774...17S, 2014NuPhA.922....1D}. 
The difference between RMF and QMF models is manifested in the distributions of $K_0$ and $M_{\rm N}^\ast / M_{\rm N}$. 
Since QMF favors a larger value of effective mass~\cite{2018ApJ...862...98Z, 2018PhRvC..97c5805Z, 2019PhRvC..99b5804Z}, the anti-correlation between effective mass and the maximum mass of a star~\cite{2018PhRvC..98f5804H} necessitates that the QMF model increases the value of $K_0$, such that its EoS can be sufficiently stiff to support a massive neutron star with the maximum mass of approximately $2\Msun$.

\section{Conclusions} 
\label{sec:conc}
Both RMF and QMF models have been successful in describing the properties of nuclear matter, finite nuclei, and neutron stars. In the RMF model, effective nucleon-nucleon interactions are mediated by various mesons (such as $\sigma$, $\omega$, and $\rho$), and nucleon-meson couplings are adjusted to reproduce experimental or empirical data. Accordingly, the QMF model inherits most of the RMF's framework. However, a unique feature of the QMF model is the introduction of an additional procedure to calculate nucleon properties (such as the mass and radius of a nucleon) in the medium. Consequently, the QMF model is capable of self-consistently producing both nucleon properties and nucleon-meson couplings from the quark confinement and quark-meson couplings. By considering the similarities and differences between the RMF and QMF models, we systematically compare these models and discuss their capabilities for describing nuclear matter, neutron stars, and Bayesian inference, while taking into account various experimental and observational data.

We undertook a theoretical comparison of the formulation of RMF and QMF, with particular emphasis on their effective mass expressions. This is a crucial difference in the nucleon-nucleon interaction between these two models. In RMF, the effective mass linearly depends on the scalar meson field $\sigma$, whereas in QMF, it also depends on the confinement and nucleon properties, and has a nontrivial relationship with $\sigma$ as well. To perform a self-consistent comparison of the models' descriptions of symmetric nuclear matter, we first ensured that both models reproduced the same results at the saturation density of symmetric nuclear matter, i.e., $n_0$, $E/A$, $K_0$, $J_0$, $L_0$, and $M_{\rm N}^\ast/M_{\rm N}$. To accomplish this, we adjusted the coupling constants and expressed them in terms of these quantities at the saturation density. Both models were able to satisfy the constraints from experimental data. However, the models predict different pressure and symmetry energy beyond $0.4\ {\rm fm}^{-3}$ (or $\approx 2.5 n_0$) due to their different effective mass formulations.

We also conducted a Bayesian inference analysis to compare the statistical performance of both the RMF and QMF models. In this analysis, we set the saturation properties as the parameters, and their posterior distributions were inferred by considering the GW170817/AT2017gfo, NICER, and neutron-skin data. Both models showed similar posterior distributions of the EoS and the mass-radius relation of neutron stars. However, the posteriors of $K_0$ and $M_{\rm N}^\ast$ exhibited model dependence. The QMF model favored a larger $M_{\rm N}^\ast$ and accordingly increased the value of $K_0$ to satisfy astrophysical constraints that result from the anti-correlation between $M_{\rm N}^\ast$ and the stiffness of the EoS. On the other hand, we also compared the results with and without the data of neutron-skin. The comparison showed that the astrophysical data preferred a smaller $L_0$ than the neutron-skin data, which led to smaller radii of neutron stars.

\acknowledgments
We appreciate discussions with Yingxun Zhang and the XMU neutron star group.
The work is supported by the National SKA Program of China (No.~2020SKA0120300), the National Natural Science Foundation of China (grant Nos.~12273028, 12203033, 12175109,~11873040) and the China postdoctoral science foundation (No.~BX20220207 and 2022M712086).
The Bayesian sampling was performed on Siyuan Mark-I at Shanghai Jiao Tong University.


\bibliographystyle{apsrev4-1}
\bibliography{EOS_QMFvs.RMF.bib}

\begin{thebibliography}{111}%
\makeatletter
\providecommand \@ifxundefined [1]{%
 \@ifx{#1\undefined}
}%
\providecommand \@ifnum [1]{%
 \ifnum #1\expandafter \@firstoftwo
 \else \expandafter \@secondoftwo
 \fi
}%
\providecommand \@ifx [1]{%
 \ifx #1\expandafter \@firstoftwo
 \else \expandafter \@secondoftwo
 \fi
}%
\providecommand \natexlab [1]{#1}%
\providecommand \enquote  [1]{``#1''}%
\providecommand \bibnamefont  [1]{#1}%
\providecommand \bibfnamefont [1]{#1}%
\providecommand \citenamefont [1]{#1}%
\providecommand \href@noop [0]{\@secondoftwo}%
\providecommand \href [0]{\begingroup \@sanitize@url \@href}%
\providecommand \@href[1]{\@@startlink{#1}\@@href}%
\providecommand \@@href[1]{\endgroup#1\@@endlink}%
\providecommand \@sanitize@url [0]{\catcode `\\12\catcode `\$12\catcode
  `\&12\catcode `\#12\catcode `\^12\catcode `\_12\catcode `\%12\relax}%
\providecommand \@@startlink[1]{}%
\providecommand \@@endlink[0]{}%
\providecommand \url  [0]{\begingroup\@sanitize@url \@url }%
\providecommand \@url [1]{\endgroup\@href {#1}{\urlprefix }}%
\providecommand \urlprefix  [0]{URL }%
\providecommand \Eprint [0]{\href }%
\providecommand \doibase [0]{http://dx.doi.org/}%
\providecommand \selectlanguage [0]{\@gobble}%
\providecommand \bibinfo  [0]{\@secondoftwo}%
\providecommand \bibfield  [0]{\@secondoftwo}%
\providecommand \translation [1]{[#1]}%
\providecommand \BibitemOpen [0]{}%
\providecommand \bibitemStop [0]{}%
\providecommand \bibitemNoStop [0]{.\EOS\space}%
\providecommand \EOS [0]{\spacefactor3000\relax}%
\providecommand \BibitemShut  [1]{\csname bibitem#1\endcsname}%
\let\auto@bib@innerbib\@empty
\bibitem [{\citenamefont {{Toki}}\ \emph {et~al.}(1998)\citenamefont {{Toki}},
  \citenamefont {{Meyer}}, \citenamefont {{Faessler}},\ and\ \citenamefont
  {{Brockmann}}}]{1998PhRvC..58.3749T}%
  \BibitemOpen
  \bibfield  {author} {\bibinfo {author} {\bibfnamefont {H.}~\bibnamefont
  {{Toki}}}, \bibinfo {author} {\bibfnamefont {U.}~\bibnamefont {{Meyer}}},
  \bibinfo {author} {\bibfnamefont {A.}~\bibnamefont {{Faessler}}}, \ and\
  \bibinfo {author} {\bibfnamefont {R.}~\bibnamefont {{Brockmann}}},\ }\href
  {\doibase 10.1103/PhysRevC.58.3749} {\bibfield  {journal} {\bibinfo
  {journal} {\prc}\ }\textbf {\bibinfo {volume} {58}},\ \bibinfo {pages} {3749}
  (\bibinfo {year} {1998})}\BibitemShut {NoStop}%
\bibitem [{\citenamefont {{Shen}}\ and\ \citenamefont
  {{Toki}}(2000)}]{2000PhRvC..61d5205S}%
  \BibitemOpen
  \bibfield  {author} {\bibinfo {author} {\bibfnamefont {H.}~\bibnamefont
  {{Shen}}}\ and\ \bibinfo {author} {\bibfnamefont {H.}~\bibnamefont
  {{Toki}}},\ }\href {\doibase 10.1103/PhysRevC.61.045205} {\bibfield
  {journal} {\bibinfo  {journal} {\prc}\ }\textbf {\bibinfo {volume} {61}},\
  \bibinfo {eid} {045205} (\bibinfo {year} {2000})},\ \Eprint
  {http://arxiv.org/abs/nucl-th/9911046} {arXiv:nucl-th/9911046 [nucl-th]}
  \BibitemShut {NoStop}%
\bibitem [{\citenamefont {{Shen}}\ and\ \citenamefont
  {{Toki}}(2002)}]{2002NuPhA.707..469S}%
  \BibitemOpen
  \bibfield  {author} {\bibinfo {author} {\bibfnamefont {H.}~\bibnamefont
  {{Shen}}}\ and\ \bibinfo {author} {\bibfnamefont {H.}~\bibnamefont
  {{Toki}}},\ }\href {\doibase 10.1016/S0375-9474(02)00961-2} {\bibfield
  {journal} {\bibinfo  {journal} {\nphysa}\ }\textbf {\bibinfo {volume}
  {707}},\ \bibinfo {pages} {469} (\bibinfo {year} {2002})},\ \Eprint
  {http://arxiv.org/abs/nucl-th/0104072} {arXiv:nucl-th/0104072 [nucl-th]}
  \BibitemShut {NoStop}%
\bibitem [{\citenamefont {{Hu}}\ \emph
  {et~al.}(2014{\natexlab{a}})\citenamefont {{Hu}}, \citenamefont {{Li}},
  \citenamefont {{Shen}},\ and\ \citenamefont {{Toki}}}]{2014PTEP.2014a3D02H}%
  \BibitemOpen
  \bibfield  {author} {\bibinfo {author} {\bibfnamefont {J.~N.}\ \bibnamefont
  {{Hu}}}, \bibinfo {author} {\bibfnamefont {A.}~\bibnamefont {{Li}}}, \bibinfo
  {author} {\bibfnamefont {H.}~\bibnamefont {{Shen}}}, \ and\ \bibinfo {author}
  {\bibfnamefont {H.}~\bibnamefont {{Toki}}},\ }\href {\doibase
  10.1093/ptep/ptt119} {\bibfield  {journal} {\bibinfo  {journal} {Progress of
  Theoretical and Experimental Physics}\ }\textbf {\bibinfo {volume} {2014}},\
  \bibinfo {eid} {013D02} (\bibinfo {year} {2014}{\natexlab{a}})},\ \Eprint
  {http://arxiv.org/abs/1310.3602} {arXiv:1310.3602 [nucl-th]} \BibitemShut
  {NoStop}%
\bibitem [{\citenamefont {{Xing}}\ \emph {et~al.}(2016)\citenamefont {{Xing}},
  \citenamefont {{Hu}},\ and\ \citenamefont {{Shen}}}]{2016PhRvC..94d4308X}%
  \BibitemOpen
  \bibfield  {author} {\bibinfo {author} {\bibfnamefont {X.}~\bibnamefont
  {{Xing}}}, \bibinfo {author} {\bibfnamefont {J.}~\bibnamefont {{Hu}}}, \ and\
  \bibinfo {author} {\bibfnamefont {H.}~\bibnamefont {{Shen}}},\ }\href
  {\doibase 10.1103/PhysRevC.94.044308} {\bibfield  {journal} {\bibinfo
  {journal} {\prc}\ }\textbf {\bibinfo {volume} {94}},\ \bibinfo {eid} {044308}
  (\bibinfo {year} {2016})},\ \Eprint {http://arxiv.org/abs/1609.04911}
  {arXiv:1609.04911 [nucl-th]} \BibitemShut {NoStop}%
\bibitem [{\citenamefont {{Hu}}\ and\ \citenamefont
  {{Shen}}(2017)}]{2017PhRvC..96e4304H}%
  \BibitemOpen
  \bibfield  {author} {\bibinfo {author} {\bibfnamefont {J.}~\bibnamefont
  {{Hu}}}\ and\ \bibinfo {author} {\bibfnamefont {H.}~\bibnamefont {{Shen}}},\
  }\href {\doibase 10.1103/PhysRevC.96.054304} {\bibfield  {journal} {\bibinfo
  {journal} {\prc}\ }\textbf {\bibinfo {volume} {96}},\ \bibinfo {eid} {054304}
  (\bibinfo {year} {2017})},\ \Eprint {http://arxiv.org/abs/1710.08613}
  {arXiv:1710.08613 [nucl-th]} \BibitemShut {NoStop}%
\bibitem [{\citenamefont {{Xing}}\ \emph {et~al.}(2017)\citenamefont {{Xing}},
  \citenamefont {{Hu}},\ and\ \citenamefont {{Shen}}}]{2017PhRvC..95e4310X}%
  \BibitemOpen
  \bibfield  {author} {\bibinfo {author} {\bibfnamefont {X.}~\bibnamefont
  {{Xing}}}, \bibinfo {author} {\bibfnamefont {J.}~\bibnamefont {{Hu}}}, \ and\
  \bibinfo {author} {\bibfnamefont {H.}~\bibnamefont {{Shen}}},\ }\href
  {\doibase 10.1103/PhysRevC.95.054310} {\bibfield  {journal} {\bibinfo
  {journal} {\prc}\ }\textbf {\bibinfo {volume} {95}},\ \bibinfo {eid} {054310}
  (\bibinfo {year} {2017})},\ \Eprint {http://arxiv.org/abs/1704.08884}
  {arXiv:1704.08884 [nucl-th]} \BibitemShut {NoStop}%
\bibitem [{\citenamefont {{Hu}}\ \emph
  {et~al.}(2014{\natexlab{b}})\citenamefont {{Hu}}, \citenamefont {{Li}},
  \citenamefont {{Toki}},\ and\ \citenamefont {{Zuo}}}]{2014PhRvC..89b5802H}%
  \BibitemOpen
  \bibfield  {author} {\bibinfo {author} {\bibfnamefont {J.~N.}\ \bibnamefont
  {{Hu}}}, \bibinfo {author} {\bibfnamefont {A.}~\bibnamefont {{Li}}}, \bibinfo
  {author} {\bibfnamefont {H.}~\bibnamefont {{Toki}}}, \ and\ \bibinfo {author}
  {\bibfnamefont {W.}~\bibnamefont {{Zuo}}},\ }\href {\doibase
  10.1103/PhysRevC.89.025802} {\bibfield  {journal} {\bibinfo  {journal}
  {\prc}\ }\textbf {\bibinfo {volume} {89}},\ \bibinfo {eid} {025802} (\bibinfo
  {year} {2014}{\natexlab{b}})},\ \Eprint {http://arxiv.org/abs/1307.4154}
  {arXiv:1307.4154 [nucl-th]} \BibitemShut {NoStop}%
\bibitem [{\citenamefont {{Zhu}}\ \emph {et~al.}(2018)\citenamefont {{Zhu}},
  \citenamefont {{Zhou}},\ and\ \citenamefont {{Li}}}]{2018ApJ...862...98Z}%
  \BibitemOpen
  \bibfield  {author} {\bibinfo {author} {\bibfnamefont {Z.-Y.}\ \bibnamefont
  {{Zhu}}}, \bibinfo {author} {\bibfnamefont {E.-P.}\ \bibnamefont {{Zhou}}}, \
  and\ \bibinfo {author} {\bibfnamefont {A.}~\bibnamefont {{Li}}},\ }\href
  {\doibase 10.3847/1538-4357/aacc28} {\bibfield  {journal} {\bibinfo
  {journal} {\apj}\ }\textbf {\bibinfo {volume} {862}},\ \bibinfo {eid} {98}
  (\bibinfo {year} {2018})},\ \Eprint {http://arxiv.org/abs/1802.05510}
  {arXiv:1802.05510 [nucl-th]} \BibitemShut {NoStop}%
\bibitem [{\citenamefont {{Zhu}}\ and\ \citenamefont
  {{Li}}(2018)}]{2018PhRvC..97c5805Z}%
  \BibitemOpen
  \bibfield  {author} {\bibinfo {author} {\bibfnamefont {Z.-Y.}\ \bibnamefont
  {{Zhu}}}\ and\ \bibinfo {author} {\bibfnamefont {A.}~\bibnamefont {{Li}}},\
  }\href {\doibase 10.1103/PhysRevC.97.035805} {\bibfield  {journal} {\bibinfo
  {journal} {\prc}\ }\textbf {\bibinfo {volume} {97}},\ \bibinfo {eid} {035805}
  (\bibinfo {year} {2018})},\ \Eprint {http://arxiv.org/abs/1802.07441}
  {arXiv:1802.07441 [nucl-th]} \BibitemShut {NoStop}%
\bibitem [{\citenamefont {{Zhu}}\ \emph {et~al.}(2019)\citenamefont {{Zhu}},
  \citenamefont {{Li}}, \citenamefont {{Hu}},\ and\ \citenamefont
  {{Shen}}}]{2019PhRvC..99b5804Z}%
  \BibitemOpen
  \bibfield  {author} {\bibinfo {author} {\bibfnamefont {Z.-Y.}\ \bibnamefont
  {{Zhu}}}, \bibinfo {author} {\bibfnamefont {A.}~\bibnamefont {{Li}}},
  \bibinfo {author} {\bibfnamefont {J.-N.}\ \bibnamefont {{Hu}}}, \ and\
  \bibinfo {author} {\bibfnamefont {H.}~\bibnamefont {{Shen}}},\ }\href
  {\doibase 10.1103/PhysRevC.99.025804} {\bibfield  {journal} {\bibinfo
  {journal} {\prc}\ }\textbf {\bibinfo {volume} {99}},\ \bibinfo {eid} {025804}
  (\bibinfo {year} {2019})},\ \Eprint {http://arxiv.org/abs/1805.04678}
  {arXiv:1805.04678 [nucl-th]} \BibitemShut {NoStop}%
\bibitem [{\citenamefont {{Li}}\ and\ \citenamefont
  {{Zhu}}(2019)}]{2019AIPC.2127b0010L}%
  \BibitemOpen
  \bibfield  {author} {\bibinfo {author} {\bibfnamefont {A.}~\bibnamefont
  {{Li}}}\ and\ \bibinfo {author} {\bibfnamefont {Z.-Y.}\ \bibnamefont
  {{Zhu}}},\ }in\ \href {\doibase 10.1063/1.5117800} {\emph {\bibinfo
  {booktitle} {Xiamen-CUSTIPEN Workshop on the Equation of State of Dense
  Neutron-Rich Matter in the Era of Gravitational Wave Astronomy}}},\ \bibinfo
  {series} {American Institute of Physics Conference Series}, Vol.\ \bibinfo
  {volume} {2127}\ (\bibinfo {year} {2019})\ p.\ \bibinfo {pages} {020010},\
  \Eprint {http://arxiv.org/abs/1903.01280} {arXiv:1903.01280 [nucl-th]}
  \BibitemShut {NoStop}%
\bibitem [{\citenamefont {{Walecka}}(1974)}]{1974AnPhy..83..491W}%
  \BibitemOpen
  \bibfield  {author} {\bibinfo {author} {\bibfnamefont {J.~D.}\ \bibnamefont
  {{Walecka}}},\ }\href {\doibase 10.1016/0003-4916(74)90208-5} {\bibfield
  {journal} {\bibinfo  {journal} {Annals of Physics}\ }\textbf {\bibinfo
  {volume} {83}},\ \bibinfo {pages} {491} (\bibinfo {year} {1974})}\BibitemShut
  {NoStop}%
\bibitem [{\citenamefont {{Guichon}}(1988)}]{1988PhLB..200..235G}%
  \BibitemOpen
  \bibfield  {author} {\bibinfo {author} {\bibfnamefont {P.~A.~M.}\
  \bibnamefont {{Guichon}}},\ }\href {\doibase 10.1016/0370-2693(88)90762-9}
  {\bibfield  {journal} {\bibinfo  {journal} {Physics Letters B}\ }\textbf
  {\bibinfo {volume} {200}},\ \bibinfo {pages} {235} (\bibinfo {year}
  {1988})}\BibitemShut {NoStop}%
\bibitem [{\citenamefont {{Guichon}}\ \emph {et~al.}(1996)\citenamefont
  {{Guichon}}, \citenamefont {{Saito}}, \citenamefont {{Rodionov}},\ and\
  \citenamefont {{Thomas}}}]{1996NuPhA.601..349G}%
  \BibitemOpen
  \bibfield  {author} {\bibinfo {author} {\bibfnamefont {P.~A.~M.}\
  \bibnamefont {{Guichon}}}, \bibinfo {author} {\bibfnamefont {K.}~\bibnamefont
  {{Saito}}}, \bibinfo {author} {\bibfnamefont {E.}~\bibnamefont {{Rodionov}}},
  \ and\ \bibinfo {author} {\bibfnamefont {A.~W.}\ \bibnamefont {{Thomas}}},\
  }\href {\doibase 10.1016/0375-9474(96)00033-4} {\bibfield  {journal}
  {\bibinfo  {journal} {\nphysa}\ }\textbf {\bibinfo {volume} {601}},\ \bibinfo
  {pages} {349} (\bibinfo {year} {1996})},\ \Eprint
  {http://arxiv.org/abs/nucl-th/9509034} {arXiv:nucl-th/9509034 [nucl-th]}
  \BibitemShut {NoStop}%
\bibitem [{\citenamefont {{Panda}}\ \emph {et~al.}(2002)\citenamefont
  {{Panda}}, \citenamefont {{Bracco}}, \citenamefont {{Chiapparini}},
  \citenamefont {{Conte}},\ and\ \citenamefont
  {{Krein}}}]{2002PhRvC..65f5206P}%
  \BibitemOpen
  \bibfield  {author} {\bibinfo {author} {\bibfnamefont {P.~K.}\ \bibnamefont
  {{Panda}}}, \bibinfo {author} {\bibfnamefont {M.~E.}\ \bibnamefont
  {{Bracco}}}, \bibinfo {author} {\bibfnamefont {M.}~\bibnamefont
  {{Chiapparini}}}, \bibinfo {author} {\bibfnamefont {E.}~\bibnamefont
  {{Conte}}}, \ and\ \bibinfo {author} {\bibfnamefont {G.}~\bibnamefont
  {{Krein}}},\ }\href {\doibase 10.1103/PhysRevC.65.065206} {\bibfield
  {journal} {\bibinfo  {journal} {\prc}\ }\textbf {\bibinfo {volume} {65}},\
  \bibinfo {eid} {065206} (\bibinfo {year} {2002})},\ \Eprint
  {http://arxiv.org/abs/nucl-th/0205051} {arXiv:nucl-th/0205051 [nucl-th]}
  \BibitemShut {NoStop}%
\bibitem [{\citenamefont {{Miyatsu}}\ \emph {et~al.}(2012)\citenamefont
  {{Miyatsu}}, \citenamefont {{Katayama}},\ and\ \citenamefont
  {{Saito}}}]{2012PhLB..709..242M}%
  \BibitemOpen
  \bibfield  {author} {\bibinfo {author} {\bibfnamefont {T.}~\bibnamefont
  {{Miyatsu}}}, \bibinfo {author} {\bibfnamefont {T.}~\bibnamefont
  {{Katayama}}}, \ and\ \bibinfo {author} {\bibfnamefont {K.}~\bibnamefont
  {{Saito}}},\ }\href {\doibase 10.1016/j.physletb.2012.02.009} {\bibfield
  {journal} {\bibinfo  {journal} {Physics Letters B}\ }\textbf {\bibinfo
  {volume} {709}},\ \bibinfo {pages} {242} (\bibinfo {year} {2012})},\ \Eprint
  {http://arxiv.org/abs/1110.3868} {arXiv:1110.3868 [nucl-th]} \BibitemShut
  {NoStop}%
\bibitem [{\citenamefont {{Panda}}\ \emph {et~al.}(2012)\citenamefont
  {{Panda}}, \citenamefont {{Santos}}, \citenamefont {{Menezes}},\ and\
  \citenamefont {{Provid{\^e}ncia}}}]{2012PhRvC..85e5802P}%
  \BibitemOpen
  \bibfield  {author} {\bibinfo {author} {\bibfnamefont {P.~K.}\ \bibnamefont
  {{Panda}}}, \bibinfo {author} {\bibfnamefont {A.~M.~S.}\ \bibnamefont
  {{Santos}}}, \bibinfo {author} {\bibfnamefont {D.~P.}\ \bibnamefont
  {{Menezes}}}, \ and\ \bibinfo {author} {\bibfnamefont {C.}~\bibnamefont
  {{Provid{\^e}ncia}}},\ }\href {\doibase 10.1103/PhysRevC.85.055802}
  {\bibfield  {journal} {\bibinfo  {journal} {\prc}\ }\textbf {\bibinfo
  {volume} {85}},\ \bibinfo {eid} {055802} (\bibinfo {year} {2012})},\ \Eprint
  {http://arxiv.org/abs/1110.4708} {arXiv:1110.4708 [nucl-th]} \BibitemShut
  {NoStop}%
\bibitem [{\citenamefont {{Barik}}\ and\ \citenamefont
  {{Dash}}(1986)}]{1986PhRvD..33.1925B}%
  \BibitemOpen
  \bibfield  {author} {\bibinfo {author} {\bibfnamefont {N.}~\bibnamefont
  {{Barik}}}\ and\ \bibinfo {author} {\bibfnamefont {B.~K.}\ \bibnamefont
  {{Dash}}},\ }\href {\doibase 10.1103/PhysRevD.33.1925} {\bibfield  {journal}
  {\bibinfo  {journal} {\prd}\ }\textbf {\bibinfo {volume} {33}},\ \bibinfo
  {pages} {1925} (\bibinfo {year} {1986})}\BibitemShut {NoStop}%
\bibitem [{\citenamefont {Frederico}\ \emph {et~al.}(1989)\citenamefont
  {Frederico}, \citenamefont {Carlson}, \citenamefont {Rego},\ and\
  \citenamefont {Hussein}}]{1989JPhG...15..297F}%
  \BibitemOpen
  \bibfield  {author} {\bibinfo {author} {\bibfnamefont {T.}~\bibnamefont
  {Frederico}}, \bibinfo {author} {\bibfnamefont {B.~V.}\ \bibnamefont
  {Carlson}}, \bibinfo {author} {\bibfnamefont {R.~A.}\ \bibnamefont {Rego}}, \
  and\ \bibinfo {author} {\bibfnamefont {M.~S.}\ \bibnamefont {Hussein}},\
  }\href {\doibase 10.1088/0954-3899/15/3/007} {\bibfield  {journal} {\bibinfo
  {journal} {Journal of Physics G: Nuclear and Particle Physics}\ }\textbf
  {\bibinfo {volume} {15}},\ \bibinfo {pages} {297} (\bibinfo {year}
  {1989})}\BibitemShut {NoStop}%
\bibitem [{\citenamefont {{Batista}}\ \emph {et~al.}(2002)\citenamefont
  {{Batista}}, \citenamefont {{Carlson}},\ and\ \citenamefont
  {{Frederico}}}]{2002NuPhA.697..469B}%
  \BibitemOpen
  \bibfield  {author} {\bibinfo {author} {\bibfnamefont {E.~F.}\ \bibnamefont
  {{Batista}}}, \bibinfo {author} {\bibfnamefont {B.~V.}\ \bibnamefont
  {{Carlson}}}, \ and\ \bibinfo {author} {\bibfnamefont {T.}~\bibnamefont
  {{Frederico}}},\ }\href {\doibase 10.1016/S0375-9474(01)01250-7} {\bibfield
  {journal} {\bibinfo  {journal} {\nphysa}\ }\textbf {\bibinfo {volume}
  {697}},\ \bibinfo {pages} {469} (\bibinfo {year} {2002})}\BibitemShut
  {NoStop}%
\bibitem [{\citenamefont {Barik}\ \emph {et~al.}(2013)\citenamefont {Barik},
  \citenamefont {Mishra}, \citenamefont {Mohanty}, \citenamefont {Panda},\ and\
  \citenamefont {Frederico}}]{2013PhRvC..88a5206B}%
  \BibitemOpen
  \bibfield  {author} {\bibinfo {author} {\bibfnamefont {N.}~\bibnamefont
  {Barik}}, \bibinfo {author} {\bibfnamefont {R.~N.}\ \bibnamefont {Mishra}},
  \bibinfo {author} {\bibfnamefont {D.~K.}\ \bibnamefont {Mohanty}}, \bibinfo
  {author} {\bibfnamefont {P.~K.}\ \bibnamefont {Panda}}, \ and\ \bibinfo
  {author} {\bibfnamefont {T.}~\bibnamefont {Frederico}},\ }\href {\doibase
  10.1103/PhysRevC.88.015206} {\bibfield  {journal} {\bibinfo  {journal} {Phys.
  Rev. C}\ }\textbf {\bibinfo {volume} {88}},\ \bibinfo {pages} {015206}
  (\bibinfo {year} {2013})}\BibitemShut {NoStop}%
\bibitem [{\citenamefont {{Mishra}}\ \emph {et~al.}(2015)\citenamefont
  {{Mishra}}, \citenamefont {{Sahoo}}, \citenamefont {{Panda}}, \citenamefont
  {{Barik}},\ and\ \citenamefont {{Frederico}}}]{2015PhRvC..92d5203M}%
  \BibitemOpen
  \bibfield  {author} {\bibinfo {author} {\bibfnamefont {R.~N.}\ \bibnamefont
  {{Mishra}}}, \bibinfo {author} {\bibfnamefont {H.~S.}\ \bibnamefont
  {{Sahoo}}}, \bibinfo {author} {\bibfnamefont {P.~K.}\ \bibnamefont
  {{Panda}}}, \bibinfo {author} {\bibfnamefont {N.}~\bibnamefont {{Barik}}}, \
  and\ \bibinfo {author} {\bibfnamefont {T.}~\bibnamefont {{Frederico}}},\
  }\href {\doibase 10.1103/PhysRevC.92.045203} {\bibfield  {journal} {\bibinfo
  {journal} {\prc}\ }\textbf {\bibinfo {volume} {92}},\ \bibinfo {eid} {045203}
  (\bibinfo {year} {2015})},\ \Eprint {http://arxiv.org/abs/1512.02395}
  {arXiv:1512.02395 [nucl-th]} \BibitemShut {NoStop}%
\bibitem [{\citenamefont {{Mishra}}\ \emph {et~al.}(2016)\citenamefont
  {{Mishra}}, \citenamefont {{Sahoo}}, \citenamefont {{Panda}}, \citenamefont
  {{Barik}},\ and\ \citenamefont {{Frederico}}}]{2016PhRvC..94c5805M}%
  \BibitemOpen
  \bibfield  {author} {\bibinfo {author} {\bibfnamefont {R.~N.}\ \bibnamefont
  {{Mishra}}}, \bibinfo {author} {\bibfnamefont {H.~S.}\ \bibnamefont
  {{Sahoo}}}, \bibinfo {author} {\bibfnamefont {P.~K.}\ \bibnamefont
  {{Panda}}}, \bibinfo {author} {\bibfnamefont {N.}~\bibnamefont {{Barik}}}, \
  and\ \bibinfo {author} {\bibfnamefont {T.}~\bibnamefont {{Frederico}}},\
  }\href {\doibase 10.1103/PhysRevC.94.035805} {\bibfield  {journal} {\bibinfo
  {journal} {\prc}\ }\textbf {\bibinfo {volume} {94}},\ \bibinfo {eid} {035805}
  (\bibinfo {year} {2016})},\ \Eprint {http://arxiv.org/abs/1609.02708}
  {arXiv:1609.02708 [nucl-th]} \BibitemShut {NoStop}%
\bibitem [{\citenamefont {{Wu}}\ \emph {et~al.}(2020)\citenamefont {{Wu}},
  \citenamefont {{Hu}},\ and\ \citenamefont {{Shen}}}]{2020PhRvC.101b4303W}%
  \BibitemOpen
  \bibfield  {author} {\bibinfo {author} {\bibfnamefont {L.}~\bibnamefont
  {{Wu}}}, \bibinfo {author} {\bibfnamefont {J.}~\bibnamefont {{Hu}}}, \ and\
  \bibinfo {author} {\bibfnamefont {H.}~\bibnamefont {{Shen}}},\ }\href
  {\doibase 10.1103/PhysRevC.101.024303} {\bibfield  {journal} {\bibinfo
  {journal} {\prc}\ }\textbf {\bibinfo {volume} {101}},\ \bibinfo {eid}
  {024303} (\bibinfo {year} {2020})},\ \Eprint
  {http://arxiv.org/abs/2001.08882} {arXiv:2001.08882 [nucl-th]} \BibitemShut
  {NoStop}%
\bibitem [{\citenamefont {{Demorest}}\ \emph {et~al.}(2010)\citenamefont
  {{Demorest}}, \citenamefont {{Pennucci}}, \citenamefont {{Ransom}},
  \citenamefont {{Roberts}},\ and\ \citenamefont
  {{Hessels}}}]{2010Natur.467.1081D}%
  \BibitemOpen
  \bibfield  {author} {\bibinfo {author} {\bibfnamefont {P.~B.}\ \bibnamefont
  {{Demorest}}}, \bibinfo {author} {\bibfnamefont {T.}~\bibnamefont
  {{Pennucci}}}, \bibinfo {author} {\bibfnamefont {S.~M.}\ \bibnamefont
  {{Ransom}}}, \bibinfo {author} {\bibfnamefont {M.~S.~E.}\ \bibnamefont
  {{Roberts}}}, \ and\ \bibinfo {author} {\bibfnamefont {J.~W.~T.}\
  \bibnamefont {{Hessels}}},\ }\href {\doibase 10.1038/nature09466} {\bibfield
  {journal} {\bibinfo  {journal} {Nature}\ }\textbf {\bibinfo {volume} {467}},\
  \bibinfo {pages} {1081} (\bibinfo {year} {2010})},\ \Eprint
  {http://arxiv.org/abs/1010.5788} {arXiv:1010.5788 [astro-ph.HE]} \BibitemShut
  {NoStop}%
\bibitem [{\citenamefont {{Antoniadis}}\ \emph {et~al.}(2013)\citenamefont
  {{Antoniadis}}, \citenamefont {{Freire}}, \citenamefont {{Wex}},
  \citenamefont {{Tauris}}, \citenamefont {{Lynch}},\ and\ \citenamefont {{et
  al.}}}]{2013Sci...340..448A}%
  \BibitemOpen
  \bibfield  {author} {\bibinfo {author} {\bibfnamefont {J.}~\bibnamefont
  {{Antoniadis}}}, \bibinfo {author} {\bibfnamefont {P.~C.~C.}\ \bibnamefont
  {{Freire}}}, \bibinfo {author} {\bibfnamefont {N.}~\bibnamefont {{Wex}}},
  \bibinfo {author} {\bibfnamefont {T.~M.}\ \bibnamefont {{Tauris}}}, \bibinfo
  {author} {\bibfnamefont {R.~S.}\ \bibnamefont {{Lynch}}}, \ and\ \bibinfo
  {author} {\bibnamefont {{et al.}}},\ }\href {\doibase
  10.1126/science.1233232} {\bibfield  {journal} {\bibinfo  {journal}
  {Science}\ }\textbf {\bibinfo {volume} {340}},\ \bibinfo {pages} {448}
  (\bibinfo {year} {2013})},\ \Eprint {http://arxiv.org/abs/1304.6875}
  {arXiv:1304.6875 [astro-ph.HE]} \BibitemShut {NoStop}%
\bibitem [{\citenamefont {{Fonseca}}\ \emph {et~al.}(2016)\citenamefont
  {{Fonseca}}, \citenamefont {{Pennucci}}, \citenamefont {{Ellis}},
  \citenamefont {{Stairs}}, \citenamefont {{Nice}}, \citenamefont {{Ransom}},
  \citenamefont {{Demorest}}, \citenamefont {{Arzoumanian}}, \citenamefont
  {{Crowter}}, \citenamefont {{Dolch}}, \citenamefont {{Ferdman}},
  \citenamefont {{Gonzalez}}, \citenamefont {{Jones}}, \citenamefont {{Jones}},
  \citenamefont {{Lam}}, \citenamefont {{Levin}}, \citenamefont {{McLaughlin}},
  \citenamefont {{Stovall}}, \citenamefont {{Swiggum}},\ and\ \citenamefont
  {{Zhu}}}]{2016ApJ...832..167F}%
  \BibitemOpen
  \bibfield  {author} {\bibinfo {author} {\bibfnamefont {E.}~\bibnamefont
  {{Fonseca}}}, \bibinfo {author} {\bibfnamefont {T.~T.}\ \bibnamefont
  {{Pennucci}}}, \bibinfo {author} {\bibfnamefont {J.~A.}\ \bibnamefont
  {{Ellis}}}, \bibinfo {author} {\bibfnamefont {I.~H.}\ \bibnamefont
  {{Stairs}}}, \bibinfo {author} {\bibfnamefont {D.~J.}\ \bibnamefont
  {{Nice}}}, \bibinfo {author} {\bibfnamefont {S.~M.}\ \bibnamefont
  {{Ransom}}}, \bibinfo {author} {\bibfnamefont {P.~B.}\ \bibnamefont
  {{Demorest}}}, \bibinfo {author} {\bibfnamefont {Z.}~\bibnamefont
  {{Arzoumanian}}}, \bibinfo {author} {\bibfnamefont {K.}~\bibnamefont
  {{Crowter}}}, \bibinfo {author} {\bibfnamefont {T.}~\bibnamefont {{Dolch}}},
  \bibinfo {author} {\bibfnamefont {R.~D.}\ \bibnamefont {{Ferdman}}}, \bibinfo
  {author} {\bibfnamefont {M.~E.}\ \bibnamefont {{Gonzalez}}}, \bibinfo
  {author} {\bibfnamefont {G.}~\bibnamefont {{Jones}}}, \bibinfo {author}
  {\bibfnamefont {M.~L.}\ \bibnamefont {{Jones}}}, \bibinfo {author}
  {\bibfnamefont {M.~T.}\ \bibnamefont {{Lam}}}, \bibinfo {author}
  {\bibfnamefont {L.}~\bibnamefont {{Levin}}}, \bibinfo {author} {\bibfnamefont
  {M.~A.}\ \bibnamefont {{McLaughlin}}}, \bibinfo {author} {\bibfnamefont
  {K.}~\bibnamefont {{Stovall}}}, \bibinfo {author} {\bibfnamefont {J.~K.}\
  \bibnamefont {{Swiggum}}}, \ and\ \bibinfo {author} {\bibfnamefont
  {W.}~\bibnamefont {{Zhu}}},\ }\href {\doibase 10.3847/0004-637X/832/2/167}
  {\bibfield  {journal} {\bibinfo  {journal} {\apj}\ }\textbf {\bibinfo
  {volume} {832}},\ \bibinfo {eid} {167} (\bibinfo {year} {2016})},\ \Eprint
  {http://arxiv.org/abs/1603.00545} {arXiv:1603.00545 [astro-ph.HE]}
  \BibitemShut {NoStop}%
\bibitem [{\citenamefont {{Abbott}}\ \emph {et~al.}(2017)\citenamefont
  {{Abbott}}, \citenamefont {{Abbott}}, \citenamefont {{Abbott}}, \citenamefont
  {{Acernese}}, \citenamefont {{Ackley}}, \citenamefont {{Adams}},
  \citenamefont {{Adams}}, \citenamefont {{Addesso}}, \citenamefont
  {{Adhikari}}, \citenamefont {{Adya}},\ and\ \citenamefont
  {et~al.}}]{2017PhRvL.119p1101A}%
  \BibitemOpen
  \bibfield  {author} {\bibinfo {author} {\bibfnamefont {B.~P.}\ \bibnamefont
  {{Abbott}}}, \bibinfo {author} {\bibfnamefont {R.}~\bibnamefont {{Abbott}}},
  \bibinfo {author} {\bibfnamefont {T.~D.}\ \bibnamefont {{Abbott}}}, \bibinfo
  {author} {\bibfnamefont {F.}~\bibnamefont {{Acernese}}}, \bibinfo {author}
  {\bibfnamefont {K.}~\bibnamefont {{Ackley}}}, \bibinfo {author}
  {\bibfnamefont {C.}~\bibnamefont {{Adams}}}, \bibinfo {author} {\bibfnamefont
  {T.}~\bibnamefont {{Adams}}}, \bibinfo {author} {\bibfnamefont
  {P.}~\bibnamefont {{Addesso}}}, \bibinfo {author} {\bibfnamefont {R.~X.}\
  \bibnamefont {{Adhikari}}}, \bibinfo {author} {\bibfnamefont {V.~B.}\
  \bibnamefont {{Adya}}}, \ and\ \bibinfo {author} {\bibnamefont {et~al.}}
  (\bibinfo {collaboration} {LIGO Scientific Collaboration and Virgo
  Collaboration}),\ }\href {\doibase 10.1103/PhysRevLett.119.161101} {\bibfield
   {journal} {\bibinfo  {journal} {Phys. Rev. Lett.}\ }\textbf {\bibinfo
  {volume} {119}},\ \bibinfo {eid} {161101} (\bibinfo {year} {2017})},\ \Eprint
  {http://arxiv.org/abs/1710.05832} {arXiv:1710.05832 [gr-qc]} \BibitemShut
  {NoStop}%
\bibitem [{\citenamefont {{Flanagan}}\ and\ \citenamefont
  {{Hinderer}}(2008)}]{2008PhRvD..77b1502F}%
  \BibitemOpen
  \bibfield  {author} {\bibinfo {author} {\bibfnamefont {{\'E}.~{\'E}.}\
  \bibnamefont {{Flanagan}}}\ and\ \bibinfo {author} {\bibfnamefont
  {T.}~\bibnamefont {{Hinderer}}},\ }\href {\doibase
  10.1103/PhysRevD.77.021502} {\bibfield  {journal} {\bibinfo  {journal}
  {Physical Review D}\ }\textbf {\bibinfo {volume} {77}},\ \bibinfo {eid}
  {021502} (\bibinfo {year} {2008})},\ \Eprint {http://arxiv.org/abs/0709.1915}
  {arXiv:0709.1915 [astro-ph]} \BibitemShut {NoStop}%
\bibitem [{\citenamefont {{Hinderer}}(2008)}]{2008ApJ...677.1216H}%
  \BibitemOpen
  \bibfield  {author} {\bibinfo {author} {\bibfnamefont {T.}~\bibnamefont
  {{Hinderer}}},\ }\href {\doibase 10.1086/533487} {\bibfield  {journal}
  {\bibinfo  {journal} {Astrophys. J.}\ }\textbf {\bibinfo {volume} {677}},\
  \bibinfo {pages} {1216} (\bibinfo {year} {2008})},\ \Eprint
  {http://arxiv.org/abs/0711.2420} {arXiv:0711.2420} \BibitemShut {NoStop}%
\bibitem [{\citenamefont {Hinderer}\ \emph {et~al.}(2010)\citenamefont
  {Hinderer}, \citenamefont {Lackey}, \citenamefont {Lang},\ and\ \citenamefont
  {Read}}]{2010PhRvD..81l3016H}%
  \BibitemOpen
  \bibfield  {author} {\bibinfo {author} {\bibfnamefont {T.}~\bibnamefont
  {Hinderer}}, \bibinfo {author} {\bibfnamefont {B.~D.}\ \bibnamefont
  {Lackey}}, \bibinfo {author} {\bibfnamefont {R.~N.}\ \bibnamefont {Lang}}, \
  and\ \bibinfo {author} {\bibfnamefont {J.~S.}\ \bibnamefont {Read}},\ }\href
  {\doibase 10.1103/PhysRevD.81.123016} {\bibfield  {journal} {\bibinfo
  {journal} {Phys. Rev. D}\ }\textbf {\bibinfo {volume} {81}},\ \bibinfo
  {pages} {123016} (\bibinfo {year} {2010})},\ \Eprint
  {http://arxiv.org/abs/0911.3535} {arXiv:0911.3535 [astro-ph.HE]} \BibitemShut
  {NoStop}%
\bibitem [{\citenamefont {Vines}\ \emph {et~al.}(2011)\citenamefont {Vines},
  \citenamefont {Flanagan},\ and\ \citenamefont
  {Hinderer}}]{2011PhRvD..83h4051V}%
  \BibitemOpen
  \bibfield  {author} {\bibinfo {author} {\bibfnamefont {J.}~\bibnamefont
  {Vines}}, \bibinfo {author} {\bibfnamefont {E.~E.}\ \bibnamefont {Flanagan}},
  \ and\ \bibinfo {author} {\bibfnamefont {T.}~\bibnamefont {Hinderer}},\
  }\href {\doibase 10.1103/PhysRevD.83.084051} {\bibfield  {journal} {\bibinfo
  {journal} {Phys. Rev. D}\ }\textbf {\bibinfo {volume} {83}},\ \bibinfo
  {pages} {084051} (\bibinfo {year} {2011})},\ \Eprint
  {http://arxiv.org/abs/1101.1673} {arXiv:1101.1673 [gr-qc]} \BibitemShut
  {NoStop}%
\bibitem [{\citenamefont {{Harry}}\ and\ \citenamefont
  {{Hinderer}}(2018)}]{2018CQGra..35n5010H}%
  \BibitemOpen
  \bibfield  {author} {\bibinfo {author} {\bibfnamefont {I.}~\bibnamefont
  {{Harry}}}\ and\ \bibinfo {author} {\bibfnamefont {T.}~\bibnamefont
  {{Hinderer}}},\ }\href {\doibase 10.1088/1361-6382/aac7e3} {\bibfield
  {journal} {\bibinfo  {journal} {Classical and Quantum Gravity}\ }\textbf
  {\bibinfo {volume} {35}},\ \bibinfo {eid} {145010} (\bibinfo {year}
  {2018})},\ \Eprint {http://arxiv.org/abs/1801.09972} {arXiv:1801.09972
  [gr-qc]} \BibitemShut {NoStop}%
\bibitem [{\citenamefont {{Abbott}}\ \emph {et~al.}(2018)\citenamefont
  {{Abbott}}, \citenamefont {{Abbott}}, \citenamefont {{Abbott}}, \citenamefont
  {{Acernese}}, \citenamefont {{Ackley}}, \citenamefont {{Adams}},
  \citenamefont {{Adams}}, \citenamefont {{Addesso}}, \citenamefont
  {{Adhikari}}, \citenamefont {{Adya}},\ and\ \citenamefont
  {et~al.}}]{2018PhRvL.121p1101A}%
  \BibitemOpen
  \bibfield  {author} {\bibinfo {author} {\bibfnamefont {B.~P.}\ \bibnamefont
  {{Abbott}}}, \bibinfo {author} {\bibfnamefont {R.}~\bibnamefont {{Abbott}}},
  \bibinfo {author} {\bibfnamefont {T.~D.}\ \bibnamefont {{Abbott}}}, \bibinfo
  {author} {\bibfnamefont {F.}~\bibnamefont {{Acernese}}}, \bibinfo {author}
  {\bibfnamefont {K.}~\bibnamefont {{Ackley}}}, \bibinfo {author}
  {\bibfnamefont {C.}~\bibnamefont {{Adams}}}, \bibinfo {author} {\bibfnamefont
  {T.}~\bibnamefont {{Adams}}}, \bibinfo {author} {\bibfnamefont
  {P.}~\bibnamefont {{Addesso}}}, \bibinfo {author} {\bibfnamefont {R.~X.}\
  \bibnamefont {{Adhikari}}}, \bibinfo {author} {\bibfnamefont {V.~B.}\
  \bibnamefont {{Adya}}}, \ and\ \bibinfo {author} {\bibnamefont {et~al.}}
  (\bibinfo {collaboration} {LIGO Scientific Collaboration and Virgo
  Collaboration}),\ }\href {\doibase 10.1103/PhysRevLett.121.161101} {\bibfield
   {journal} {\bibinfo  {journal} {Physical Review Letters}\ }\textbf {\bibinfo
  {volume} {121}},\ \bibinfo {eid} {161101} (\bibinfo {year} {2018})},\ \Eprint
  {http://arxiv.org/abs/1805.11581} {arXiv:1805.11581 [gr-qc]} \BibitemShut
  {NoStop}%
\bibitem [{\citenamefont {{Zhou}}\ \emph {et~al.}(2018)\citenamefont {{Zhou}},
  \citenamefont {{Zhou}},\ and\ \citenamefont {{Li}}}]{2018PhRvD..97h3015Z}%
  \BibitemOpen
  \bibfield  {author} {\bibinfo {author} {\bibfnamefont {E.-P.}\ \bibnamefont
  {{Zhou}}}, \bibinfo {author} {\bibfnamefont {X.}~\bibnamefont {{Zhou}}}, \
  and\ \bibinfo {author} {\bibfnamefont {A.}~\bibnamefont {{Li}}},\ }\href
  {\doibase 10.1103/PhysRevD.97.083015} {\bibfield  {journal} {\bibinfo
  {journal} {\prd}\ }\textbf {\bibinfo {volume} {97}},\ \bibinfo {eid} {083015}
  (\bibinfo {year} {2018})},\ \Eprint {http://arxiv.org/abs/1711.04312}
  {arXiv:1711.04312 [astro-ph.HE]} \BibitemShut {NoStop}%
\bibitem [{\citenamefont {{Andreoni}}\ \emph {et~al.}(2017)\citenamefont
  {{Andreoni}}, \citenamefont {{Ackley}}, \citenamefont {{Cooke}},
  \citenamefont {{Acharyya}}, \citenamefont {{Allison}}, \citenamefont
  {{Anderson}}, \citenamefont {{Ashley}}, \citenamefont {{Baade}},\ and\
  \citenamefont {et~al.}}]{2017PASA...34...69A}%
  \BibitemOpen
  \bibfield  {author} {\bibinfo {author} {\bibfnamefont {I.}~\bibnamefont
  {{Andreoni}}}, \bibinfo {author} {\bibfnamefont {K.}~\bibnamefont
  {{Ackley}}}, \bibinfo {author} {\bibfnamefont {J.}~\bibnamefont {{Cooke}}},
  \bibinfo {author} {\bibfnamefont {A.}~\bibnamefont {{Acharyya}}}, \bibinfo
  {author} {\bibfnamefont {J.~R.}\ \bibnamefont {{Allison}}}, \bibinfo {author}
  {\bibfnamefont {G.~E.}\ \bibnamefont {{Anderson}}}, \bibinfo {author}
  {\bibfnamefont {M.~C.~B.}\ \bibnamefont {{Ashley}}}, \bibinfo {author}
  {\bibfnamefont {D.}~\bibnamefont {{Baade}}}, \ and\ \bibinfo {author}
  {\bibnamefont {et~al.}},\ }\href {\doibase 10.1017/pasa.2017.65} {\bibfield
  {journal} {\bibinfo  {journal} {PASA}\ }\textbf {\bibinfo {volume} {34}},\
  \bibinfo {eid} {e069} (\bibinfo {year} {2017})},\ \Eprint
  {http://arxiv.org/abs/1710.05846} {arXiv:1710.05846 [astro-ph.HE]}
  \BibitemShut {NoStop}%
\bibitem [{\citenamefont {{Arcavi}}\ \emph {et~al.}(2017)\citenamefont
  {{Arcavi}}, \citenamefont {{Hosseinzadeh}}, \citenamefont {{Howell}},
  \citenamefont {{McCully}}, \citenamefont {{Poznanski}}, \citenamefont
  {{Kasen}}, \citenamefont {{Barnes}}, \citenamefont {{Zaltzman}},
  \citenamefont {{Vasylyev}}, \citenamefont {{Maoz}},\ and\ \citenamefont
  {{Valenti}}}]{2017Natur.551...64A}%
  \BibitemOpen
  \bibfield  {author} {\bibinfo {author} {\bibfnamefont {I.}~\bibnamefont
  {{Arcavi}}}, \bibinfo {author} {\bibfnamefont {G.}~\bibnamefont
  {{Hosseinzadeh}}}, \bibinfo {author} {\bibfnamefont {D.~A.}\ \bibnamefont
  {{Howell}}}, \bibinfo {author} {\bibfnamefont {C.}~\bibnamefont {{McCully}}},
  \bibinfo {author} {\bibfnamefont {D.}~\bibnamefont {{Poznanski}}}, \bibinfo
  {author} {\bibfnamefont {D.}~\bibnamefont {{Kasen}}}, \bibinfo {author}
  {\bibfnamefont {J.}~\bibnamefont {{Barnes}}}, \bibinfo {author}
  {\bibfnamefont {M.}~\bibnamefont {{Zaltzman}}}, \bibinfo {author}
  {\bibfnamefont {S.}~\bibnamefont {{Vasylyev}}}, \bibinfo {author}
  {\bibfnamefont {D.}~\bibnamefont {{Maoz}}}, \ and\ \bibinfo {author}
  {\bibfnamefont {S.}~\bibnamefont {{Valenti}}},\ }\href {\doibase
  10.1038/nature24291} {\bibfield  {journal} {\bibinfo  {journal} {\nat}\
  }\textbf {\bibinfo {volume} {551}},\ \bibinfo {pages} {64} (\bibinfo {year}
  {2017})},\ \Eprint {http://arxiv.org/abs/1710.05843} {arXiv:1710.05843
  [astro-ph.HE]} \BibitemShut {NoStop}%
\bibitem [{\citenamefont {{Coulter}}\ \emph {et~al.}(2017)\citenamefont
  {{Coulter}}, \citenamefont {{Foley}}, \citenamefont {{Kilpatrick}},
  \citenamefont {{Drout}}, \citenamefont {{Piro}}, \citenamefont {{Shappee}},
  \citenamefont {{Siebert}}, \citenamefont {{Simon}}, \citenamefont {{Ulloa}},
  \citenamefont {{Kasen}}, \citenamefont {{Madore}}, \citenamefont
  {{Murguia-Berthier}}, \citenamefont {{Pan}}, \citenamefont {{Prochaska}},
  \citenamefont {{Ramirez-Ruiz}}, \citenamefont {{Rest}},\ and\ \citenamefont
  {{Rojas-Bravo}}}]{2017Sci...358.1556C}%
  \BibitemOpen
  \bibfield  {author} {\bibinfo {author} {\bibfnamefont {D.~A.}\ \bibnamefont
  {{Coulter}}}, \bibinfo {author} {\bibfnamefont {R.~J.}\ \bibnamefont
  {{Foley}}}, \bibinfo {author} {\bibfnamefont {C.~D.}\ \bibnamefont
  {{Kilpatrick}}}, \bibinfo {author} {\bibfnamefont {M.~R.}\ \bibnamefont
  {{Drout}}}, \bibinfo {author} {\bibfnamefont {A.~L.}\ \bibnamefont {{Piro}}},
  \bibinfo {author} {\bibfnamefont {B.~J.}\ \bibnamefont {{Shappee}}}, \bibinfo
  {author} {\bibfnamefont {M.~R.}\ \bibnamefont {{Siebert}}}, \bibinfo {author}
  {\bibfnamefont {J.~D.}\ \bibnamefont {{Simon}}}, \bibinfo {author}
  {\bibfnamefont {N.}~\bibnamefont {{Ulloa}}}, \bibinfo {author} {\bibfnamefont
  {D.}~\bibnamefont {{Kasen}}}, \bibinfo {author} {\bibfnamefont {B.~F.}\
  \bibnamefont {{Madore}}}, \bibinfo {author} {\bibfnamefont {A.}~\bibnamefont
  {{Murguia-Berthier}}}, \bibinfo {author} {\bibfnamefont {Y.~C.}\ \bibnamefont
  {{Pan}}}, \bibinfo {author} {\bibfnamefont {J.~X.}\ \bibnamefont
  {{Prochaska}}}, \bibinfo {author} {\bibfnamefont {E.}~\bibnamefont
  {{Ramirez-Ruiz}}}, \bibinfo {author} {\bibfnamefont {A.}~\bibnamefont
  {{Rest}}}, \ and\ \bibinfo {author} {\bibfnamefont {C.}~\bibnamefont
  {{Rojas-Bravo}}},\ }\href {\doibase 10.1126/science.aap9811} {\bibfield
  {journal} {\bibinfo  {journal} {Science}\ }\textbf {\bibinfo {volume}
  {358}},\ \bibinfo {pages} {1556} (\bibinfo {year} {2017})},\ \Eprint
  {http://arxiv.org/abs/1710.05452} {arXiv:1710.05452 [astro-ph.HE]}
  \BibitemShut {NoStop}%
\bibitem [{\citenamefont {{Cowperthwaite}}\ \emph {et~al.}(2017)\citenamefont
  {{Cowperthwaite}}, \citenamefont {{Berger}}, \citenamefont {{Villar}},
  \citenamefont {{Metzger}}, \citenamefont {{Nicholl}}, \citenamefont
  {{Chornock}}, \citenamefont {{Blanchard}}, \citenamefont {{Fong}},
  \citenamefont {{Margutti}}, \citenamefont {{Soares-Santos}}, \citenamefont
  {{Alexander}}, \citenamefont {{Allam}}, \citenamefont {{Annis}},
  \citenamefont {{Brout}}, \citenamefont {{Brown}}, \citenamefont {{Butler}}
  \emph {et~al.}}]{2017ApJ...848L..17C}%
  \BibitemOpen
  \bibfield  {author} {\bibinfo {author} {\bibfnamefont {P.~S.}\ \bibnamefont
  {{Cowperthwaite}}}, \bibinfo {author} {\bibfnamefont {E.}~\bibnamefont
  {{Berger}}}, \bibinfo {author} {\bibfnamefont {V.~A.}\ \bibnamefont
  {{Villar}}}, \bibinfo {author} {\bibfnamefont {B.~D.}\ \bibnamefont
  {{Metzger}}}, \bibinfo {author} {\bibfnamefont {M.}~\bibnamefont
  {{Nicholl}}}, \bibinfo {author} {\bibfnamefont {R.}~\bibnamefont
  {{Chornock}}}, \bibinfo {author} {\bibfnamefont {P.~K.}\ \bibnamefont
  {{Blanchard}}}, \bibinfo {author} {\bibfnamefont {W.}~\bibnamefont {{Fong}}},
  \bibinfo {author} {\bibfnamefont {R.}~\bibnamefont {{Margutti}}}, \bibinfo
  {author} {\bibfnamefont {M.}~\bibnamefont {{Soares-Santos}}}, \bibinfo
  {author} {\bibfnamefont {K.~D.}\ \bibnamefont {{Alexander}}}, \bibinfo
  {author} {\bibfnamefont {S.}~\bibnamefont {{Allam}}}, \bibinfo {author}
  {\bibfnamefont {J.}~\bibnamefont {{Annis}}}, \bibinfo {author} {\bibfnamefont
  {D.}~\bibnamefont {{Brout}}}, \bibinfo {author} {\bibfnamefont {D.~A.}\
  \bibnamefont {{Brown}}}, \bibinfo {author} {\bibfnamefont {R.~E.}\
  \bibnamefont {{Butler}}},  \emph {et~al.},\ }\href {\doibase
  10.3847/2041-8213/aa8fc7} {\bibfield  {journal} {\bibinfo  {journal} {\apjl}\
  }\textbf {\bibinfo {volume} {848}},\ \bibinfo {eid} {L17} (\bibinfo {year}
  {2017})},\ \Eprint {http://arxiv.org/abs/1710.05840} {arXiv:1710.05840
  [astro-ph.HE]} \BibitemShut {NoStop}%
\bibitem [{\citenamefont {{D{\'\i}az}}\ \emph {et~al.}(2017)\citenamefont
  {{D{\'\i}az}}, \citenamefont {{Macri}}, \citenamefont {{Garcia Lambas}},
  \citenamefont {{Mendes de Oliveira}}, \citenamefont {{Nilo Castell{\'o}n}},
  \citenamefont {{Ribeiro}}, \citenamefont {{S{\'a}nchez}}, \citenamefont
  {{Schoenell}}, \citenamefont {{Abramo}}, \citenamefont {{Akras}},
  \citenamefont {{Alcaniz}}, \citenamefont {{Artola}}, \citenamefont
  {{Beroiz}}, \citenamefont {{Bonoli}}, \citenamefont {{Cabral}} \emph
  {et~al.}}]{2017ApJ...848L..29D}%
  \BibitemOpen
  \bibfield  {author} {\bibinfo {author} {\bibfnamefont {M.~C.}\ \bibnamefont
  {{D{\'\i}az}}}, \bibinfo {author} {\bibfnamefont {L.~M.}\ \bibnamefont
  {{Macri}}}, \bibinfo {author} {\bibfnamefont {D.}~\bibnamefont {{Garcia
  Lambas}}}, \bibinfo {author} {\bibfnamefont {C.}~\bibnamefont {{Mendes de
  Oliveira}}}, \bibinfo {author} {\bibfnamefont {J.~L.}\ \bibnamefont {{Nilo
  Castell{\'o}n}}}, \bibinfo {author} {\bibfnamefont {T.}~\bibnamefont
  {{Ribeiro}}}, \bibinfo {author} {\bibfnamefont {B.}~\bibnamefont
  {{S{\'a}nchez}}}, \bibinfo {author} {\bibfnamefont {W.}~\bibnamefont
  {{Schoenell}}}, \bibinfo {author} {\bibfnamefont {L.~R.}\ \bibnamefont
  {{Abramo}}}, \bibinfo {author} {\bibfnamefont {S.}~\bibnamefont {{Akras}}},
  \bibinfo {author} {\bibfnamefont {J.~S.}\ \bibnamefont {{Alcaniz}}}, \bibinfo
  {author} {\bibfnamefont {R.}~\bibnamefont {{Artola}}}, \bibinfo {author}
  {\bibfnamefont {M.}~\bibnamefont {{Beroiz}}}, \bibinfo {author}
  {\bibfnamefont {S.}~\bibnamefont {{Bonoli}}}, \bibinfo {author}
  {\bibfnamefont {J.}~\bibnamefont {{Cabral}}},  \emph {et~al.},\ }\href
  {\doibase 10.3847/2041-8213/aa9060} {\bibfield  {journal} {\bibinfo
  {journal} {\apjl}\ }\textbf {\bibinfo {volume} {848}},\ \bibinfo {eid} {L29}
  (\bibinfo {year} {2017})},\ \Eprint {http://arxiv.org/abs/1710.05844}
  {arXiv:1710.05844 [astro-ph.HE]} \BibitemShut {NoStop}%
\bibitem [{\citenamefont {{Drout}}\ \emph {et~al.}(2017)\citenamefont
  {{Drout}}, \citenamefont {{Piro}}, \citenamefont {{Shappee}}, \citenamefont
  {{Kilpatrick}}, \citenamefont {{Simon}}, \citenamefont {{Contreras}},
  \citenamefont {{Coulter}}, \citenamefont {{Foley}}, \citenamefont
  {{Siebert}}, \citenamefont {{Morrell}}, \citenamefont {{Boutsia}},
  \citenamefont {{Di Mille}}, \citenamefont {{Holoien}}, \citenamefont
  {{Kasen}}, \citenamefont {{Kollmeier}} \emph {et~al.}}]{2017Sci...358.1570D}%
  \BibitemOpen
  \bibfield  {author} {\bibinfo {author} {\bibfnamefont {M.~R.}\ \bibnamefont
  {{Drout}}}, \bibinfo {author} {\bibfnamefont {A.~L.}\ \bibnamefont {{Piro}}},
  \bibinfo {author} {\bibfnamefont {B.~J.}\ \bibnamefont {{Shappee}}}, \bibinfo
  {author} {\bibfnamefont {C.~D.}\ \bibnamefont {{Kilpatrick}}}, \bibinfo
  {author} {\bibfnamefont {J.~D.}\ \bibnamefont {{Simon}}}, \bibinfo {author}
  {\bibfnamefont {C.}~\bibnamefont {{Contreras}}}, \bibinfo {author}
  {\bibfnamefont {D.~A.}\ \bibnamefont {{Coulter}}}, \bibinfo {author}
  {\bibfnamefont {R.~J.}\ \bibnamefont {{Foley}}}, \bibinfo {author}
  {\bibfnamefont {M.~R.}\ \bibnamefont {{Siebert}}}, \bibinfo {author}
  {\bibfnamefont {N.}~\bibnamefont {{Morrell}}}, \bibinfo {author}
  {\bibfnamefont {K.}~\bibnamefont {{Boutsia}}}, \bibinfo {author}
  {\bibfnamefont {F.}~\bibnamefont {{Di Mille}}}, \bibinfo {author}
  {\bibfnamefont {T.~W.~S.}\ \bibnamefont {{Holoien}}}, \bibinfo {author}
  {\bibfnamefont {D.}~\bibnamefont {{Kasen}}}, \bibinfo {author} {\bibfnamefont
  {J.~A.}\ \bibnamefont {{Kollmeier}}},  \emph {et~al.},\ }\href {\doibase
  10.1126/science.aaq0049} {\bibfield  {journal} {\bibinfo  {journal}
  {Science}\ }\textbf {\bibinfo {volume} {358}},\ \bibinfo {pages} {1570}
  (\bibinfo {year} {2017})},\ \Eprint {http://arxiv.org/abs/1710.05443}
  {arXiv:1710.05443 [astro-ph.HE]} \BibitemShut {NoStop}%
\bibitem [{\citenamefont {{Evans}}\ \emph {et~al.}(2017)\citenamefont
  {{Evans}}, \citenamefont {{Cenko}}, \citenamefont {{Kennea}}, \citenamefont
  {{Emery}}, \citenamefont {{Kuin}}, \citenamefont {{Korobkin}}, \citenamefont
  {{Wollaeger}}, \citenamefont {{Fryer}}, \citenamefont {{Madsen}},
  \citenamefont {{Harrison}}, \citenamefont {{Xu}}, \citenamefont {{Nakar}},
  \citenamefont {{Hotokezaka}}, \citenamefont {{Lien}}, \citenamefont
  {{Campana}} \emph {et~al.}}]{2017Sci...358.1565E}%
  \BibitemOpen
  \bibfield  {author} {\bibinfo {author} {\bibfnamefont {P.~A.}\ \bibnamefont
  {{Evans}}}, \bibinfo {author} {\bibfnamefont {S.~B.}\ \bibnamefont
  {{Cenko}}}, \bibinfo {author} {\bibfnamefont {J.~A.}\ \bibnamefont
  {{Kennea}}}, \bibinfo {author} {\bibfnamefont {S.~W.~K.}\ \bibnamefont
  {{Emery}}}, \bibinfo {author} {\bibfnamefont {N.~P.~M.}\ \bibnamefont
  {{Kuin}}}, \bibinfo {author} {\bibfnamefont {O.}~\bibnamefont {{Korobkin}}},
  \bibinfo {author} {\bibfnamefont {R.~T.}\ \bibnamefont {{Wollaeger}}},
  \bibinfo {author} {\bibfnamefont {C.~L.}\ \bibnamefont {{Fryer}}}, \bibinfo
  {author} {\bibfnamefont {K.~K.}\ \bibnamefont {{Madsen}}}, \bibinfo {author}
  {\bibfnamefont {F.~A.}\ \bibnamefont {{Harrison}}}, \bibinfo {author}
  {\bibfnamefont {Y.}~\bibnamefont {{Xu}}}, \bibinfo {author} {\bibfnamefont
  {E.}~\bibnamefont {{Nakar}}}, \bibinfo {author} {\bibfnamefont
  {K.}~\bibnamefont {{Hotokezaka}}}, \bibinfo {author} {\bibfnamefont
  {A.}~\bibnamefont {{Lien}}}, \bibinfo {author} {\bibfnamefont
  {S.}~\bibnamefont {{Campana}}},  \emph {et~al.},\ }\href {\doibase
  10.1126/science.aap9580} {\bibfield  {journal} {\bibinfo  {journal}
  {Science}\ }\textbf {\bibinfo {volume} {358}},\ \bibinfo {pages} {1565}
  (\bibinfo {year} {2017})},\ \Eprint {http://arxiv.org/abs/1710.05437}
  {arXiv:1710.05437 [astro-ph.HE]} \BibitemShut {NoStop}%
\bibitem [{\citenamefont {{Hu}}\ \emph {et~al.}(2017)\citenamefont {{Hu}},
  \citenamefont {{Wu}}, \citenamefont {{Andreoni}}, \citenamefont {{Ashley}},
  \citenamefont {{Cooke}}, \citenamefont {{Cui}}, \citenamefont {{Du}},
  \citenamefont {{Dai}}, \citenamefont {{Gu}}, \citenamefont {{Hu}},
  \citenamefont {{Lu}}, \citenamefont {{Li}}, \citenamefont {{Li}},
  \citenamefont {{Liang}}, \citenamefont {{Liu}} \emph
  {et~al.}}]{2017SciBu..62.1433H}%
  \BibitemOpen
  \bibfield  {author} {\bibinfo {author} {\bibfnamefont {L.}~\bibnamefont
  {{Hu}}}, \bibinfo {author} {\bibfnamefont {X.}~\bibnamefont {{Wu}}}, \bibinfo
  {author} {\bibfnamefont {I.}~\bibnamefont {{Andreoni}}}, \bibinfo {author}
  {\bibfnamefont {M.~C.~B.}\ \bibnamefont {{Ashley}}}, \bibinfo {author}
  {\bibfnamefont {J.}~\bibnamefont {{Cooke}}}, \bibinfo {author} {\bibfnamefont
  {X.}~\bibnamefont {{Cui}}}, \bibinfo {author} {\bibfnamefont
  {F.}~\bibnamefont {{Du}}}, \bibinfo {author} {\bibfnamefont {Z.}~\bibnamefont
  {{Dai}}}, \bibinfo {author} {\bibfnamefont {B.}~\bibnamefont {{Gu}}},
  \bibinfo {author} {\bibfnamefont {Y.}~\bibnamefont {{Hu}}}, \bibinfo {author}
  {\bibfnamefont {H.}~\bibnamefont {{Lu}}}, \bibinfo {author} {\bibfnamefont
  {X.}~\bibnamefont {{Li}}}, \bibinfo {author} {\bibfnamefont {Z.}~\bibnamefont
  {{Li}}}, \bibinfo {author} {\bibfnamefont {E.}~\bibnamefont {{Liang}}},
  \bibinfo {author} {\bibfnamefont {L.}~\bibnamefont {{Liu}}},  \emph
  {et~al.},\ }\href {\doibase 10.1016/j.scib.2017.10.006} {\bibfield  {journal}
  {\bibinfo  {journal} {Science Bulletin}\ }\textbf {\bibinfo {volume} {62}},\
  \bibinfo {pages} {1433} (\bibinfo {year} {2017})},\ \Eprint
  {http://arxiv.org/abs/1710.05462} {arXiv:1710.05462 [astro-ph.HE]}
  \BibitemShut {NoStop}%
\bibitem [{\citenamefont {{Kasliwal}}\ \emph {et~al.}(2017)\citenamefont
  {{Kasliwal}}, \citenamefont {{Nakar}}, \citenamefont {{Singer}},
  \citenamefont {{Kaplan}}, \citenamefont {{Cook}}, \citenamefont {{Van
  Sistine}}, \citenamefont {{Lau}}, \citenamefont {{Fremling}}, \citenamefont
  {{Gottlieb}}, \citenamefont {{Jencson}}, \citenamefont {{Adams}},
  \citenamefont {{Feindt}}, \citenamefont {{Hotokezaka}}, \citenamefont
  {{Ghosh}}, \citenamefont {{Perley}}, \citenamefont {{Yu}} \emph
  {et~al.}}]{2017Sci...358.1559K}%
  \BibitemOpen
  \bibfield  {author} {\bibinfo {author} {\bibfnamefont {M.~M.}\ \bibnamefont
  {{Kasliwal}}}, \bibinfo {author} {\bibfnamefont {E.}~\bibnamefont {{Nakar}}},
  \bibinfo {author} {\bibfnamefont {L.~P.}\ \bibnamefont {{Singer}}}, \bibinfo
  {author} {\bibfnamefont {D.~L.}\ \bibnamefont {{Kaplan}}}, \bibinfo {author}
  {\bibfnamefont {D.~O.}\ \bibnamefont {{Cook}}}, \bibinfo {author}
  {\bibfnamefont {A.}~\bibnamefont {{Van Sistine}}}, \bibinfo {author}
  {\bibfnamefont {R.~M.}\ \bibnamefont {{Lau}}}, \bibinfo {author}
  {\bibfnamefont {C.}~\bibnamefont {{Fremling}}}, \bibinfo {author}
  {\bibfnamefont {O.}~\bibnamefont {{Gottlieb}}}, \bibinfo {author}
  {\bibfnamefont {J.~E.}\ \bibnamefont {{Jencson}}}, \bibinfo {author}
  {\bibfnamefont {S.~M.}\ \bibnamefont {{Adams}}}, \bibinfo {author}
  {\bibfnamefont {U.}~\bibnamefont {{Feindt}}}, \bibinfo {author}
  {\bibfnamefont {K.}~\bibnamefont {{Hotokezaka}}}, \bibinfo {author}
  {\bibfnamefont {S.}~\bibnamefont {{Ghosh}}}, \bibinfo {author} {\bibfnamefont
  {D.~A.}\ \bibnamefont {{Perley}}}, \bibinfo {author} {\bibfnamefont {P.~C.}\
  \bibnamefont {{Yu}}},  \emph {et~al.},\ }\href {\doibase
  10.1126/science.aap9455} {\bibfield  {journal} {\bibinfo  {journal}
  {Science}\ }\textbf {\bibinfo {volume} {358}},\ \bibinfo {pages} {1559}
  (\bibinfo {year} {2017})},\ \Eprint {http://arxiv.org/abs/1710.05436}
  {arXiv:1710.05436 [astro-ph.HE]} \BibitemShut {NoStop}%
\bibitem [{\citenamefont {{Lipunov}}\ \emph {et~al.}(2017)\citenamefont
  {{Lipunov}}, \citenamefont {{Gorbovskoy}}, \citenamefont {{Kornilov}},
  \citenamefont {{. Tyurina}}, \citenamefont {{Balanutsa}}, \citenamefont
  {{Kuznetsov}}, \citenamefont {{Vlasenko}}, \citenamefont {{Kuvshinov}},
  \citenamefont {{Gorbunov}}, \citenamefont {{Buckley}}, \citenamefont
  {{Krylov}}, \citenamefont {{Podesta}}, \citenamefont {{Lopez}}, \citenamefont
  {{Podesta}}, \citenamefont {{Levato}} \emph {et~al.}}]{2017ApJ...850L...1L}%
  \BibitemOpen
  \bibfield  {author} {\bibinfo {author} {\bibfnamefont {V.~M.}\ \bibnamefont
  {{Lipunov}}}, \bibinfo {author} {\bibfnamefont {E.}~\bibnamefont
  {{Gorbovskoy}}}, \bibinfo {author} {\bibfnamefont {V.~G.}\ \bibnamefont
  {{Kornilov}}}, \bibinfo {author} {\bibfnamefont {N.}~\bibnamefont {{.
  Tyurina}}}, \bibinfo {author} {\bibfnamefont {P.}~\bibnamefont
  {{Balanutsa}}}, \bibinfo {author} {\bibfnamefont {A.}~\bibnamefont
  {{Kuznetsov}}}, \bibinfo {author} {\bibfnamefont {D.}~\bibnamefont
  {{Vlasenko}}}, \bibinfo {author} {\bibfnamefont {D.}~\bibnamefont
  {{Kuvshinov}}}, \bibinfo {author} {\bibfnamefont {I.}~\bibnamefont
  {{Gorbunov}}}, \bibinfo {author} {\bibfnamefont {D.~A.~H.}\ \bibnamefont
  {{Buckley}}}, \bibinfo {author} {\bibfnamefont {A.~V.}\ \bibnamefont
  {{Krylov}}}, \bibinfo {author} {\bibfnamefont {R.}~\bibnamefont {{Podesta}}},
  \bibinfo {author} {\bibfnamefont {C.}~\bibnamefont {{Lopez}}}, \bibinfo
  {author} {\bibfnamefont {F.}~\bibnamefont {{Podesta}}}, \bibinfo {author}
  {\bibfnamefont {H.}~\bibnamefont {{Levato}}},  \emph {et~al.},\ }\href
  {\doibase 10.3847/2041-8213/aa92c0} {\bibfield  {journal} {\bibinfo
  {journal} {\apjl}\ }\textbf {\bibinfo {volume} {850}},\ \bibinfo {eid} {L1}
  (\bibinfo {year} {2017})},\ \Eprint {http://arxiv.org/abs/1710.05461}
  {arXiv:1710.05461 [astro-ph.HE]} \BibitemShut {NoStop}%
\bibitem [{\citenamefont {{Pian}}\ \emph {et~al.}(2017)\citenamefont {{Pian}},
  \citenamefont {{D'Avanzo}}, \citenamefont {{Benetti}}, \citenamefont
  {{Branchesi}}, \citenamefont {{Brocato}}, \citenamefont {{Campana}},
  \citenamefont {{Cappellaro}}, \citenamefont {{Covino}}, \citenamefont
  {{D'Elia}}, \citenamefont {{Fynbo}}, \citenamefont {{Getman}}, \citenamefont
  {{Ghirlanda}}, \citenamefont {{Ghisellini}}, \citenamefont {{Grado}},
  \citenamefont {{Greco}} \emph {et~al.}}]{2017Natur.551...67P}%
  \BibitemOpen
  \bibfield  {author} {\bibinfo {author} {\bibfnamefont {E.}~\bibnamefont
  {{Pian}}}, \bibinfo {author} {\bibfnamefont {P.}~\bibnamefont {{D'Avanzo}}},
  \bibinfo {author} {\bibfnamefont {S.}~\bibnamefont {{Benetti}}}, \bibinfo
  {author} {\bibfnamefont {M.}~\bibnamefont {{Branchesi}}}, \bibinfo {author}
  {\bibfnamefont {E.}~\bibnamefont {{Brocato}}}, \bibinfo {author}
  {\bibfnamefont {S.}~\bibnamefont {{Campana}}}, \bibinfo {author}
  {\bibfnamefont {E.}~\bibnamefont {{Cappellaro}}}, \bibinfo {author}
  {\bibfnamefont {S.}~\bibnamefont {{Covino}}}, \bibinfo {author}
  {\bibfnamefont {V.}~\bibnamefont {{D'Elia}}}, \bibinfo {author}
  {\bibfnamefont {J.~P.~U.}\ \bibnamefont {{Fynbo}}}, \bibinfo {author}
  {\bibfnamefont {F.}~\bibnamefont {{Getman}}}, \bibinfo {author}
  {\bibfnamefont {G.}~\bibnamefont {{Ghirlanda}}}, \bibinfo {author}
  {\bibfnamefont {G.}~\bibnamefont {{Ghisellini}}}, \bibinfo {author}
  {\bibfnamefont {A.}~\bibnamefont {{Grado}}}, \bibinfo {author} {\bibfnamefont
  {G.}~\bibnamefont {{Greco}}},  \emph {et~al.},\ }\href {\doibase
  10.1038/nature24298} {\bibfield  {journal} {\bibinfo  {journal} {\nat}\
  }\textbf {\bibinfo {volume} {551}},\ \bibinfo {pages} {67} (\bibinfo {year}
  {2017})},\ \Eprint {http://arxiv.org/abs/1710.05858} {arXiv:1710.05858
  [astro-ph.HE]} \BibitemShut {NoStop}%
\bibitem [{\citenamefont {{Pozanenko}}\ \emph {et~al.}(2018)\citenamefont
  {{Pozanenko}}, \citenamefont {{Barkov}}, \citenamefont {{Minaev}},
  \citenamefont {{Volnova}}, \citenamefont {{Mazaeva}}, \citenamefont
  {{Moskvitin}}, \citenamefont {{Krugov}}, \citenamefont {{Samodurov}},
  \citenamefont {{Loznikov}},\ and\ \citenamefont
  {{Lyutikov}}}]{2018ApJ...852L..30P}%
  \BibitemOpen
  \bibfield  {author} {\bibinfo {author} {\bibfnamefont {A.~S.}\ \bibnamefont
  {{Pozanenko}}}, \bibinfo {author} {\bibfnamefont {M.~V.}\ \bibnamefont
  {{Barkov}}}, \bibinfo {author} {\bibfnamefont {P.~Y.}\ \bibnamefont
  {{Minaev}}}, \bibinfo {author} {\bibfnamefont {A.~A.}\ \bibnamefont
  {{Volnova}}}, \bibinfo {author} {\bibfnamefont {E.~D.}\ \bibnamefont
  {{Mazaeva}}}, \bibinfo {author} {\bibfnamefont {A.~S.}\ \bibnamefont
  {{Moskvitin}}}, \bibinfo {author} {\bibfnamefont {M.~A.}\ \bibnamefont
  {{Krugov}}}, \bibinfo {author} {\bibfnamefont {V.~A.}\ \bibnamefont
  {{Samodurov}}}, \bibinfo {author} {\bibfnamefont {V.~M.}\ \bibnamefont
  {{Loznikov}}}, \ and\ \bibinfo {author} {\bibfnamefont {M.}~\bibnamefont
  {{Lyutikov}}},\ }\href {\doibase 10.3847/2041-8213/aaa2f6} {\bibfield
  {journal} {\bibinfo  {journal} {\apjl}\ }\textbf {\bibinfo {volume} {852}},\
  \bibinfo {eid} {L30} (\bibinfo {year} {2018})},\ \Eprint
  {http://arxiv.org/abs/1710.05448} {arXiv:1710.05448 [astro-ph.HE]}
  \BibitemShut {NoStop}%
\bibitem [{\citenamefont {{Shappee}}\ \emph {et~al.}(2017)\citenamefont
  {{Shappee}}, \citenamefont {{Simon}}, \citenamefont {{Drout}}, \citenamefont
  {{Piro}}, \citenamefont {{Morrell}}, \citenamefont {{Prieto}}, \citenamefont
  {{Kasen}}, \citenamefont {{Holoien}}, \citenamefont {{Kollmeier}},
  \citenamefont {{Kelson}}, \citenamefont {{Coulter}}, \citenamefont {{Foley}},
  \citenamefont {{Kilpatrick}}, \citenamefont {{Siebert}}, \citenamefont
  {{Madore}} \emph {et~al.}}]{2017Sci...358.1574S}%
  \BibitemOpen
  \bibfield  {author} {\bibinfo {author} {\bibfnamefont {B.~J.}\ \bibnamefont
  {{Shappee}}}, \bibinfo {author} {\bibfnamefont {J.~D.}\ \bibnamefont
  {{Simon}}}, \bibinfo {author} {\bibfnamefont {M.~R.}\ \bibnamefont
  {{Drout}}}, \bibinfo {author} {\bibfnamefont {A.~L.}\ \bibnamefont {{Piro}}},
  \bibinfo {author} {\bibfnamefont {N.}~\bibnamefont {{Morrell}}}, \bibinfo
  {author} {\bibfnamefont {J.~L.}\ \bibnamefont {{Prieto}}}, \bibinfo {author}
  {\bibfnamefont {D.}~\bibnamefont {{Kasen}}}, \bibinfo {author} {\bibfnamefont
  {T.~W.~S.}\ \bibnamefont {{Holoien}}}, \bibinfo {author} {\bibfnamefont
  {J.~A.}\ \bibnamefont {{Kollmeier}}}, \bibinfo {author} {\bibfnamefont
  {D.~D.}\ \bibnamefont {{Kelson}}}, \bibinfo {author} {\bibfnamefont {D.~A.}\
  \bibnamefont {{Coulter}}}, \bibinfo {author} {\bibfnamefont {R.~J.}\
  \bibnamefont {{Foley}}}, \bibinfo {author} {\bibfnamefont {C.~D.}\
  \bibnamefont {{Kilpatrick}}}, \bibinfo {author} {\bibfnamefont {M.~R.}\
  \bibnamefont {{Siebert}}}, \bibinfo {author} {\bibfnamefont {B.~F.}\
  \bibnamefont {{Madore}}},  \emph {et~al.},\ }\href {\doibase
  10.1126/science.aaq0186} {\bibfield  {journal} {\bibinfo  {journal}
  {Science}\ }\textbf {\bibinfo {volume} {358}},\ \bibinfo {pages} {1574}
  (\bibinfo {year} {2017})},\ \Eprint {http://arxiv.org/abs/1710.05432}
  {arXiv:1710.05432 [astro-ph.HE]} \BibitemShut {NoStop}%
\bibitem [{\citenamefont {{Smartt}}\ \emph {et~al.}(2017)\citenamefont
  {{Smartt}}, \citenamefont {{Chen}}, \citenamefont {{Jerkstrand}},
  \citenamefont {{Coughlin}}, \citenamefont {{Kankare}}, \citenamefont {{Sim}},
  \citenamefont {{Fraser}}, \citenamefont {{Inserra}}, \citenamefont
  {{Maguire}}, \citenamefont {{Chambers}}, \citenamefont {{Huber}},
  \citenamefont {{Kr{\"u}hler}}, \citenamefont {{Leloudas}}, \citenamefont
  {{Magee}}, \citenamefont {{Shingles}} \emph {et~al.}}]{2017Natur.551...75S}%
  \BibitemOpen
  \bibfield  {author} {\bibinfo {author} {\bibfnamefont {S.~J.}\ \bibnamefont
  {{Smartt}}}, \bibinfo {author} {\bibfnamefont {T.~W.}\ \bibnamefont
  {{Chen}}}, \bibinfo {author} {\bibfnamefont {A.}~\bibnamefont
  {{Jerkstrand}}}, \bibinfo {author} {\bibfnamefont {M.}~\bibnamefont
  {{Coughlin}}}, \bibinfo {author} {\bibfnamefont {E.}~\bibnamefont
  {{Kankare}}}, \bibinfo {author} {\bibfnamefont {S.~A.}\ \bibnamefont
  {{Sim}}}, \bibinfo {author} {\bibfnamefont {M.}~\bibnamefont {{Fraser}}},
  \bibinfo {author} {\bibfnamefont {C.}~\bibnamefont {{Inserra}}}, \bibinfo
  {author} {\bibfnamefont {K.}~\bibnamefont {{Maguire}}}, \bibinfo {author}
  {\bibfnamefont {K.~C.}\ \bibnamefont {{Chambers}}}, \bibinfo {author}
  {\bibfnamefont {M.~E.}\ \bibnamefont {{Huber}}}, \bibinfo {author}
  {\bibfnamefont {T.}~\bibnamefont {{Kr{\"u}hler}}}, \bibinfo {author}
  {\bibfnamefont {G.}~\bibnamefont {{Leloudas}}}, \bibinfo {author}
  {\bibfnamefont {M.}~\bibnamefont {{Magee}}}, \bibinfo {author} {\bibfnamefont
  {L.~J.}\ \bibnamefont {{Shingles}}},  \emph {et~al.},\ }\href {\doibase
  10.1038/nature24303} {\bibfield  {journal} {\bibinfo  {journal} {\nat}\
  }\textbf {\bibinfo {volume} {551}},\ \bibinfo {pages} {75} (\bibinfo {year}
  {2017})},\ \Eprint {http://arxiv.org/abs/1710.05841} {arXiv:1710.05841
  [astro-ph.HE]} \BibitemShut {NoStop}%
\bibitem [{\citenamefont {{Tanvir}}\ \emph {et~al.}(2017)\citenamefont
  {{Tanvir}}, \citenamefont {{Levan}}, \citenamefont
  {{Gonz{\'a}lez-Fern{\'a}ndez}}, \citenamefont {{Korobkin}}, \citenamefont
  {{Mandel}}, \citenamefont {{Rosswog}}, \citenamefont {{Hjorth}},
  \citenamefont {{D'Avanzo}}, \citenamefont {{Fruchter}}, \citenamefont
  {{Fryer}}, \citenamefont {{Kangas}}, \citenamefont {{Milvang-Jensen}},
  \citenamefont {{Rosetti}}, \citenamefont {{Steeghs}}, \citenamefont
  {{Wollaeger}} \emph {et~al.}}]{2017ApJ...848L..27T}%
  \BibitemOpen
  \bibfield  {author} {\bibinfo {author} {\bibfnamefont {N.~R.}\ \bibnamefont
  {{Tanvir}}}, \bibinfo {author} {\bibfnamefont {A.~J.}\ \bibnamefont
  {{Levan}}}, \bibinfo {author} {\bibfnamefont {C.}~\bibnamefont
  {{Gonz{\'a}lez-Fern{\'a}ndez}}}, \bibinfo {author} {\bibfnamefont
  {O.}~\bibnamefont {{Korobkin}}}, \bibinfo {author} {\bibfnamefont
  {I.}~\bibnamefont {{Mandel}}}, \bibinfo {author} {\bibfnamefont
  {S.}~\bibnamefont {{Rosswog}}}, \bibinfo {author} {\bibfnamefont
  {J.}~\bibnamefont {{Hjorth}}}, \bibinfo {author} {\bibfnamefont
  {P.}~\bibnamefont {{D'Avanzo}}}, \bibinfo {author} {\bibfnamefont {A.~S.}\
  \bibnamefont {{Fruchter}}}, \bibinfo {author} {\bibfnamefont {C.~L.}\
  \bibnamefont {{Fryer}}}, \bibinfo {author} {\bibfnamefont {T.}~\bibnamefont
  {{Kangas}}}, \bibinfo {author} {\bibfnamefont {B.}~\bibnamefont
  {{Milvang-Jensen}}}, \bibinfo {author} {\bibfnamefont {S.}~\bibnamefont
  {{Rosetti}}}, \bibinfo {author} {\bibfnamefont {D.}~\bibnamefont
  {{Steeghs}}}, \bibinfo {author} {\bibfnamefont {R.~T.}\ \bibnamefont
  {{Wollaeger}}},  \emph {et~al.},\ }\href {\doibase 10.3847/2041-8213/aa90b6}
  {\bibfield  {journal} {\bibinfo  {journal} {\apjl}\ }\textbf {\bibinfo
  {volume} {848}},\ \bibinfo {eid} {L27} (\bibinfo {year} {2017})},\ \Eprint
  {http://arxiv.org/abs/1710.05455} {arXiv:1710.05455 [astro-ph.HE]}
  \BibitemShut {NoStop}%
\bibitem [{\citenamefont {{Troja}}\ \emph {et~al.}(2017)\citenamefont
  {{Troja}}, \citenamefont {{Piro}}, \citenamefont {{van Eerten}},
  \citenamefont {{Wollaeger}}, \citenamefont {{Im}}, \citenamefont {{Fox}},
  \citenamefont {{Butler}}, \citenamefont {{Cenko}}, \citenamefont
  {{Sakamoto}}, \citenamefont {{Fryer}}, \citenamefont {{Ricci}}, \citenamefont
  {{Lien}}, \citenamefont {{Ryan}}, \citenamefont {{Korobkin}}, \citenamefont
  {{Lee}} \emph {et~al.}}]{2017Natur.551...71T}%
  \BibitemOpen
  \bibfield  {author} {\bibinfo {author} {\bibfnamefont {E.}~\bibnamefont
  {{Troja}}}, \bibinfo {author} {\bibfnamefont {L.}~\bibnamefont {{Piro}}},
  \bibinfo {author} {\bibfnamefont {H.}~\bibnamefont {{van Eerten}}}, \bibinfo
  {author} {\bibfnamefont {R.~T.}\ \bibnamefont {{Wollaeger}}}, \bibinfo
  {author} {\bibfnamefont {M.}~\bibnamefont {{Im}}}, \bibinfo {author}
  {\bibfnamefont {O.~D.}\ \bibnamefont {{Fox}}}, \bibinfo {author}
  {\bibfnamefont {N.~R.}\ \bibnamefont {{Butler}}}, \bibinfo {author}
  {\bibfnamefont {S.~B.}\ \bibnamefont {{Cenko}}}, \bibinfo {author}
  {\bibfnamefont {T.}~\bibnamefont {{Sakamoto}}}, \bibinfo {author}
  {\bibfnamefont {C.~L.}\ \bibnamefont {{Fryer}}}, \bibinfo {author}
  {\bibfnamefont {R.}~\bibnamefont {{Ricci}}}, \bibinfo {author} {\bibfnamefont
  {A.}~\bibnamefont {{Lien}}}, \bibinfo {author} {\bibfnamefont {R.~E.}\
  \bibnamefont {{Ryan}}}, \bibinfo {author} {\bibfnamefont {O.}~\bibnamefont
  {{Korobkin}}}, \bibinfo {author} {\bibfnamefont {S.~K.}\ \bibnamefont
  {{Lee}}},  \emph {et~al.},\ }\href {\doibase 10.1038/nature24290} {\bibfield
  {journal} {\bibinfo  {journal} {\nat}\ }\textbf {\bibinfo {volume} {551}},\
  \bibinfo {pages} {71} (\bibinfo {year} {2017})},\ \Eprint
  {http://arxiv.org/abs/1710.05433} {arXiv:1710.05433 [astro-ph.HE]}
  \BibitemShut {NoStop}%
\bibitem [{\citenamefont {{Utsumi}}\ \emph {et~al.}(2017)\citenamefont
  {{Utsumi}}, \citenamefont {{Tanaka}}, \citenamefont {{Tominaga}},
  \citenamefont {{Yoshida}}, \citenamefont {{Barway}}, \citenamefont
  {{Nagayama}}, \citenamefont {{Zenko}}, \citenamefont {{Aoki}}, \citenamefont
  {{Fujiyoshi}}, \citenamefont {{Furusawa}}, \citenamefont {{Kawabata}},
  \citenamefont {{Koshida}}, \citenamefont {{Lee}}, \citenamefont {{Morokuma}},
  \citenamefont {{Motohara}} \emph {et~al.}}]{2017PASJ...69..101U}%
  \BibitemOpen
  \bibfield  {author} {\bibinfo {author} {\bibfnamefont {Y.}~\bibnamefont
  {{Utsumi}}}, \bibinfo {author} {\bibfnamefont {M.}~\bibnamefont {{Tanaka}}},
  \bibinfo {author} {\bibfnamefont {N.}~\bibnamefont {{Tominaga}}}, \bibinfo
  {author} {\bibfnamefont {M.}~\bibnamefont {{Yoshida}}}, \bibinfo {author}
  {\bibfnamefont {S.}~\bibnamefont {{Barway}}}, \bibinfo {author}
  {\bibfnamefont {T.}~\bibnamefont {{Nagayama}}}, \bibinfo {author}
  {\bibfnamefont {T.}~\bibnamefont {{Zenko}}}, \bibinfo {author} {\bibfnamefont
  {K.}~\bibnamefont {{Aoki}}}, \bibinfo {author} {\bibfnamefont
  {T.}~\bibnamefont {{Fujiyoshi}}}, \bibinfo {author} {\bibfnamefont
  {H.}~\bibnamefont {{Furusawa}}}, \bibinfo {author} {\bibfnamefont {K.~S.}\
  \bibnamefont {{Kawabata}}}, \bibinfo {author} {\bibfnamefont
  {S.}~\bibnamefont {{Koshida}}}, \bibinfo {author} {\bibfnamefont {C.-H.}\
  \bibnamefont {{Lee}}}, \bibinfo {author} {\bibfnamefont {T.}~\bibnamefont
  {{Morokuma}}}, \bibinfo {author} {\bibfnamefont {K.}~\bibnamefont
  {{Motohara}}},  \emph {et~al.},\ }\href {\doibase 10.1093/pasj/psx118}
  {\bibfield  {journal} {\bibinfo  {journal} {PASJ}\ }\textbf {\bibinfo
  {volume} {69}},\ \bibinfo {eid} {101} (\bibinfo {year} {2017})},\ \Eprint
  {http://arxiv.org/abs/1710.05848} {arXiv:1710.05848 [astro-ph.HE]}
  \BibitemShut {NoStop}%
\bibitem [{\citenamefont {{Valenti}}\ \emph {et~al.}(2017)\citenamefont
  {{Valenti}}, \citenamefont {{Sand}}, \citenamefont {{Yang}}, \citenamefont
  {{Cappellaro}}, \citenamefont {{Tartaglia}}, \citenamefont {{Corsi}},
  \citenamefont {{Jha}}, \citenamefont {{Reichart}}, \citenamefont
  {{Haislip}},\ and\ \citenamefont {{Kouprianov}}}]{2017ApJ...848L..24V}%
  \BibitemOpen
  \bibfield  {author} {\bibinfo {author} {\bibfnamefont {S.}~\bibnamefont
  {{Valenti}}}, \bibinfo {author} {\bibfnamefont {D.~J.}\ \bibnamefont
  {{Sand}}}, \bibinfo {author} {\bibfnamefont {S.}~\bibnamefont {{Yang}}},
  \bibinfo {author} {\bibfnamefont {E.}~\bibnamefont {{Cappellaro}}}, \bibinfo
  {author} {\bibfnamefont {L.}~\bibnamefont {{Tartaglia}}}, \bibinfo {author}
  {\bibfnamefont {A.}~\bibnamefont {{Corsi}}}, \bibinfo {author} {\bibfnamefont
  {S.~W.}\ \bibnamefont {{Jha}}}, \bibinfo {author} {\bibfnamefont {D.~E.}\
  \bibnamefont {{Reichart}}}, \bibinfo {author} {\bibfnamefont
  {J.}~\bibnamefont {{Haislip}}}, \ and\ \bibinfo {author} {\bibfnamefont
  {V.}~\bibnamefont {{Kouprianov}}},\ }\href {\doibase
  10.3847/2041-8213/aa8edf} {\bibfield  {journal} {\bibinfo  {journal} {\apjl}\
  }\textbf {\bibinfo {volume} {848}},\ \bibinfo {eid} {L24} (\bibinfo {year}
  {2017})},\ \Eprint {http://arxiv.org/abs/1710.05854} {arXiv:1710.05854
  [astro-ph.HE]} \BibitemShut {NoStop}%
\bibitem [{\citenamefont {{Margalit}}\ and\ \citenamefont
  {{Metzger}}(2017)}]{2017ApJ...850L..19M}%
  \BibitemOpen
  \bibfield  {author} {\bibinfo {author} {\bibfnamefont {B.}~\bibnamefont
  {{Margalit}}}\ and\ \bibinfo {author} {\bibfnamefont {B.~D.}\ \bibnamefont
  {{Metzger}}},\ }\href {\doibase 10.3847/2041-8213/aa991c} {\bibfield
  {journal} {\bibinfo  {journal} {Astrophys. J. Lett.}\ }\textbf {\bibinfo
  {volume} {850}},\ \bibinfo {eid} {L19} (\bibinfo {year} {2017})},\ \Eprint
  {http://arxiv.org/abs/1710.05938} {arXiv:1710.05938 [astro-ph.HE]}
  \BibitemShut {NoStop}%
\bibitem [{\citenamefont {{Radice}}\ \emph {et~al.}(2018)\citenamefont
  {{Radice}}, \citenamefont {{Perego}}, \citenamefont {{Zappa}},\ and\
  \citenamefont {{Bernuzzi}}}]{2018ApJ...852L..29R}%
  \BibitemOpen
  \bibfield  {author} {\bibinfo {author} {\bibfnamefont {D.}~\bibnamefont
  {{Radice}}}, \bibinfo {author} {\bibfnamefont {A.}~\bibnamefont {{Perego}}},
  \bibinfo {author} {\bibfnamefont {F.}~\bibnamefont {{Zappa}}}, \ and\
  \bibinfo {author} {\bibfnamefont {S.}~\bibnamefont {{Bernuzzi}}},\ }\href
  {\doibase 10.3847/2041-8213/aaa402} {\bibfield  {journal} {\bibinfo
  {journal} {Astrophys. J. Lett.}\ }\textbf {\bibinfo {volume} {852}},\
  \bibinfo {eid} {L29} (\bibinfo {year} {2018})},\ \Eprint
  {http://arxiv.org/abs/1711.03647} {arXiv:1711.03647 [astro-ph.HE]}
  \BibitemShut {NoStop}%
\bibitem [{\citenamefont {{Coughlin}}\ \emph {et~al.}(2019)\citenamefont
  {{Coughlin}}, \citenamefont {{Dietrich}}, \citenamefont {{Margalit}},\ and\
  \citenamefont {{Metzger}}}]{2019MNRAS.489L..91C}%
  \BibitemOpen
  \bibfield  {author} {\bibinfo {author} {\bibfnamefont {M.~W.}\ \bibnamefont
  {{Coughlin}}}, \bibinfo {author} {\bibfnamefont {T.}~\bibnamefont
  {{Dietrich}}}, \bibinfo {author} {\bibfnamefont {B.}~\bibnamefont
  {{Margalit}}}, \ and\ \bibinfo {author} {\bibfnamefont {B.~D.}\ \bibnamefont
  {{Metzger}}},\ }\href {\doibase 10.1093/mnrasl/slz133} {\bibfield  {journal}
  {\bibinfo  {journal} {\mnras}\ }\textbf {\bibinfo {volume} {489}},\ \bibinfo
  {pages} {L91} (\bibinfo {year} {2019})},\ \Eprint
  {http://arxiv.org/abs/1812.04803} {arXiv:1812.04803 [astro-ph.HE]}
  \BibitemShut {NoStop}%
\bibitem [{\citenamefont {{Breschi}}\ \emph {et~al.}(2021)\citenamefont
  {{Breschi}}, \citenamefont {{Perego}}, \citenamefont {{Bernuzzi}},
  \citenamefont {{Del Pozzo}}, \citenamefont {{Nedora}}, \citenamefont
  {{Radice}},\ and\ \citenamefont {{Vescovi}}}]{2021MNRAS.505.1661B}%
  \BibitemOpen
  \bibfield  {author} {\bibinfo {author} {\bibfnamefont {M.}~\bibnamefont
  {{Breschi}}}, \bibinfo {author} {\bibfnamefont {A.}~\bibnamefont {{Perego}}},
  \bibinfo {author} {\bibfnamefont {S.}~\bibnamefont {{Bernuzzi}}}, \bibinfo
  {author} {\bibfnamefont {W.}~\bibnamefont {{Del Pozzo}}}, \bibinfo {author}
  {\bibfnamefont {V.}~\bibnamefont {{Nedora}}}, \bibinfo {author}
  {\bibfnamefont {D.}~\bibnamefont {{Radice}}}, \ and\ \bibinfo {author}
  {\bibfnamefont {D.}~\bibnamefont {{Vescovi}}},\ }\href {\doibase
  10.1093/mnras/stab1287} {\bibfield  {journal} {\bibinfo  {journal} {\mnras}\
  }\textbf {\bibinfo {volume} {505}},\ \bibinfo {pages} {1661} (\bibinfo {year}
  {2021})},\ \Eprint {http://arxiv.org/abs/2101.01201} {arXiv:2101.01201
  [astro-ph.HE]} \BibitemShut {NoStop}%
\bibitem [{\citenamefont {{Nedora}}\ \emph {et~al.}(2021)\citenamefont
  {{Nedora}}, \citenamefont {{Bernuzzi}}, \citenamefont {{Radice}},
  \citenamefont {{Daszuta}}, \citenamefont {{Endrizzi}}, \citenamefont
  {{Perego}}, \citenamefont {{Prakash}}, \citenamefont {{Safarzadeh}},
  \citenamefont {{Schianchi}},\ and\ \citenamefont
  {{Logoteta}}}]{2021ApJ...906...98N}%
  \BibitemOpen
  \bibfield  {author} {\bibinfo {author} {\bibfnamefont {V.}~\bibnamefont
  {{Nedora}}}, \bibinfo {author} {\bibfnamefont {S.}~\bibnamefont
  {{Bernuzzi}}}, \bibinfo {author} {\bibfnamefont {D.}~\bibnamefont
  {{Radice}}}, \bibinfo {author} {\bibfnamefont {B.}~\bibnamefont {{Daszuta}}},
  \bibinfo {author} {\bibfnamefont {A.}~\bibnamefont {{Endrizzi}}}, \bibinfo
  {author} {\bibfnamefont {A.}~\bibnamefont {{Perego}}}, \bibinfo {author}
  {\bibfnamefont {A.}~\bibnamefont {{Prakash}}}, \bibinfo {author}
  {\bibfnamefont {M.}~\bibnamefont {{Safarzadeh}}}, \bibinfo {author}
  {\bibfnamefont {F.}~\bibnamefont {{Schianchi}}}, \ and\ \bibinfo {author}
  {\bibfnamefont {D.}~\bibnamefont {{Logoteta}}},\ }\href {\doibase
  10.3847/1538-4357/abc9be} {\bibfield  {journal} {\bibinfo  {journal} {\apj}\
  }\textbf {\bibinfo {volume} {906}},\ \bibinfo {eid} {98} (\bibinfo {year}
  {2021})},\ \Eprint {http://arxiv.org/abs/2008.04333} {arXiv:2008.04333
  [astro-ph.HE]} \BibitemShut {NoStop}%
\bibitem [{\citenamefont {{Holmbeck}}\ \emph {et~al.}(2022)\citenamefont
  {{Holmbeck}}, \citenamefont {{O'Shaughnessy}}, \citenamefont {{Delfavero}},\
  and\ \citenamefont {{Belczynski}}}]{2022ApJ...926..196H}%
  \BibitemOpen
  \bibfield  {author} {\bibinfo {author} {\bibfnamefont {E.~M.}\ \bibnamefont
  {{Holmbeck}}}, \bibinfo {author} {\bibfnamefont {R.}~\bibnamefont
  {{O'Shaughnessy}}}, \bibinfo {author} {\bibfnamefont {V.}~\bibnamefont
  {{Delfavero}}}, \ and\ \bibinfo {author} {\bibfnamefont {K.}~\bibnamefont
  {{Belczynski}}},\ }\href {\doibase 10.3847/1538-4357/ac490e} {\bibfield
  {journal} {\bibinfo  {journal} {\apj}\ }\textbf {\bibinfo {volume} {926}},\
  \bibinfo {eid} {196} (\bibinfo {year} {2022})},\ \Eprint
  {http://arxiv.org/abs/2110.06432} {arXiv:2110.06432 [astro-ph.HE]}
  \BibitemShut {NoStop}%
\bibitem [{\citenamefont {{Zhu}}\ \emph {et~al.}(2023)\citenamefont {{Zhu}},
  \citenamefont {{Li}},\ and\ \citenamefont {{Liu}}}]{2023ApJ...943..163Z}%
  \BibitemOpen
  \bibfield  {author} {\bibinfo {author} {\bibfnamefont {Z.}~\bibnamefont
  {{Zhu}}}, \bibinfo {author} {\bibfnamefont {A.}~\bibnamefont {{Li}}}, \ and\
  \bibinfo {author} {\bibfnamefont {T.}~\bibnamefont {{Liu}}},\ }\href
  {\doibase 10.3847/1538-4357/acac1f} {\bibfield  {journal} {\bibinfo
  {journal} {\apj}\ }\textbf {\bibinfo {volume} {943}},\ \bibinfo {eid} {163}
  (\bibinfo {year} {2023})},\ \Eprint {http://arxiv.org/abs/2211.02007}
  {arXiv:2211.02007 [astro-ph.HE]} \BibitemShut {NoStop}%
\bibitem [{\citenamefont {{Miller}}\ \emph {et~al.}(2019)\citenamefont
  {{Miller}}, \citenamefont {{Lamb}}, \citenamefont {{Dittmann}}, \citenamefont
  {{Bogdanov}}, \citenamefont {{Arzoumanian}}, \citenamefont {{Gendreau}},
  \citenamefont {{Guillot}}, \citenamefont {{Harding}}, \citenamefont {{Ho}},
  \citenamefont {{Lattimer}}, \citenamefont {{Ludlam}}, \citenamefont
  {{Mahmoodifar}}, \citenamefont {{Morsink}}, \citenamefont {{Ray}},
  \citenamefont {{Strohmayer}}, \citenamefont {{Wood}}, \citenamefont
  {{Enoto}}, \citenamefont {{Foster}}, \citenamefont {{Okajima}}, \citenamefont
  {{Prigozhin}},\ and\ \citenamefont {{Soong}}}]{2019ApJ...887L..24M}%
  \BibitemOpen
  \bibfield  {author} {\bibinfo {author} {\bibfnamefont {M.~C.}\ \bibnamefont
  {{Miller}}}, \bibinfo {author} {\bibfnamefont {F.~K.}\ \bibnamefont
  {{Lamb}}}, \bibinfo {author} {\bibfnamefont {A.~J.}\ \bibnamefont
  {{Dittmann}}}, \bibinfo {author} {\bibfnamefont {S.}~\bibnamefont
  {{Bogdanov}}}, \bibinfo {author} {\bibfnamefont {Z.}~\bibnamefont
  {{Arzoumanian}}}, \bibinfo {author} {\bibfnamefont {K.~C.}\ \bibnamefont
  {{Gendreau}}}, \bibinfo {author} {\bibfnamefont {S.}~\bibnamefont
  {{Guillot}}}, \bibinfo {author} {\bibfnamefont {A.~K.}\ \bibnamefont
  {{Harding}}}, \bibinfo {author} {\bibfnamefont {W.~C.~G.}\ \bibnamefont
  {{Ho}}}, \bibinfo {author} {\bibfnamefont {J.~M.}\ \bibnamefont
  {{Lattimer}}}, \bibinfo {author} {\bibfnamefont {R.~M.}\ \bibnamefont
  {{Ludlam}}}, \bibinfo {author} {\bibfnamefont {S.}~\bibnamefont
  {{Mahmoodifar}}}, \bibinfo {author} {\bibfnamefont {S.~M.}\ \bibnamefont
  {{Morsink}}}, \bibinfo {author} {\bibfnamefont {P.~S.}\ \bibnamefont
  {{Ray}}}, \bibinfo {author} {\bibfnamefont {T.~E.}\ \bibnamefont
  {{Strohmayer}}}, \bibinfo {author} {\bibfnamefont {K.~S.}\ \bibnamefont
  {{Wood}}}, \bibinfo {author} {\bibfnamefont {T.}~\bibnamefont {{Enoto}}},
  \bibinfo {author} {\bibfnamefont {R.}~\bibnamefont {{Foster}}}, \bibinfo
  {author} {\bibfnamefont {T.}~\bibnamefont {{Okajima}}}, \bibinfo {author}
  {\bibfnamefont {G.}~\bibnamefont {{Prigozhin}}}, \ and\ \bibinfo {author}
  {\bibfnamefont {Y.}~\bibnamefont {{Soong}}},\ }\href {\doibase
  10.3847/2041-8213/ab50c5} {\bibfield  {journal} {\bibinfo  {journal}
  {Astrophys. J. Lett.}\ }\textbf {\bibinfo {volume} {887}},\ \bibinfo {eid}
  {L24} (\bibinfo {year} {2019})},\ \Eprint {http://arxiv.org/abs/1912.05705}
  {arXiv:1912.05705 [astro-ph.HE]} \BibitemShut {NoStop}%
\bibitem [{\citenamefont {{Riley}}\ \emph
  {et~al.}(2019{\natexlab{a}})\citenamefont {{Riley}}, \citenamefont {{Watts}},
  \citenamefont {{Bogdanov}}, \citenamefont {{Ray}}, \citenamefont {{Ludlam}},
  \citenamefont {{Guillot}}, \citenamefont {{Arzoumanian}}, \citenamefont
  {{Baker}}, \citenamefont {{Bilous}}, \citenamefont {{Chakrabarty}},
  \citenamefont {{Gendreau}}, \citenamefont {{Harding}}, \citenamefont {{Ho}},
  \citenamefont {{Lattimer}}, \citenamefont {{Morsink}},\ and\ \citenamefont
  {{Strohmayer}}}]{2019ApJ...887L..21R}%
  \BibitemOpen
  \bibfield  {author} {\bibinfo {author} {\bibfnamefont {T.~E.}\ \bibnamefont
  {{Riley}}}, \bibinfo {author} {\bibfnamefont {A.~L.}\ \bibnamefont
  {{Watts}}}, \bibinfo {author} {\bibfnamefont {S.}~\bibnamefont {{Bogdanov}}},
  \bibinfo {author} {\bibfnamefont {P.~S.}\ \bibnamefont {{Ray}}}, \bibinfo
  {author} {\bibfnamefont {R.~M.}\ \bibnamefont {{Ludlam}}}, \bibinfo {author}
  {\bibfnamefont {S.}~\bibnamefont {{Guillot}}}, \bibinfo {author}
  {\bibfnamefont {Z.}~\bibnamefont {{Arzoumanian}}}, \bibinfo {author}
  {\bibfnamefont {C.~L.}\ \bibnamefont {{Baker}}}, \bibinfo {author}
  {\bibfnamefont {A.~V.}\ \bibnamefont {{Bilous}}}, \bibinfo {author}
  {\bibfnamefont {D.}~\bibnamefont {{Chakrabarty}}}, \bibinfo {author}
  {\bibfnamefont {K.~C.}\ \bibnamefont {{Gendreau}}}, \bibinfo {author}
  {\bibfnamefont {A.~K.}\ \bibnamefont {{Harding}}}, \bibinfo {author}
  {\bibfnamefont {W.~C.~G.}\ \bibnamefont {{Ho}}}, \bibinfo {author}
  {\bibfnamefont {J.~M.}\ \bibnamefont {{Lattimer}}}, \bibinfo {author}
  {\bibfnamefont {S.~M.}\ \bibnamefont {{Morsink}}}, \ and\ \bibinfo {author}
  {\bibfnamefont {T.~E.}\ \bibnamefont {{Strohmayer}}},\ }\href {\doibase
  10.3847/2041-8213/ab481c} {\bibfield  {journal} {\bibinfo  {journal}
  {Astrophys. J. Lett.}\ }\textbf {\bibinfo {volume} {887}},\ \bibinfo {eid}
  {L21} (\bibinfo {year} {2019}{\natexlab{a}})},\ \Eprint
  {http://arxiv.org/abs/1912.05702} {arXiv:1912.05702 [astro-ph.HE]}
  \BibitemShut {NoStop}%
\bibitem [{\citenamefont {{Miller}}\ \emph {et~al.}(2021)\citenamefont
  {{Miller}}, \citenamefont {{Lamb}}, \citenamefont {{Dittmann}}, \citenamefont
  {{Bogdanov}}, \citenamefont {{Arzoumanian}}, \citenamefont {{Gendreau}},
  \citenamefont {{Guillot}}, \citenamefont {{Ho}}, \citenamefont {{Lattimer}},
  \citenamefont {{Loewenstein}}, \citenamefont {{Morsink}}, \citenamefont
  {{Ray}}, \citenamefont {{Wolff}}, \citenamefont {{Baker}}, \citenamefont
  {{Cazeau}}, \citenamefont {{Manthripragada}}, \citenamefont {{Markwardt}},
  \citenamefont {{Okajima}}, \citenamefont {{Pollard}}, \citenamefont
  {{Cognard}}, \citenamefont {{Cromartie}}, \citenamefont {{Fonseca}},
  \citenamefont {{Guillemot}}, \citenamefont {{Kerr}}, \citenamefont
  {{Parthasarathy}}, \citenamefont {{Pennucci}}, \citenamefont {{Ransom}},\
  and\ \citenamefont {{Stairs}}}]{2021ApJ...918L..28M}%
  \BibitemOpen
  \bibfield  {author} {\bibinfo {author} {\bibfnamefont {M.~C.}\ \bibnamefont
  {{Miller}}}, \bibinfo {author} {\bibfnamefont {F.~K.}\ \bibnamefont
  {{Lamb}}}, \bibinfo {author} {\bibfnamefont {A.~J.}\ \bibnamefont
  {{Dittmann}}}, \bibinfo {author} {\bibfnamefont {S.}~\bibnamefont
  {{Bogdanov}}}, \bibinfo {author} {\bibfnamefont {Z.}~\bibnamefont
  {{Arzoumanian}}}, \bibinfo {author} {\bibfnamefont {K.~C.}\ \bibnamefont
  {{Gendreau}}}, \bibinfo {author} {\bibfnamefont {S.}~\bibnamefont
  {{Guillot}}}, \bibinfo {author} {\bibfnamefont {W.~C.~G.}\ \bibnamefont
  {{Ho}}}, \bibinfo {author} {\bibfnamefont {J.~M.}\ \bibnamefont
  {{Lattimer}}}, \bibinfo {author} {\bibfnamefont {M.}~\bibnamefont
  {{Loewenstein}}}, \bibinfo {author} {\bibfnamefont {S.~M.}\ \bibnamefont
  {{Morsink}}}, \bibinfo {author} {\bibfnamefont {P.~S.}\ \bibnamefont
  {{Ray}}}, \bibinfo {author} {\bibfnamefont {M.~T.}\ \bibnamefont {{Wolff}}},
  \bibinfo {author} {\bibfnamefont {C.~L.}\ \bibnamefont {{Baker}}}, \bibinfo
  {author} {\bibfnamefont {T.}~\bibnamefont {{Cazeau}}}, \bibinfo {author}
  {\bibfnamefont {S.}~\bibnamefont {{Manthripragada}}}, \bibinfo {author}
  {\bibfnamefont {C.~B.}\ \bibnamefont {{Markwardt}}}, \bibinfo {author}
  {\bibfnamefont {T.}~\bibnamefont {{Okajima}}}, \bibinfo {author}
  {\bibfnamefont {S.}~\bibnamefont {{Pollard}}}, \bibinfo {author}
  {\bibfnamefont {I.}~\bibnamefont {{Cognard}}}, \bibinfo {author}
  {\bibfnamefont {H.~T.}\ \bibnamefont {{Cromartie}}}, \bibinfo {author}
  {\bibfnamefont {E.}~\bibnamefont {{Fonseca}}}, \bibinfo {author}
  {\bibfnamefont {L.}~\bibnamefont {{Guillemot}}}, \bibinfo {author}
  {\bibfnamefont {M.}~\bibnamefont {{Kerr}}}, \bibinfo {author} {\bibfnamefont
  {A.}~\bibnamefont {{Parthasarathy}}}, \bibinfo {author} {\bibfnamefont
  {T.~T.}\ \bibnamefont {{Pennucci}}}, \bibinfo {author} {\bibfnamefont
  {S.}~\bibnamefont {{Ransom}}}, \ and\ \bibinfo {author} {\bibfnamefont
  {I.}~\bibnamefont {{Stairs}}},\ }\href {\doibase 10.3847/2041-8213/ac089b}
  {\bibfield  {journal} {\bibinfo  {journal} {Astrophys. J. Lett.}\ }\textbf
  {\bibinfo {volume} {918}},\ \bibinfo {eid} {L28} (\bibinfo {year} {2021})},\
  \Eprint {http://arxiv.org/abs/2105.06979} {arXiv:2105.06979 [astro-ph.HE]}
  \BibitemShut {NoStop}%
\bibitem [{\citenamefont {{Riley}}\ \emph
  {et~al.}(2021{\natexlab{a}})\citenamefont {{Riley}}, \citenamefont {{Watts}},
  \citenamefont {{Ray}}, \citenamefont {{Bogdanov}}, \citenamefont {{Guillot}},
  \citenamefont {{Morsink}}, \citenamefont {{Bilous}}, \citenamefont
  {{Arzoumanian}}, \citenamefont {{Choudhury}}, \citenamefont {{Deneva}},
  \citenamefont {{Gendreau}}, \citenamefont {{Harding}}, \citenamefont {{Ho}},
  \citenamefont {{Lattimer}}, \citenamefont {{Loewenstein}}, \citenamefont
  {{Ludlam}}, \citenamefont {{Markwardt}}, \citenamefont {{Okajima}},
  \citenamefont {{Prescod-Weinstein}}, \citenamefont {{Remillard}},
  \citenamefont {{Wolff}}, \citenamefont {{Fonseca}}, \citenamefont
  {{Cromartie}}, \citenamefont {{Kerr}}, \citenamefont {{Pennucci}},
  \citenamefont {{Parthasarathy}}, \citenamefont {{Ransom}}, \citenamefont
  {{Stairs}}, \citenamefont {{Guillemot}},\ and\ \citenamefont
  {{Cognard}}}]{2021ApJ...918L..27R}%
  \BibitemOpen
  \bibfield  {author} {\bibinfo {author} {\bibfnamefont {T.~E.}\ \bibnamefont
  {{Riley}}}, \bibinfo {author} {\bibfnamefont {A.~L.}\ \bibnamefont
  {{Watts}}}, \bibinfo {author} {\bibfnamefont {P.~S.}\ \bibnamefont {{Ray}}},
  \bibinfo {author} {\bibfnamefont {S.}~\bibnamefont {{Bogdanov}}}, \bibinfo
  {author} {\bibfnamefont {S.}~\bibnamefont {{Guillot}}}, \bibinfo {author}
  {\bibfnamefont {S.~M.}\ \bibnamefont {{Morsink}}}, \bibinfo {author}
  {\bibfnamefont {A.~V.}\ \bibnamefont {{Bilous}}}, \bibinfo {author}
  {\bibfnamefont {Z.}~\bibnamefont {{Arzoumanian}}}, \bibinfo {author}
  {\bibfnamefont {D.}~\bibnamefont {{Choudhury}}}, \bibinfo {author}
  {\bibfnamefont {J.~S.}\ \bibnamefont {{Deneva}}}, \bibinfo {author}
  {\bibfnamefont {K.~C.}\ \bibnamefont {{Gendreau}}}, \bibinfo {author}
  {\bibfnamefont {A.~K.}\ \bibnamefont {{Harding}}}, \bibinfo {author}
  {\bibfnamefont {W.~C.~G.}\ \bibnamefont {{Ho}}}, \bibinfo {author}
  {\bibfnamefont {J.~M.}\ \bibnamefont {{Lattimer}}}, \bibinfo {author}
  {\bibfnamefont {M.}~\bibnamefont {{Loewenstein}}}, \bibinfo {author}
  {\bibfnamefont {R.~M.}\ \bibnamefont {{Ludlam}}}, \bibinfo {author}
  {\bibfnamefont {C.~B.}\ \bibnamefont {{Markwardt}}}, \bibinfo {author}
  {\bibfnamefont {T.}~\bibnamefont {{Okajima}}}, \bibinfo {author}
  {\bibfnamefont {C.}~\bibnamefont {{Prescod-Weinstein}}}, \bibinfo {author}
  {\bibfnamefont {R.~A.}\ \bibnamefont {{Remillard}}}, \bibinfo {author}
  {\bibfnamefont {M.~T.}\ \bibnamefont {{Wolff}}}, \bibinfo {author}
  {\bibfnamefont {E.}~\bibnamefont {{Fonseca}}}, \bibinfo {author}
  {\bibfnamefont {H.~T.}\ \bibnamefont {{Cromartie}}}, \bibinfo {author}
  {\bibfnamefont {M.}~\bibnamefont {{Kerr}}}, \bibinfo {author} {\bibfnamefont
  {T.~T.}\ \bibnamefont {{Pennucci}}}, \bibinfo {author} {\bibfnamefont
  {A.}~\bibnamefont {{Parthasarathy}}}, \bibinfo {author} {\bibfnamefont
  {S.}~\bibnamefont {{Ransom}}}, \bibinfo {author} {\bibfnamefont
  {I.}~\bibnamefont {{Stairs}}}, \bibinfo {author} {\bibfnamefont
  {L.}~\bibnamefont {{Guillemot}}}, \ and\ \bibinfo {author} {\bibfnamefont
  {I.}~\bibnamefont {{Cognard}}},\ }\href {\doibase 10.3847/2041-8213/ac0a81}
  {\bibfield  {journal} {\bibinfo  {journal} {Astrophys. J. Lett.}\ }\textbf
  {\bibinfo {volume} {918}},\ \bibinfo {eid} {L27} (\bibinfo {year}
  {2021}{\natexlab{a}})},\ \Eprint {http://arxiv.org/abs/2105.06980}
  {arXiv:2105.06980 [astro-ph.HE]} \BibitemShut {NoStop}%
\bibitem [{\citenamefont {{Adhikari}}\ \emph {et~al.}(2021)\citenamefont
  {{Adhikari}}, \citenamefont {{Albataineh}}, \citenamefont {{Androic}},
  \citenamefont {{Aniol}}, \citenamefont {{Armstrong}}, \citenamefont
  {{Averett}}, \citenamefont {{Ayerbe Gayoso}}, \citenamefont {{Barcus}},
  \citenamefont {{Bellini}}, \citenamefont {{Beminiwattha}}, \citenamefont
  {{Benesch}}, \citenamefont {{Bhatt}}, \citenamefont {{Bhatta Pathak}},
  \citenamefont {{Bhetuwal}}, \citenamefont {{Blaikie}},\ and\ \citenamefont
  {\textit{et al}. ({PREX Collaboration})}}]{2021PhRvL.126q2502A}%
  \BibitemOpen
  \bibfield  {author} {\bibinfo {author} {\bibfnamefont {D.}~\bibnamefont
  {{Adhikari}}}, \bibinfo {author} {\bibfnamefont {H.}~\bibnamefont
  {{Albataineh}}}, \bibinfo {author} {\bibfnamefont {D.}~\bibnamefont
  {{Androic}}}, \bibinfo {author} {\bibfnamefont {K.}~\bibnamefont {{Aniol}}},
  \bibinfo {author} {\bibfnamefont {D.~S.}\ \bibnamefont {{Armstrong}}},
  \bibinfo {author} {\bibfnamefont {T.}~\bibnamefont {{Averett}}}, \bibinfo
  {author} {\bibfnamefont {C.}~\bibnamefont {{Ayerbe Gayoso}}}, \bibinfo
  {author} {\bibfnamefont {S.}~\bibnamefont {{Barcus}}}, \bibinfo {author}
  {\bibfnamefont {V.}~\bibnamefont {{Bellini}}}, \bibinfo {author}
  {\bibfnamefont {R.~S.}\ \bibnamefont {{Beminiwattha}}}, \bibinfo {author}
  {\bibfnamefont {J.~F.}\ \bibnamefont {{Benesch}}}, \bibinfo {author}
  {\bibfnamefont {H.}~\bibnamefont {{Bhatt}}}, \bibinfo {author} {\bibfnamefont
  {D.}~\bibnamefont {{Bhatta Pathak}}}, \bibinfo {author} {\bibfnamefont
  {D.}~\bibnamefont {{Bhetuwal}}}, \bibinfo {author} {\bibfnamefont
  {B.}~\bibnamefont {{Blaikie}}}, \ and\ \bibinfo {author} {\bibnamefont
  {\textit{et al}. ({PREX Collaboration})}},\ }\href {\doibase
  10.1103/PhysRevLett.126.172502} {\bibfield  {journal} {\bibinfo  {journal}
  {\prl}\ }\textbf {\bibinfo {volume} {126}},\ \bibinfo {eid} {172502}
  (\bibinfo {year} {2021})},\ \Eprint {http://arxiv.org/abs/2102.10767}
  {arXiv:2102.10767 [nucl-ex]} \BibitemShut {NoStop}%
\bibitem [{\citenamefont {{Hu}}\ \emph {et~al.}(2022)\citenamefont {{Hu}},
  \citenamefont {{Jiang}}, \citenamefont {{Miyagi}}, \citenamefont {{Sun}},
  \citenamefont {{Ekstr{\"o}m}}, \citenamefont {{Forss{\'e}n}}, \citenamefont
  {{Hagen}}, \citenamefont {{Holt}}, \citenamefont {{Papenbrock}},
  \citenamefont {{Stroberg}},\ and\ \citenamefont
  {{Vernon}}}]{2022NatPh..18.1196H}%
  \BibitemOpen
  \bibfield  {author} {\bibinfo {author} {\bibfnamefont {B.}~\bibnamefont
  {{Hu}}}, \bibinfo {author} {\bibfnamefont {W.}~\bibnamefont {{Jiang}}},
  \bibinfo {author} {\bibfnamefont {T.}~\bibnamefont {{Miyagi}}}, \bibinfo
  {author} {\bibfnamefont {Z.}~\bibnamefont {{Sun}}}, \bibinfo {author}
  {\bibfnamefont {A.}~\bibnamefont {{Ekstr{\"o}m}}}, \bibinfo {author}
  {\bibfnamefont {C.}~\bibnamefont {{Forss{\'e}n}}}, \bibinfo {author}
  {\bibfnamefont {G.}~\bibnamefont {{Hagen}}}, \bibinfo {author} {\bibfnamefont
  {J.~D.}\ \bibnamefont {{Holt}}}, \bibinfo {author} {\bibfnamefont
  {T.}~\bibnamefont {{Papenbrock}}}, \bibinfo {author} {\bibfnamefont {S.~R.}\
  \bibnamefont {{Stroberg}}}, \ and\ \bibinfo {author} {\bibfnamefont
  {I.}~\bibnamefont {{Vernon}}},\ }\href {\doibase 10.1038/s41567-022-01715-8}
  {\bibfield  {journal} {\bibinfo  {journal} {Nature Physics}\ }\textbf
  {\bibinfo {volume} {18}},\ \bibinfo {pages} {1196} (\bibinfo {year}
  {2022})},\ \Eprint {http://arxiv.org/abs/2112.01125} {arXiv:2112.01125
  [nucl-th]} \BibitemShut {NoStop}%
\bibitem [{\citenamefont {{Adhikari}}\ \emph {et~al.}(2022)\citenamefont
  {{Adhikari}}, \citenamefont {{Albataineh}}, \citenamefont {{Androic}},
  \citenamefont {{Aniol}}, \citenamefont {{Armstrong}}, \citenamefont
  {{Averett}}, \citenamefont {{Ayerbe Gayoso}}, \citenamefont {{Barcus}},
  \citenamefont {{Bellini}}, \citenamefont {{Beminiwattha}}, \citenamefont
  {{Benesch}}, \citenamefont {{Bhatt}}, \citenamefont {{Bhatta Pathak}},
  \citenamefont {{Bhetuwal}}, \citenamefont {{Blaikie}},\ and\ \citenamefont
  {\textit{et al}. ({CREX Collaboration})}}]{2022PhRvL.129d2501A}%
  \BibitemOpen
  \bibfield  {author} {\bibinfo {author} {\bibfnamefont {D.}~\bibnamefont
  {{Adhikari}}}, \bibinfo {author} {\bibfnamefont {H.}~\bibnamefont
  {{Albataineh}}}, \bibinfo {author} {\bibfnamefont {D.}~\bibnamefont
  {{Androic}}}, \bibinfo {author} {\bibfnamefont {K.~A.}\ \bibnamefont
  {{Aniol}}}, \bibinfo {author} {\bibfnamefont {D.~S.}\ \bibnamefont
  {{Armstrong}}}, \bibinfo {author} {\bibfnamefont {T.}~\bibnamefont
  {{Averett}}}, \bibinfo {author} {\bibfnamefont {C.}~\bibnamefont {{Ayerbe
  Gayoso}}}, \bibinfo {author} {\bibfnamefont {S.~K.}\ \bibnamefont
  {{Barcus}}}, \bibinfo {author} {\bibfnamefont {V.}~\bibnamefont {{Bellini}}},
  \bibinfo {author} {\bibfnamefont {R.~S.}\ \bibnamefont {{Beminiwattha}}},
  \bibinfo {author} {\bibfnamefont {J.~F.}\ \bibnamefont {{Benesch}}}, \bibinfo
  {author} {\bibfnamefont {H.}~\bibnamefont {{Bhatt}}}, \bibinfo {author}
  {\bibfnamefont {D.}~\bibnamefont {{Bhatta Pathak}}}, \bibinfo {author}
  {\bibfnamefont {D.}~\bibnamefont {{Bhetuwal}}}, \bibinfo {author}
  {\bibfnamefont {B.}~\bibnamefont {{Blaikie}}}, \ and\ \bibinfo {author}
  {\bibnamefont {\textit{et al}. ({CREX Collaboration})}},\ }\href {\doibase
  10.1103/PhysRevLett.129.042501} {\bibfield  {journal} {\bibinfo  {journal}
  {\prl}\ }\textbf {\bibinfo {volume} {129}},\ \bibinfo {eid} {042501}
  (\bibinfo {year} {2022})},\ \Eprint {http://arxiv.org/abs/2205.11593}
  {arXiv:2205.11593 [nucl-ex]} \BibitemShut {NoStop}%
\bibitem [{\citenamefont {{Todd-Rutel}}\ and\ \citenamefont
  {{Piekarewicz}}(2005)}]{2005PhRvL..95l2501T}%
  \BibitemOpen
  \bibfield  {author} {\bibinfo {author} {\bibfnamefont {B.~G.}\ \bibnamefont
  {{Todd-Rutel}}}\ and\ \bibinfo {author} {\bibfnamefont {J.}~\bibnamefont
  {{Piekarewicz}}},\ }\href {\doibase 10.1103/PhysRevLett.95.122501} {\bibfield
   {journal} {\bibinfo  {journal} {\prl}\ }\textbf {\bibinfo {volume} {95}},\
  \bibinfo {eid} {122501} (\bibinfo {year} {2005})},\ \Eprint
  {http://arxiv.org/abs/nucl-th/0504034} {arXiv:nucl-th/0504034 [astro-ph]}
  \BibitemShut {NoStop}%
\bibitem [{\citenamefont {{Centelles}}\ \emph {et~al.}(2009)\citenamefont
  {{Centelles}}, \citenamefont {{Roca-Maza}}, \citenamefont {{Vi{\~n}as}},\
  and\ \citenamefont {{Warda}}}]{2009PhRvL.102l2502C}%
  \BibitemOpen
  \bibfield  {author} {\bibinfo {author} {\bibfnamefont {M.}~\bibnamefont
  {{Centelles}}}, \bibinfo {author} {\bibfnamefont {X.}~\bibnamefont
  {{Roca-Maza}}}, \bibinfo {author} {\bibfnamefont {X.}~\bibnamefont
  {{Vi{\~n}as}}}, \ and\ \bibinfo {author} {\bibfnamefont {M.}~\bibnamefont
  {{Warda}}},\ }\href {\doibase 10.1103/PhysRevLett.102.122502} {\bibfield
  {journal} {\bibinfo  {journal} {\prl}\ }\textbf {\bibinfo {volume} {102}},\
  \bibinfo {eid} {122502} (\bibinfo {year} {2009})},\ \Eprint
  {http://arxiv.org/abs/0806.2886} {arXiv:0806.2886 [nucl-th]} \BibitemShut
  {NoStop}%
\bibitem [{\citenamefont {{Chen}}\ \emph {et~al.}(2010)\citenamefont {{Chen}},
  \citenamefont {{Ko}}, \citenamefont {{Li}},\ and\ \citenamefont
  {{Xu}}}]{2010PhRvC..82b4321C}%
  \BibitemOpen
  \bibfield  {author} {\bibinfo {author} {\bibfnamefont {L.-W.}\ \bibnamefont
  {{Chen}}}, \bibinfo {author} {\bibfnamefont {C.~M.}\ \bibnamefont {{Ko}}},
  \bibinfo {author} {\bibfnamefont {B.-A.}\ \bibnamefont {{Li}}}, \ and\
  \bibinfo {author} {\bibfnamefont {J.}~\bibnamefont {{Xu}}},\ }\href {\doibase
  10.1103/PhysRevC.82.024321} {\bibfield  {journal} {\bibinfo  {journal}
  {\prc}\ }\textbf {\bibinfo {volume} {82}},\ \bibinfo {eid} {024321} (\bibinfo
  {year} {2010})},\ \Eprint {http://arxiv.org/abs/1004.4672} {arXiv:1004.4672
  [nucl-th]} \BibitemShut {NoStop}%
\bibitem [{\citenamefont {{Roca-Maza}}\ \emph {et~al.}(2011)\citenamefont
  {{Roca-Maza}}, \citenamefont {{Centelles}}, \citenamefont {{Vi{\~n}as}},\
  and\ \citenamefont {{Warda}}}]{2011PhRvL.106y2501R}%
  \BibitemOpen
  \bibfield  {author} {\bibinfo {author} {\bibfnamefont {X.}~\bibnamefont
  {{Roca-Maza}}}, \bibinfo {author} {\bibfnamefont {M.}~\bibnamefont
  {{Centelles}}}, \bibinfo {author} {\bibfnamefont {X.}~\bibnamefont
  {{Vi{\~n}as}}}, \ and\ \bibinfo {author} {\bibfnamefont {M.}~\bibnamefont
  {{Warda}}},\ }\href {\doibase 10.1103/PhysRevLett.106.252501} {\bibfield
  {journal} {\bibinfo  {journal} {\prl}\ }\textbf {\bibinfo {volume} {106}},\
  \bibinfo {eid} {252501} (\bibinfo {year} {2011})},\ \Eprint
  {http://arxiv.org/abs/1103.1762} {arXiv:1103.1762 [nucl-th]} \BibitemShut
  {NoStop}%
\bibitem [{\citenamefont {{Fattoyev}}\ \emph {et~al.}(2018)\citenamefont
  {{Fattoyev}}, \citenamefont {{Piekarewicz}},\ and\ \citenamefont
  {{Horowitz}}}]{2018PhRvL.120q2702F}%
  \BibitemOpen
  \bibfield  {author} {\bibinfo {author} {\bibfnamefont {F.~J.}\ \bibnamefont
  {{Fattoyev}}}, \bibinfo {author} {\bibfnamefont {J.}~\bibnamefont
  {{Piekarewicz}}}, \ and\ \bibinfo {author} {\bibfnamefont {C.~J.}\
  \bibnamefont {{Horowitz}}},\ }\href {\doibase 10.1103/PhysRevLett.120.172702}
  {\bibfield  {journal} {\bibinfo  {journal} {\prl}\ }\textbf {\bibinfo
  {volume} {120}},\ \bibinfo {eid} {172702} (\bibinfo {year} {2018})},\ \Eprint
  {http://arxiv.org/abs/1711.06615} {arXiv:1711.06615 [nucl-th]} \BibitemShut
  {NoStop}%
\bibitem [{\citenamefont {{Reed}}\ \emph {et~al.}(2021)\citenamefont {{Reed}},
  \citenamefont {{Fattoyev}}, \citenamefont {{Horowitz}},\ and\ \citenamefont
  {{Piekarewicz}}}]{2021PhRvL.126q2503R}%
  \BibitemOpen
  \bibfield  {author} {\bibinfo {author} {\bibfnamefont {B.~T.}\ \bibnamefont
  {{Reed}}}, \bibinfo {author} {\bibfnamefont {F.~J.}\ \bibnamefont
  {{Fattoyev}}}, \bibinfo {author} {\bibfnamefont {C.~J.}\ \bibnamefont
  {{Horowitz}}}, \ and\ \bibinfo {author} {\bibfnamefont {J.}~\bibnamefont
  {{Piekarewicz}}},\ }\href {\doibase 10.1103/PhysRevLett.126.172503}
  {\bibfield  {journal} {\bibinfo  {journal} {\prl}\ }\textbf {\bibinfo
  {volume} {126}},\ \bibinfo {eid} {172503} (\bibinfo {year} {2021})},\ \Eprint
  {http://arxiv.org/abs/2101.03193} {arXiv:2101.03193 [nucl-th]} \BibitemShut
  {NoStop}%
\bibitem [{\citenamefont {{Essick}}\ \emph {et~al.}(2021)\citenamefont
  {{Essick}}, \citenamefont {{Tews}}, \citenamefont {{Landry}},\ and\
  \citenamefont {{Schwenk}}}]{2021PhRvL.127s2701E}%
  \BibitemOpen
  \bibfield  {author} {\bibinfo {author} {\bibfnamefont {R.}~\bibnamefont
  {{Essick}}}, \bibinfo {author} {\bibfnamefont {I.}~\bibnamefont {{Tews}}},
  \bibinfo {author} {\bibfnamefont {P.}~\bibnamefont {{Landry}}}, \ and\
  \bibinfo {author} {\bibfnamefont {A.}~\bibnamefont {{Schwenk}}},\ }\href
  {\doibase 10.1103/PhysRevLett.127.192701} {\bibfield  {journal} {\bibinfo
  {journal} {\prl}\ }\textbf {\bibinfo {volume} {127}},\ \bibinfo {eid}
  {192701} (\bibinfo {year} {2021})},\ \Eprint
  {http://arxiv.org/abs/2102.10074} {arXiv:2102.10074 [nucl-th]} \BibitemShut
  {NoStop}%
\bibitem [{\citenamefont {{Lynch}}\ and\ \citenamefont
  {{Tsang}}(2022)}]{2022PhLB..83037098L}%
  \BibitemOpen
  \bibfield  {author} {\bibinfo {author} {\bibfnamefont {W.~G.}\ \bibnamefont
  {{Lynch}}}\ and\ \bibinfo {author} {\bibfnamefont {M.~B.}\ \bibnamefont
  {{Tsang}}},\ }\href {\doibase 10.1016/j.physletb.2022.137098} {\bibfield
  {journal} {\bibinfo  {journal} {Physics Letters B}\ }\textbf {\bibinfo
  {volume} {830}},\ \bibinfo {eid} {137098} (\bibinfo {year} {2022})},\ \Eprint
  {http://arxiv.org/abs/2106.10119} {arXiv:2106.10119 [nucl-th]} \BibitemShut
  {NoStop}%
\bibitem [{\citenamefont {{Mondal}}\ and\ \citenamefont
  {{Gulminelli}}(2023)}]{2023PhRvC.107a5801M}%
  \BibitemOpen
  \bibfield  {author} {\bibinfo {author} {\bibfnamefont {C.}~\bibnamefont
  {{Mondal}}}\ and\ \bibinfo {author} {\bibfnamefont {F.}~\bibnamefont
  {{Gulminelli}}},\ }\href {\doibase 10.1103/PhysRevC.107.015801} {\bibfield
  {journal} {\bibinfo  {journal} {\prc}\ }\textbf {\bibinfo {volume} {107}},\
  \bibinfo {eid} {015801} (\bibinfo {year} {2023})},\ \Eprint
  {http://arxiv.org/abs/2209.05177} {arXiv:2209.05177 [nucl-th]} \BibitemShut
  {NoStop}%
\bibitem [{\citenamefont {{Salinas}}\ and\ \citenamefont
  {{Piekarewicz}}(2023)}]{2023PhRvC.107d5802S}%
  \BibitemOpen
  \bibfield  {author} {\bibinfo {author} {\bibfnamefont {M.}~\bibnamefont
  {{Salinas}}}\ and\ \bibinfo {author} {\bibfnamefont {J.}~\bibnamefont
  {{Piekarewicz}}},\ }\href {\doibase 10.1103/PhysRevC.107.045802} {\bibfield
  {journal} {\bibinfo  {journal} {\prc}\ }\textbf {\bibinfo {volume} {107}},\
  \bibinfo {eid} {045802} (\bibinfo {year} {2023})},\ \Eprint
  {http://arxiv.org/abs/2301.09692} {arXiv:2301.09692 [nucl-th]} \BibitemShut
  {NoStop}%
\bibitem [{\citenamefont {{Carlson}}\ \emph {et~al.}(2023)\citenamefont
  {{Carlson}}, \citenamefont {{Dutra}}, \citenamefont {{Louren{\c{c}}o}},\ and\
  \citenamefont {{Margueron}}}]{2023PhRvC.107c5805C}%
  \BibitemOpen
  \bibfield  {author} {\bibinfo {author} {\bibfnamefont {B.~V.}\ \bibnamefont
  {{Carlson}}}, \bibinfo {author} {\bibfnamefont {M.}~\bibnamefont {{Dutra}}},
  \bibinfo {author} {\bibfnamefont {O.}~\bibnamefont {{Louren{\c{c}}o}}}, \
  and\ \bibinfo {author} {\bibfnamefont {J.}~\bibnamefont {{Margueron}}},\
  }\href {\doibase 10.1103/PhysRevC.107.035805} {\bibfield  {journal} {\bibinfo
   {journal} {\prc}\ }\textbf {\bibinfo {volume} {107}},\ \bibinfo {eid}
  {035805} (\bibinfo {year} {2023})},\ \Eprint
  {http://arxiv.org/abs/2209.03257} {arXiv:2209.03257 [nucl-th]} \BibitemShut
  {NoStop}%
\bibitem [{\citenamefont {{Miyatsu}}\ \emph {et~al.}(2023)\citenamefont
  {{Miyatsu}}, \citenamefont {{Cheoun}}, \citenamefont {{Kim}},\ and\
  \citenamefont {{Saito}}}]{2023arXiv230314763M}%
  \BibitemOpen
  \bibfield  {author} {\bibinfo {author} {\bibfnamefont {T.}~\bibnamefont
  {{Miyatsu}}}, \bibinfo {author} {\bibfnamefont {M.-K.}\ \bibnamefont
  {{Cheoun}}}, \bibinfo {author} {\bibfnamefont {K.}~\bibnamefont {{Kim}}}, \
  and\ \bibinfo {author} {\bibfnamefont {K.}~\bibnamefont {{Saito}}},\ }\href
  {\doibase 10.48550/arXiv.2303.14763} {\bibfield  {journal} {\bibinfo
  {journal} {arXiv e-prints}\ ,\ \bibinfo {eid} {arXiv:2303.14763}} (\bibinfo
  {year} {2023})},\ \Eprint {http://arxiv.org/abs/2303.14763} {arXiv:2303.14763
  [nucl-th]} \BibitemShut {NoStop}%
\bibitem [{\citenamefont {{Miao}}\ \emph {et~al.}(2020)\citenamefont {{Miao}},
  \citenamefont {{Li}}, \citenamefont {{Zhu}},\ and\ \citenamefont
  {{Han}}}]{2020ApJ...904..103M}%
  \BibitemOpen
  \bibfield  {author} {\bibinfo {author} {\bibfnamefont {Z.}~\bibnamefont
  {{Miao}}}, \bibinfo {author} {\bibfnamefont {A.}~\bibnamefont {{Li}}},
  \bibinfo {author} {\bibfnamefont {Z.}~\bibnamefont {{Zhu}}}, \ and\ \bibinfo
  {author} {\bibfnamefont {S.}~\bibnamefont {{Han}}},\ }\href {\doibase
  10.3847/1538-4357/abbd41} {\bibfield  {journal} {\bibinfo  {journal} {\apj}\
  }\textbf {\bibinfo {volume} {904}},\ \bibinfo {eid} {103} (\bibinfo {year}
  {2020})},\ \Eprint {http://arxiv.org/abs/2006.00839} {arXiv:2006.00839
  [nucl-th]} \BibitemShut {NoStop}%
\bibitem [{\citenamefont {{Li}}\ \emph
  {et~al.}(2021{\natexlab{a}})\citenamefont {{Li}}, \citenamefont {{Miao}},
  \citenamefont {{Han}},\ and\ \citenamefont {{Zhang}}}]{2021ApJ...913...27L}%
  \BibitemOpen
  \bibfield  {author} {\bibinfo {author} {\bibfnamefont {A.}~\bibnamefont
  {{Li}}}, \bibinfo {author} {\bibfnamefont {Z.}~\bibnamefont {{Miao}}},
  \bibinfo {author} {\bibfnamefont {S.}~\bibnamefont {{Han}}}, \ and\ \bibinfo
  {author} {\bibfnamefont {B.}~\bibnamefont {{Zhang}}},\ }\href {\doibase
  10.3847/1538-4357/abf355} {\bibfield  {journal} {\bibinfo  {journal} {\apj}\
  }\textbf {\bibinfo {volume} {913}},\ \bibinfo {eid} {27} (\bibinfo {year}
  {2021}{\natexlab{a}})},\ \Eprint {http://arxiv.org/abs/2103.15119}
  {arXiv:2103.15119 [astro-ph.HE]} \BibitemShut {NoStop}%
\bibitem [{\citenamefont {{Li}}\ \emph
  {et~al.}(2021{\natexlab{b}})\citenamefont {{Li}}, \citenamefont {{Miao}},
  \citenamefont {{Jiang}}, \citenamefont {{Tang}},\ and\ \citenamefont
  {{Xu}}}]{2021MNRAS.506.5916L}%
  \BibitemOpen
  \bibfield  {author} {\bibinfo {author} {\bibfnamefont {A.}~\bibnamefont
  {{Li}}}, \bibinfo {author} {\bibfnamefont {Z.~Q.}\ \bibnamefont {{Miao}}},
  \bibinfo {author} {\bibfnamefont {J.~L.}\ \bibnamefont {{Jiang}}}, \bibinfo
  {author} {\bibfnamefont {S.~P.}\ \bibnamefont {{Tang}}}, \ and\ \bibinfo
  {author} {\bibfnamefont {R.~X.}\ \bibnamefont {{Xu}}},\ }\href {\doibase
  10.1093/mnras/stab2029} {\bibfield  {journal} {\bibinfo  {journal} {\mnras}\
  }\textbf {\bibinfo {volume} {506}},\ \bibinfo {pages} {5916} (\bibinfo {year}
  {2021}{\natexlab{b}})},\ \Eprint {http://arxiv.org/abs/2009.12571}
  {arXiv:2009.12571 [astro-ph.HE]} \BibitemShut {NoStop}%
\bibitem [{\citenamefont {{Miao}}\ \emph
  {et~al.}(2022{\natexlab{a}})\citenamefont {{Miao}}, \citenamefont {{Li}},\
  and\ \citenamefont {{Dai}}}]{2022MNRAS.515.5071M}%
  \BibitemOpen
  \bibfield  {author} {\bibinfo {author} {\bibfnamefont {Z.}~\bibnamefont
  {{Miao}}}, \bibinfo {author} {\bibfnamefont {A.}~\bibnamefont {{Li}}}, \ and\
  \bibinfo {author} {\bibfnamefont {Z.-G.}\ \bibnamefont {{Dai}}},\ }\href
  {\doibase 10.1093/mnras/stac2015} {\bibfield  {journal} {\bibinfo  {journal}
  {\mnras}\ }\textbf {\bibinfo {volume} {515}},\ \bibinfo {pages} {5071}
  (\bibinfo {year} {2022}{\natexlab{a}})},\ \Eprint
  {http://arxiv.org/abs/2107.07979} {arXiv:2107.07979 [astro-ph.HE]}
  \BibitemShut {NoStop}%
\bibitem [{\citenamefont {{Miao}}\ \emph
  {et~al.}(2022{\natexlab{b}})\citenamefont {{Miao}}, \citenamefont {{Zhu}},
  \citenamefont {{Li}},\ and\ \citenamefont {{Huang}}}]{2022ApJ...936...69M}%
  \BibitemOpen
  \bibfield  {author} {\bibinfo {author} {\bibfnamefont {Z.}~\bibnamefont
  {{Miao}}}, \bibinfo {author} {\bibfnamefont {Y.}~\bibnamefont {{Zhu}}},
  \bibinfo {author} {\bibfnamefont {A.}~\bibnamefont {{Li}}}, \ and\ \bibinfo
  {author} {\bibfnamefont {F.}~\bibnamefont {{Huang}}},\ }\href {\doibase
  10.3847/1538-4357/ac8544} {\bibfield  {journal} {\bibinfo  {journal} {\apj}\
  }\textbf {\bibinfo {volume} {936}},\ \bibinfo {eid} {69} (\bibinfo {year}
  {2022}{\natexlab{b}})},\ \Eprint {http://arxiv.org/abs/2204.05560}
  {arXiv:2204.05560 [astro-ph.HE]} \BibitemShut {NoStop}%
\bibitem [{\citenamefont {{Sun}}\ \emph {et~al.}(2023)\citenamefont {{Sun}},
  \citenamefont {{Miao}}, \citenamefont {{Sun}},\ and\ \citenamefont
  {{Li}}}]{2023ApJ...942...55S}%
  \BibitemOpen
  \bibfield  {author} {\bibinfo {author} {\bibfnamefont {X.}~\bibnamefont
  {{Sun}}}, \bibinfo {author} {\bibfnamefont {Z.}~\bibnamefont {{Miao}}},
  \bibinfo {author} {\bibfnamefont {B.}~\bibnamefont {{Sun}}}, \ and\ \bibinfo
  {author} {\bibfnamefont {A.}~\bibnamefont {{Li}}},\ }\href {\doibase
  10.3847/1538-4357/ac9d9a} {\bibfield  {journal} {\bibinfo  {journal} {\apj}\
  }\textbf {\bibinfo {volume} {942}},\ \bibinfo {eid} {55} (\bibinfo {year}
  {2023})},\ \Eprint {http://arxiv.org/abs/2205.10631} {arXiv:2205.10631
  [astro-ph.HE]} \BibitemShut {NoStop}%
\bibitem [{\citenamefont {{Huang}}\ \emph {et~al.}(2022)\citenamefont
  {{Huang}}, \citenamefont {{Hu}}, \citenamefont {{Zhang}},\ and\ \citenamefont
  {{Shen}}}]{2022ApJ...935...88H}%
  \BibitemOpen
  \bibfield  {author} {\bibinfo {author} {\bibfnamefont {K.}~\bibnamefont
  {{Huang}}}, \bibinfo {author} {\bibfnamefont {J.}~\bibnamefont {{Hu}}},
  \bibinfo {author} {\bibfnamefont {Y.}~\bibnamefont {{Zhang}}}, \ and\
  \bibinfo {author} {\bibfnamefont {H.}~\bibnamefont {{Shen}}},\ }\href
  {\doibase 10.3847/1538-4357/ac7f3c} {\bibfield  {journal} {\bibinfo
  {journal} {\apj}\ }\textbf {\bibinfo {volume} {935}},\ \bibinfo {eid} {88}
  (\bibinfo {year} {2022})},\ \Eprint {http://arxiv.org/abs/2206.12760}
  {arXiv:2206.12760 [nucl-th]} \BibitemShut {NoStop}%
\bibitem [{\citenamefont {{Zhou}}\ \emph {et~al.}(2023)\citenamefont {{Zhou}},
  \citenamefont {{Hu}}, \citenamefont {{Zhang}},\ and\ \citenamefont
  {{Shen}}}]{2023arXiv230503323Z}%
  \BibitemOpen
  \bibfield  {author} {\bibinfo {author} {\bibfnamefont {W.}~\bibnamefont
  {{Zhou}}}, \bibinfo {author} {\bibfnamefont {J.}~\bibnamefont {{Hu}}},
  \bibinfo {author} {\bibfnamefont {Y.}~\bibnamefont {{Zhang}}}, \ and\
  \bibinfo {author} {\bibfnamefont {H.}~\bibnamefont {{Shen}}},\ }\href
  {\doibase 10.48550/arXiv.2305.03323} {\bibfield  {journal} {\bibinfo
  {journal} {arXiv e-prints}\ ,\ \bibinfo {eid} {arXiv:2305.03323}} (\bibinfo
  {year} {2023})},\ \Eprint {http://arxiv.org/abs/2305.03323} {arXiv:2305.03323
  [nucl-th]} \BibitemShut {NoStop}%
\bibitem [{\citenamefont {{Miao}}\ \emph {et~al.}(2023)\citenamefont {{Miao}},
  \citenamefont {{Zhou}},\ and\ \citenamefont {{Li}}}]{2023arXiv230508401M}%
  \BibitemOpen
  \bibfield  {author} {\bibinfo {author} {\bibfnamefont {Z.}~\bibnamefont
  {{Miao}}}, \bibinfo {author} {\bibfnamefont {E.}~\bibnamefont {{Zhou}}}, \
  and\ \bibinfo {author} {\bibfnamefont {A.}~\bibnamefont {{Li}}},\ }\href
  {\doibase 10.48550/arXiv.2305.08401} {\bibfield  {journal} {\bibinfo
  {journal} {arXiv e-prints}\ ,\ \bibinfo {eid} {arXiv:2305.08401}} (\bibinfo
  {year} {2023})},\ \Eprint {http://arxiv.org/abs/2305.08401} {arXiv:2305.08401
  [nucl-th]} \BibitemShut {NoStop}%
\bibitem [{\citenamefont {{Nik{\v{s}}i{\'c}}}\ \emph
  {et~al.}(2011)\citenamefont {{Nik{\v{s}}i{\'c}}}, \citenamefont
  {{Vretenar}},\ and\ \citenamefont {{Ring}}}]{2011PrPNP..66..519N}%
  \BibitemOpen
  \bibfield  {author} {\bibinfo {author} {\bibfnamefont {T.}~\bibnamefont
  {{Nik{\v{s}}i{\'c}}}}, \bibinfo {author} {\bibfnamefont {D.}~\bibnamefont
  {{Vretenar}}}, \ and\ \bibinfo {author} {\bibfnamefont {P.}~\bibnamefont
  {{Ring}}},\ }\href {\doibase 10.1016/j.ppnp.2011.01.055} {\bibfield
  {journal} {\bibinfo  {journal} {Progress in Particle and Nuclear Physics}\
  }\textbf {\bibinfo {volume} {66}},\ \bibinfo {pages} {519} (\bibinfo {year}
  {2011})},\ \Eprint {http://arxiv.org/abs/1102.4193} {arXiv:1102.4193
  [nucl-th]} \BibitemShut {NoStop}%
\bibitem [{\citenamefont {{Li}}\ \emph {et~al.}(2008)\citenamefont {{Li}},
  \citenamefont {{Chen}},\ and\ \citenamefont {{Ko}}}]{2008PhR...464..113L}%
  \BibitemOpen
  \bibfield  {author} {\bibinfo {author} {\bibfnamefont {B.-A.}\ \bibnamefont
  {{Li}}}, \bibinfo {author} {\bibfnamefont {L.-W.}\ \bibnamefont {{Chen}}}, \
  and\ \bibinfo {author} {\bibfnamefont {C.~M.}\ \bibnamefont {{Ko}}},\ }\href
  {\doibase 10.1016/j.physrep.2008.04.005} {\bibfield  {journal} {\bibinfo
  {journal} {\physrep}\ }\textbf {\bibinfo {volume} {464}},\ \bibinfo {pages}
  {113} (\bibinfo {year} {2008})},\ \Eprint {http://arxiv.org/abs/0804.3580}
  {arXiv:0804.3580 [nucl-th]} \BibitemShut {NoStop}%
\bibitem [{\citenamefont {{Isgur}}\ and\ \citenamefont
  {{Karl}}(1978)}]{1978PhRvD..18.4187I}%
  \BibitemOpen
  \bibfield  {author} {\bibinfo {author} {\bibfnamefont {N.}~\bibnamefont
  {{Isgur}}}\ and\ \bibinfo {author} {\bibfnamefont {G.}~\bibnamefont
  {{Karl}}},\ }\href {\doibase 10.1103/PhysRevD.18.4187} {\bibfield  {journal}
  {\bibinfo  {journal} {\prd}\ }\textbf {\bibinfo {volume} {18}},\ \bibinfo
  {pages} {4187} (\bibinfo {year} {1978})}\BibitemShut {NoStop}%
\bibitem [{\citenamefont {{Baym}}\ \emph {et~al.}(1971)\citenamefont {{Baym}},
  \citenamefont {{Pethick}},\ and\ \citenamefont
  {{Sutherland}}}]{1971ApJ...170..299B}%
  \BibitemOpen
  \bibfield  {author} {\bibinfo {author} {\bibfnamefont {G.}~\bibnamefont
  {{Baym}}}, \bibinfo {author} {\bibfnamefont {C.}~\bibnamefont {{Pethick}}}, \
  and\ \bibinfo {author} {\bibfnamefont {P.}~\bibnamefont {{Sutherland}}},\
  }\href {\doibase 10.1086/151216} {\bibfield  {journal} {\bibinfo  {journal}
  {Astrophys. J.}\ }\textbf {\bibinfo {volume} {170}},\ \bibinfo {pages} {299}
  (\bibinfo {year} {1971})}\BibitemShut {NoStop}%
\bibitem [{\citenamefont {{Negele}}\ and\ \citenamefont
  {{Vautherin}}(1973)}]{1973NuPhA.207..298N}%
  \BibitemOpen
  \bibfield  {author} {\bibinfo {author} {\bibfnamefont {J.~W.}\ \bibnamefont
  {{Negele}}}\ and\ \bibinfo {author} {\bibfnamefont {D.}~\bibnamefont
  {{Vautherin}}},\ }\href {\doibase 10.1016/0375-9474(73)90349-7} {\bibfield
  {journal} {\bibinfo  {journal} {\nphysa}\ }\textbf {\bibinfo {volume}
  {207}},\ \bibinfo {pages} {298} (\bibinfo {year} {1973})}\BibitemShut
  {NoStop}%
\bibitem [{\citenamefont {{Zhu}}\ \emph {et~al.}(2020)\citenamefont {{Zhu}},
  \citenamefont {{Li}},\ and\ \citenamefont
  {{Rezzolla}}}]{2020PhRvD.102h4058Z}%
  \BibitemOpen
  \bibfield  {author} {\bibinfo {author} {\bibfnamefont {Z.}~\bibnamefont
  {{Zhu}}}, \bibinfo {author} {\bibfnamefont {A.}~\bibnamefont {{Li}}}, \ and\
  \bibinfo {author} {\bibfnamefont {L.}~\bibnamefont {{Rezzolla}}},\ }\href
  {\doibase 10.1103/PhysRevD.102.084058} {\bibfield  {journal} {\bibinfo
  {journal} {\prd}\ }\textbf {\bibinfo {volume} {102}},\ \bibinfo {eid}
  {084058} (\bibinfo {year} {2020})},\ \Eprint
  {http://arxiv.org/abs/2005.02677} {arXiv:2005.02677 [astro-ph.HE]}
  \BibitemShut {NoStop}%
\bibitem [{\citenamefont {{Ashton}}\ \emph {et~al.}(2019)\citenamefont
  {{Ashton}}, \citenamefont {{H{\"u}bner}}, \citenamefont {{Lasky}},
  \citenamefont {{Talbot}}, \citenamefont {{Ackley}}, \citenamefont
  {{Biscoveanu}}, \citenamefont {{Chu}}, \citenamefont {{Divakarla}},
  \citenamefont {{Easter}}, \citenamefont {{Goncharov}}, \citenamefont
  {{Hernandez Vivanco}}, \citenamefont {{Harms}}, \citenamefont {{Lower}},
  \citenamefont {{Meadors}}, \citenamefont {{Melchor}} \emph
  {et~al.}}]{2019ApJS..241...27A}%
  \BibitemOpen
  \bibfield  {author} {\bibinfo {author} {\bibfnamefont {G.}~\bibnamefont
  {{Ashton}}}, \bibinfo {author} {\bibfnamefont {M.}~\bibnamefont
  {{H{\"u}bner}}}, \bibinfo {author} {\bibfnamefont {P.~D.}\ \bibnamefont
  {{Lasky}}}, \bibinfo {author} {\bibfnamefont {C.}~\bibnamefont {{Talbot}}},
  \bibinfo {author} {\bibfnamefont {K.}~\bibnamefont {{Ackley}}}, \bibinfo
  {author} {\bibfnamefont {S.}~\bibnamefont {{Biscoveanu}}}, \bibinfo {author}
  {\bibfnamefont {Q.}~\bibnamefont {{Chu}}}, \bibinfo {author} {\bibfnamefont
  {A.}~\bibnamefont {{Divakarla}}}, \bibinfo {author} {\bibfnamefont {P.~J.}\
  \bibnamefont {{Easter}}}, \bibinfo {author} {\bibfnamefont {B.}~\bibnamefont
  {{Goncharov}}}, \bibinfo {author} {\bibfnamefont {F.}~\bibnamefont
  {{Hernandez Vivanco}}}, \bibinfo {author} {\bibfnamefont {J.}~\bibnamefont
  {{Harms}}}, \bibinfo {author} {\bibfnamefont {M.~E.}\ \bibnamefont
  {{Lower}}}, \bibinfo {author} {\bibfnamefont {G.~D.}\ \bibnamefont
  {{Meadors}}}, \bibinfo {author} {\bibfnamefont {D.}~\bibnamefont
  {{Melchor}}},  \emph {et~al.},\ }\href {\doibase 10.3847/1538-4365/ab06fc}
  {\bibfield  {journal} {\bibinfo  {journal} {\apjs}\ }\textbf {\bibinfo
  {volume} {241}},\ \bibinfo {eid} {27} (\bibinfo {year} {2019})},\ \Eprint
  {http://arxiv.org/abs/1811.02042} {arXiv:1811.02042 [astro-ph.IM]}
  \BibitemShut {NoStop}%
\bibitem [{\citenamefont {{Romero-Shaw}}\ \emph {et~al.}(2020)\citenamefont
  {{Romero-Shaw}}, \citenamefont {{Talbot}}, \citenamefont {{Biscoveanu}},
  \citenamefont {{D'Emilio}}, \citenamefont {{Ashton}}, \citenamefont
  {{Berry}}, \citenamefont {{Coughlin}}, \citenamefont {{Galaudage}},
  \citenamefont {{Hoy}}, \citenamefont {{H{\"u}bner}}, \citenamefont
  {{Phukon}}, \citenamefont {{Pitkin}}, \citenamefont {{Rizzo}}, \citenamefont
  {{Sarin}}, \citenamefont {{Smith}} \emph {et~al.}}]{2020MNRAS.499.3295R}%
  \BibitemOpen
  \bibfield  {author} {\bibinfo {author} {\bibfnamefont {I.~M.}\ \bibnamefont
  {{Romero-Shaw}}}, \bibinfo {author} {\bibfnamefont {C.}~\bibnamefont
  {{Talbot}}}, \bibinfo {author} {\bibfnamefont {S.}~\bibnamefont
  {{Biscoveanu}}}, \bibinfo {author} {\bibfnamefont {V.}~\bibnamefont
  {{D'Emilio}}}, \bibinfo {author} {\bibfnamefont {G.}~\bibnamefont
  {{Ashton}}}, \bibinfo {author} {\bibfnamefont {C.~P.~L.}\ \bibnamefont
  {{Berry}}}, \bibinfo {author} {\bibfnamefont {S.}~\bibnamefont {{Coughlin}}},
  \bibinfo {author} {\bibfnamefont {S.}~\bibnamefont {{Galaudage}}}, \bibinfo
  {author} {\bibfnamefont {C.}~\bibnamefont {{Hoy}}}, \bibinfo {author}
  {\bibfnamefont {M.}~\bibnamefont {{H{\"u}bner}}}, \bibinfo {author}
  {\bibfnamefont {K.~S.}\ \bibnamefont {{Phukon}}}, \bibinfo {author}
  {\bibfnamefont {M.}~\bibnamefont {{Pitkin}}}, \bibinfo {author}
  {\bibfnamefont {M.}~\bibnamefont {{Rizzo}}}, \bibinfo {author} {\bibfnamefont
  {N.}~\bibnamefont {{Sarin}}}, \bibinfo {author} {\bibfnamefont
  {R.}~\bibnamefont {{Smith}}},  \emph {et~al.},\ }\href {\doibase
  10.1093/mnras/staa2850} {\bibfield  {journal} {\bibinfo  {journal} {\mnras}\
  }\textbf {\bibinfo {volume} {499}},\ \bibinfo {pages} {3295} (\bibinfo {year}
  {2020})},\ \Eprint {http://arxiv.org/abs/2006.00714} {arXiv:2006.00714
  [astro-ph.IM]} \BibitemShut {NoStop}%
\bibitem [{\citenamefont {{Buchner}}\ \emph {et~al.}(2014)\citenamefont
  {{Buchner}}, \citenamefont {{Georgakakis}}, \citenamefont {{Nandra}},
  \citenamefont {{Hsu}}, \citenamefont {{Rangel}}, \citenamefont {{Brightman}},
  \citenamefont {{Merloni}}, \citenamefont {{Salvato}}, \citenamefont
  {{Donley}},\ and\ \citenamefont {{Kocevski}}}]{2014A&A...564A.125B}%
  \BibitemOpen
  \bibfield  {author} {\bibinfo {author} {\bibfnamefont {J.}~\bibnamefont
  {{Buchner}}}, \bibinfo {author} {\bibfnamefont {A.}~\bibnamefont
  {{Georgakakis}}}, \bibinfo {author} {\bibfnamefont {K.}~\bibnamefont
  {{Nandra}}}, \bibinfo {author} {\bibfnamefont {L.}~\bibnamefont {{Hsu}}},
  \bibinfo {author} {\bibfnamefont {C.}~\bibnamefont {{Rangel}}}, \bibinfo
  {author} {\bibfnamefont {M.}~\bibnamefont {{Brightman}}}, \bibinfo {author}
  {\bibfnamefont {A.}~\bibnamefont {{Merloni}}}, \bibinfo {author}
  {\bibfnamefont {M.}~\bibnamefont {{Salvato}}}, \bibinfo {author}
  {\bibfnamefont {J.}~\bibnamefont {{Donley}}}, \ and\ \bibinfo {author}
  {\bibfnamefont {D.}~\bibnamefont {{Kocevski}}},\ }\href {\doibase
  10.1051/0004-6361/201322971} {\bibfield  {journal} {\bibinfo  {journal}
  {\aap}\ }\textbf {\bibinfo {volume} {564}},\ \bibinfo {eid} {A125} (\bibinfo
  {year} {2014})},\ \Eprint {http://arxiv.org/abs/1402.0004} {arXiv:1402.0004
  [astro-ph.HE]} \BibitemShut {NoStop}%
\bibitem [{\citenamefont {{Hernandez Vivanco}}\ \emph
  {et~al.}(2020)\citenamefont {{Hernandez Vivanco}}, \citenamefont {{Smith}},
  \citenamefont {{Thrane}},\ and\ \citenamefont
  {{Lasky}}}]{2020MNRAS.499.5972H}%
  \BibitemOpen
  \bibfield  {author} {\bibinfo {author} {\bibfnamefont {F.}~\bibnamefont
  {{Hernandez Vivanco}}}, \bibinfo {author} {\bibfnamefont {R.}~\bibnamefont
  {{Smith}}}, \bibinfo {author} {\bibfnamefont {E.}~\bibnamefont {{Thrane}}}, \
  and\ \bibinfo {author} {\bibfnamefont {P.~D.}\ \bibnamefont {{Lasky}}},\
  }\href {\doibase 10.1093/mnras/staa3243} {\bibfield  {journal} {\bibinfo
  {journal} {\mnras}\ }\textbf {\bibinfo {volume} {499}},\ \bibinfo {pages}
  {5972} (\bibinfo {year} {2020})},\ \Eprint {http://arxiv.org/abs/2008.05627}
  {arXiv:2008.05627 [astro-ph.HE]} \BibitemShut {NoStop}%
\bibitem [{\citenamefont {{Riley}}\ \emph
  {et~al.}(2019{\natexlab{b}})\citenamefont {{Riley}}, \citenamefont {{Watts}},
  \citenamefont {{Bogdanov}}, \citenamefont {{Ray}}, \citenamefont {{Ludlam}},
  \citenamefont {{Guillot}}, \citenamefont {{Arzoumanian}}, \citenamefont
  {{Baker}}, \citenamefont {{Bilous}}, \citenamefont {{Chakrabarty}},
  \citenamefont {{Gendreau}}, \citenamefont {{Harding}}, \citenamefont {{Ho}},
  \citenamefont {{Lattimer}}, \citenamefont {{Morsink}},\ and\ \citenamefont
  {{Strohmayer}}}]{Riley2019b}%
  \BibitemOpen
  \bibfield  {author} {\bibinfo {author} {\bibfnamefont {T.~E.}\ \bibnamefont
  {{Riley}}}, \bibinfo {author} {\bibfnamefont {A.~L.}\ \bibnamefont
  {{Watts}}}, \bibinfo {author} {\bibfnamefont {S.}~\bibnamefont {{Bogdanov}}},
  \bibinfo {author} {\bibfnamefont {P.~S.}\ \bibnamefont {{Ray}}}, \bibinfo
  {author} {\bibfnamefont {R.~M.}\ \bibnamefont {{Ludlam}}}, \bibinfo {author}
  {\bibfnamefont {S.}~\bibnamefont {{Guillot}}}, \bibinfo {author}
  {\bibfnamefont {Z.}~\bibnamefont {{Arzoumanian}}}, \bibinfo {author}
  {\bibfnamefont {C.~L.}\ \bibnamefont {{Baker}}}, \bibinfo {author}
  {\bibfnamefont {A.~V.}\ \bibnamefont {{Bilous}}}, \bibinfo {author}
  {\bibfnamefont {D.}~\bibnamefont {{Chakrabarty}}}, \bibinfo {author}
  {\bibfnamefont {K.~C.}\ \bibnamefont {{Gendreau}}}, \bibinfo {author}
  {\bibfnamefont {A.~K.}\ \bibnamefont {{Harding}}}, \bibinfo {author}
  {\bibfnamefont {W.~C.~G.}\ \bibnamefont {{Ho}}}, \bibinfo {author}
  {\bibfnamefont {J.~M.}\ \bibnamefont {{Lattimer}}}, \bibinfo {author}
  {\bibfnamefont {S.~M.}\ \bibnamefont {{Morsink}}}, \ and\ \bibinfo {author}
  {\bibfnamefont {T.~E.}\ \bibnamefont {{Strohmayer}}},\ }\href {\doibase
  10.5281/zenodo.3386449} {\enquote {\bibinfo {title} {{A NICER View of PSR
  J0030+0451: Nested Samples for Millisecond Pulsar Parameter Estimation}},}\ }
  (\bibinfo {year} {2019}{\natexlab{b}})\BibitemShut {NoStop}%
\bibitem [{\citenamefont {{Riley}}\ \emph
  {et~al.}(2021{\natexlab{b}})\citenamefont {{Riley}}, \citenamefont {{Watts}},
  \citenamefont {{Ray}}, \citenamefont {{Bogdanov}}, \citenamefont {{Guillot}},
  \citenamefont {{Morsink}}, \citenamefont {{Bilous}}, \citenamefont
  {{Arzoumanian}}, \citenamefont {{Choudhury}}, \citenamefont {{Deneva}},
  \citenamefont {{Gendreau}}, \citenamefont {{Harding}}, \citenamefont {{Ho}},
  \citenamefont {{Lattimer}}, \citenamefont {{Loewenstein}}, \citenamefont
  {{Ludlam}}, \citenamefont {{Markwardt}}, \citenamefont {{Okajima}},
  \citenamefont {{Prescod-Weinstein}}, \citenamefont {{Remillard}},
  \citenamefont {{Wolff}}, \citenamefont {{Fonseca}}, \citenamefont
  {{Cromartie}}, \citenamefont {{Kerr}}, \citenamefont {{Pennucci}},
  \citenamefont {{Parthasarathy}}, \citenamefont {{Ransom}}, \citenamefont
  {{Stairs}}, \citenamefont {{Guillemot}},\ and\ \citenamefont
  {{Cognard}}}]{Riley2021b}%
  \BibitemOpen
  \bibfield  {author} {\bibinfo {author} {\bibfnamefont {T.~E.}\ \bibnamefont
  {{Riley}}}, \bibinfo {author} {\bibfnamefont {A.~L.}\ \bibnamefont
  {{Watts}}}, \bibinfo {author} {\bibfnamefont {P.~S.}\ \bibnamefont {{Ray}}},
  \bibinfo {author} {\bibfnamefont {S.}~\bibnamefont {{Bogdanov}}}, \bibinfo
  {author} {\bibfnamefont {S.}~\bibnamefont {{Guillot}}}, \bibinfo {author}
  {\bibfnamefont {S.~M.}\ \bibnamefont {{Morsink}}}, \bibinfo {author}
  {\bibfnamefont {A.~V.}\ \bibnamefont {{Bilous}}}, \bibinfo {author}
  {\bibfnamefont {Z.}~\bibnamefont {{Arzoumanian}}}, \bibinfo {author}
  {\bibfnamefont {D.}~\bibnamefont {{Choudhury}}}, \bibinfo {author}
  {\bibfnamefont {J.~S.}\ \bibnamefont {{Deneva}}}, \bibinfo {author}
  {\bibfnamefont {K.~C.}\ \bibnamefont {{Gendreau}}}, \bibinfo {author}
  {\bibfnamefont {A.~K.}\ \bibnamefont {{Harding}}}, \bibinfo {author}
  {\bibfnamefont {W.~C.~G.}\ \bibnamefont {{Ho}}}, \bibinfo {author}
  {\bibfnamefont {J.~M.}\ \bibnamefont {{Lattimer}}}, \bibinfo {author}
  {\bibfnamefont {M.}~\bibnamefont {{Loewenstein}}}, \bibinfo {author}
  {\bibfnamefont {R.~M.}\ \bibnamefont {{Ludlam}}}, \bibinfo {author}
  {\bibfnamefont {C.~B.}\ \bibnamefont {{Markwardt}}}, \bibinfo {author}
  {\bibfnamefont {T.}~\bibnamefont {{Okajima}}}, \bibinfo {author}
  {\bibfnamefont {C.}~\bibnamefont {{Prescod-Weinstein}}}, \bibinfo {author}
  {\bibfnamefont {R.~A.}\ \bibnamefont {{Remillard}}}, \bibinfo {author}
  {\bibfnamefont {M.~T.}\ \bibnamefont {{Wolff}}}, \bibinfo {author}
  {\bibfnamefont {E.}~\bibnamefont {{Fonseca}}}, \bibinfo {author}
  {\bibfnamefont {H.~T.}\ \bibnamefont {{Cromartie}}}, \bibinfo {author}
  {\bibfnamefont {M.}~\bibnamefont {{Kerr}}}, \bibinfo {author} {\bibfnamefont
  {T.~T.}\ \bibnamefont {{Pennucci}}}, \bibinfo {author} {\bibfnamefont
  {A.}~\bibnamefont {{Parthasarathy}}}, \bibinfo {author} {\bibfnamefont
  {S.}~\bibnamefont {{Ransom}}}, \bibinfo {author} {\bibfnamefont
  {I.}~\bibnamefont {{Stairs}}}, \bibinfo {author} {\bibfnamefont
  {L.}~\bibnamefont {{Guillemot}}}, \ and\ \bibinfo {author} {\bibfnamefont
  {I.}~\bibnamefont {{Cognard}}},\ }\href {\doibase 10.5281/zenodo.4697625}
  {\enquote {\bibinfo {title} {{A NICER View of the Massive Pulsar PSR
  J0740+6620 Informed by Radio Timing and XMM-Newton Spectroscopy: Nested
  Samples for Millisecond Pulsar Parameter Estimation}},}\ } (\bibinfo {year}
  {2021}{\natexlab{b}})\BibitemShut {NoStop}%
\bibitem [{\citenamefont {{Zhang}}\ and\ \citenamefont
  {{Chen}}(2022)}]{2022arXiv220703328Z}%
  \BibitemOpen
  \bibfield  {author} {\bibinfo {author} {\bibfnamefont {Z.}~\bibnamefont
  {{Zhang}}}\ and\ \bibinfo {author} {\bibfnamefont {L.-W.}\ \bibnamefont
  {{Chen}}},\ }\href {\doibase 10.48550/arXiv.2207.03328} {\bibfield  {journal}
  {\bibinfo  {journal} {arXiv e-prints}\ ,\ \bibinfo {eid} {arXiv:2207.03328}}
  (\bibinfo {year} {2022})},\ \Eprint {http://arxiv.org/abs/2207.03328}
  {arXiv:2207.03328 [nucl-th]} \BibitemShut {NoStop}%
\bibitem [{\citenamefont {{Danielewicz}}\ \emph {et~al.}(2002)\citenamefont
  {{Danielewicz}}, \citenamefont {{Lacey}},\ and\ \citenamefont
  {{Lynch}}}]{2002Sci...298.1592D}%
  \BibitemOpen
  \bibfield  {author} {\bibinfo {author} {\bibfnamefont {P.}~\bibnamefont
  {{Danielewicz}}}, \bibinfo {author} {\bibfnamefont {R.}~\bibnamefont
  {{Lacey}}}, \ and\ \bibinfo {author} {\bibfnamefont {W.~G.}\ \bibnamefont
  {{Lynch}}},\ }\href {\doibase 10.1126/science.1078070} {\bibfield  {journal}
  {\bibinfo  {journal} {Science}\ }\textbf {\bibinfo {volume} {298}},\ \bibinfo
  {pages} {1592} (\bibinfo {year} {2002})},\ \Eprint
  {http://arxiv.org/abs/nucl-th/0208016} {arXiv:nucl-th/0208016 [nucl-th]}
  \BibitemShut {NoStop}%
\bibitem [{\citenamefont {{Fuchs}}(2006)}]{2006PrPNP..56....1F}%
  \BibitemOpen
  \bibfield  {author} {\bibinfo {author} {\bibfnamefont {C.}~\bibnamefont
  {{Fuchs}}},\ }\href {\doibase 10.1016/j.ppnp.2005.07.004} {\bibfield
  {journal} {\bibinfo  {journal} {Progress in Particle and Nuclear Physics}\
  }\textbf {\bibinfo {volume} {56}},\ \bibinfo {pages} {1} (\bibinfo {year}
  {2006})},\ \Eprint {http://arxiv.org/abs/nucl-th/0507017}
  {arXiv:nucl-th/0507017 [nucl-th]} \BibitemShut {NoStop}%
\bibitem [{\citenamefont {{Liu}}\ \emph {et~al.}(2021)\citenamefont {{Liu}},
  \citenamefont {{Wang}}, \citenamefont {{Cui}}, \citenamefont {{Xia}},
  \citenamefont {{Li}}, \citenamefont {{Chen}}, \citenamefont {{Li}},\ and\
  \citenamefont {{Zhang}}}]{2021PhRvC.103a4616L}%
  \BibitemOpen
  \bibfield  {author} {\bibinfo {author} {\bibfnamefont {Y.}~\bibnamefont
  {{Liu}}}, \bibinfo {author} {\bibfnamefont {Y.}~\bibnamefont {{Wang}}},
  \bibinfo {author} {\bibfnamefont {Y.}~\bibnamefont {{Cui}}}, \bibinfo
  {author} {\bibfnamefont {C.-J.}\ \bibnamefont {{Xia}}}, \bibinfo {author}
  {\bibfnamefont {Z.}~\bibnamefont {{Li}}}, \bibinfo {author} {\bibfnamefont
  {Y.}~\bibnamefont {{Chen}}}, \bibinfo {author} {\bibfnamefont
  {Q.}~\bibnamefont {{Li}}}, \ and\ \bibinfo {author} {\bibfnamefont
  {Y.}~\bibnamefont {{Zhang}}},\ }\href {\doibase 10.1103/PhysRevC.103.014616}
  {\bibfield  {journal} {\bibinfo  {journal} {\prc}\ }\textbf {\bibinfo
  {volume} {103}},\ \bibinfo {eid} {014616} (\bibinfo {year} {2021})},\ \Eprint
  {http://arxiv.org/abs/2006.15861} {arXiv:2006.15861 [nucl-th]} \BibitemShut
  {NoStop}%
\bibitem [{\citenamefont {{Russotto}}\ \emph {et~al.}(2016)\citenamefont
  {{Russotto}}, \citenamefont {{Gannon}}, \citenamefont {{Kupny}},
  \citenamefont {{Lasko}}, \citenamefont {{Acosta}}, \citenamefont
  {{Adamczyk}}, \citenamefont {{Al-Ajlan}}, \citenamefont {{Al-Garawi}},
  \citenamefont {{Al-Homaidhi}}, \citenamefont {{Amorini}},\ and\ \citenamefont
  {et~al.}}]{2016PhRvC..94c4608R_etal}%
  \BibitemOpen
  \bibfield  {author} {\bibinfo {author} {\bibfnamefont {P.}~\bibnamefont
  {{Russotto}}}, \bibinfo {author} {\bibfnamefont {S.}~\bibnamefont
  {{Gannon}}}, \bibinfo {author} {\bibfnamefont {S.}~\bibnamefont {{Kupny}}},
  \bibinfo {author} {\bibfnamefont {P.}~\bibnamefont {{Lasko}}}, \bibinfo
  {author} {\bibfnamefont {L.}~\bibnamefont {{Acosta}}}, \bibinfo {author}
  {\bibfnamefont {M.}~\bibnamefont {{Adamczyk}}}, \bibinfo {author}
  {\bibfnamefont {A.}~\bibnamefont {{Al-Ajlan}}}, \bibinfo {author}
  {\bibfnamefont {M.}~\bibnamefont {{Al-Garawi}}}, \bibinfo {author}
  {\bibfnamefont {S.}~\bibnamefont {{Al-Homaidhi}}}, \bibinfo {author}
  {\bibfnamefont {F.}~\bibnamefont {{Amorini}}}, \ and\ \bibinfo {author}
  {\bibnamefont {et~al.}},\ }\href {\doibase 10.1103/PhysRevC.94.034608}
  {\bibfield  {journal} {\bibinfo  {journal} {\prc}\ }\textbf {\bibinfo
  {volume} {94}},\ \bibinfo {eid} {034608} (\bibinfo {year} {2016})},\ \Eprint
  {http://arxiv.org/abs/1608.04332} {arXiv:1608.04332 [nucl-ex]} \BibitemShut
  {NoStop}%
\bibitem [{\citenamefont {{Tsang}}\ \emph {et~al.}(2009)\citenamefont
  {{Tsang}}, \citenamefont {{Zhang}}, \citenamefont {{Danielewicz}},
  \citenamefont {{Famiano}}, \citenamefont {{Li}}, \citenamefont {{Lynch}},\
  and\ \citenamefont {{Steiner}}}]{2009PhRvL.102l2701T}%
  \BibitemOpen
  \bibfield  {author} {\bibinfo {author} {\bibfnamefont {M.~B.}\ \bibnamefont
  {{Tsang}}}, \bibinfo {author} {\bibfnamefont {Y.}~\bibnamefont {{Zhang}}},
  \bibinfo {author} {\bibfnamefont {P.}~\bibnamefont {{Danielewicz}}}, \bibinfo
  {author} {\bibfnamefont {M.}~\bibnamefont {{Famiano}}}, \bibinfo {author}
  {\bibfnamefont {Z.}~\bibnamefont {{Li}}}, \bibinfo {author} {\bibfnamefont
  {W.~G.}\ \bibnamefont {{Lynch}}}, \ and\ \bibinfo {author} {\bibfnamefont
  {A.~W.}\ \bibnamefont {{Steiner}}},\ }\href {\doibase
  10.1103/PhysRevLett.102.122701} {\bibfield  {journal} {\bibinfo  {journal}
  {\prl}\ }\textbf {\bibinfo {volume} {102}},\ \bibinfo {eid} {122701}
  (\bibinfo {year} {2009})},\ \Eprint {http://arxiv.org/abs/0811.3107}
  {arXiv:0811.3107 [nucl-ex]} \BibitemShut {NoStop}%
\bibitem [{\citenamefont {{Danielewicz}}\ and\ \citenamefont
  {{Lee}}(2014)}]{2014NuPhA.922....1D}%
  \BibitemOpen
  \bibfield  {author} {\bibinfo {author} {\bibfnamefont {P.}~\bibnamefont
  {{Danielewicz}}}\ and\ \bibinfo {author} {\bibfnamefont {J.}~\bibnamefont
  {{Lee}}},\ }\href {\doibase 10.1016/j.nuclphysa.2013.11.005} {\bibfield
  {journal} {\bibinfo  {journal} {\nphysa}\ }\textbf {\bibinfo {volume}
  {922}},\ \bibinfo {pages} {1} (\bibinfo {year} {2014})},\ \Eprint
  {http://arxiv.org/abs/1307.4130} {arXiv:1307.4130 [nucl-th]} \BibitemShut
  {NoStop}%
\bibitem [{\citenamefont {{Li}}\ and\ \citenamefont
  {{Han}}(2013)}]{2013PhLB..727..276L}%
  \BibitemOpen
  \bibfield  {author} {\bibinfo {author} {\bibfnamefont {B.-A.}\ \bibnamefont
  {{Li}}}\ and\ \bibinfo {author} {\bibfnamefont {X.}~\bibnamefont {{Han}}},\
  }\href {\doibase 10.1016/j.physletb.2013.10.006} {\bibfield  {journal}
  {\bibinfo  {journal} {Physics Letters B}\ }\textbf {\bibinfo {volume}
  {727}},\ \bibinfo {pages} {276} (\bibinfo {year} {2013})},\ \Eprint
  {http://arxiv.org/abs/1304.3368} {arXiv:1304.3368 [nucl-th]} \BibitemShut
  {NoStop}%
\bibitem [{\citenamefont {{Steiner}}\ \emph {et~al.}(2013)\citenamefont
  {{Steiner}}, \citenamefont {{Hempel}},\ and\ \citenamefont
  {{Fischer}}}]{2013ApJ...774...17S}%
  \BibitemOpen
  \bibfield  {author} {\bibinfo {author} {\bibfnamefont {A.~W.}\ \bibnamefont
  {{Steiner}}}, \bibinfo {author} {\bibfnamefont {M.}~\bibnamefont {{Hempel}}},
  \ and\ \bibinfo {author} {\bibfnamefont {T.}~\bibnamefont {{Fischer}}},\
  }\href {\doibase 10.1088/0004-637X/774/1/17} {\bibfield  {journal} {\bibinfo
  {journal} {\apj}\ }\textbf {\bibinfo {volume} {774}},\ \bibinfo {eid} {17}
  (\bibinfo {year} {2013})},\ \Eprint {http://arxiv.org/abs/1207.2184}
  {arXiv:1207.2184 [astro-ph.SR]} \BibitemShut {NoStop}%
\bibitem [{\citenamefont {{Hornick}}\ \emph {et~al.}(2018)\citenamefont
  {{Hornick}}, \citenamefont {{Tolos}}, \citenamefont {{Zacchi}}, \citenamefont
  {{Christian}},\ and\ \citenamefont
  {{Schaffner-Bielich}}}]{2018PhRvC..98f5804H}%
  \BibitemOpen
  \bibfield  {author} {\bibinfo {author} {\bibfnamefont {N.}~\bibnamefont
  {{Hornick}}}, \bibinfo {author} {\bibfnamefont {L.}~\bibnamefont {{Tolos}}},
  \bibinfo {author} {\bibfnamefont {A.}~\bibnamefont {{Zacchi}}}, \bibinfo
  {author} {\bibfnamefont {J.-E.}\ \bibnamefont {{Christian}}}, \ and\ \bibinfo
  {author} {\bibfnamefont {J.}~\bibnamefont {{Schaffner-Bielich}}},\ }\href
  {\doibase 10.1103/PhysRevC.98.065804} {\bibfield  {journal} {\bibinfo
  {journal} {\prc}\ }\textbf {\bibinfo {volume} {98}},\ \bibinfo {eid} {065804}
  (\bibinfo {year} {2018})},\ \Eprint {http://arxiv.org/abs/1808.06808}
  {arXiv:1808.06808 [astro-ph.HE]} \BibitemShut {NoStop}%
\end{thebibliography}%

\end{document}